\documentclass{article} 
\usepackage{iclr2024_conference,times}

\usepackage{amsmath,amsfonts,bm}

\def\eqref#1{equation~\ref{#1}}

\def\1{\bm{1}}

\def\mH{{\bm{H}}}

\def\mS{{\bm{S}}}

\DeclareMathAlphabet{\mathsfit}{\encodingdefault}{\sfdefault}{m}{sl}
\SetMathAlphabet{\mathsfit}{bold}{\encodingdefault}{\sfdefault}{bx}{n}

\newcommand{\R}{\mathbb{R}}

\usepackage{graphicx}
\usepackage{hyperref}
\usepackage{url}
\usepackage{booktabs}
\usepackage{multicol}
\usepackage{multirow}
\usepackage{wrapfig}
\usepackage{tikz}
\usepackage{caption}
\usepackage{subcaption}
\usepackage{color, colortbl}
\usepackage{xcolor}
\usepackage{arrayjob}
\pgfkeys{/pgf/number format/.cd, fixed, precision=2}
\colorlet{mono}{blue!10}
\colorlet{regional}{teal!10}
\colorlet{multi}{yellow!10}
\colorlet{euro}{orange!10}

\title{Multi-resolution HuBERT: Multi-resolution Speech Self-Supervised Learning with Masked Unit Prediction}

\iclrfinalcopy

\author{Jiatong Shi$^{1}$\thanks{The work was conducted by Jiatong Shi during his summer internship at Meta.} , Hirofumi Inaguma$^{2}$, Xutai Ma$^{2}$, Ilia Kulikov$^{2}$, Anna Sun$^{2}$ \\
$^1$ Language Technologies Institute, Carnegie Mellon University; $^2$  Meta AI \\
\texttt{jiatongs@cs.cmu.edu} \\
\texttt{\{hirofumii, xutaima, kulikov, annaysun\}@meta.com}
}

\begin{document}

\maketitle

\begin{abstract}

Existing Self-Supervised Learning (SSL) models for speech typically process speech signals at a fixed resolution of 20 milliseconds. This approach overlooks the varying informational content present at different resolutions in speech signals. In contrast, this paper aims to incorporate multi-resolution information into speech self-supervised representation learning. We introduce a SSL model that leverages a hierarchical Transformer architecture, complemented by HuBERT-style masked prediction objectives, to process speech at multiple resolutions. Experimental results indicate that the proposed model not only achieves more efficient inference but also exhibits superior or comparable performance to the original HuBERT model over various tasks. Specifically, significant performance improvements over the original HuBERT have been observed in fine-tuning experiments on the LibriSpeech speech recognition benchmark as well as in evaluations using the Speech Universal PERformance Benchmark (SUPERB) and Multilingual SUPERB (ML-SUPERB).
\end{abstract}

\section{Introduction}
\label{sec: intro}

In physics, speech is defined as a vibration that propagates as an acoustic wave through a transmission medium \citep{fitz2007fundamentals}. In the field of speech processing, speech signals are stored using techniques such as sampling and quantization. This results in a discretized abstraction of the original waveform, in both time and amplitude \citep{roberts1987digital}.

In practical real-world scenarios, the sampling rate for speech signals can vary between 8 kHz and 48 kHz. High sampling rates can pose challenges for processing due to complications in analyzing long sequences. 
Typically, speech signals exhibit short-term stationarity within intervals ranging from 10 to 30ms \citep{zhu2000use}. Taking these factors into account, past research has recommended frame-wise processing of speech signals, with frames being extracted over localized sample points \citep{huang2001spoken}. Traditional spectral feature extraction methods, often based on psychoacoustics, utilize short-term Fourier transform over windows ranging from 20 to 40ms, with shifts between 10 and 30ms \citep{huang2001spoken, davis1980comparison, hermansky1990perceptual}.

While these conventional spectral features exhibit properties that align well with human psychoacoustics, speech processing systems relying on these features require large volumes of transcribed audio data to achieve high performance \citep{yu2016automatic}. In contrast, Self-Supervised Learning (SSL) speech models utilize unlabeled speech data to generate contextualized speech representations \citep{oord2018representation, liu2020mockingjay, baevski2020wav2vec, hsu2021hubert, chung2021w2v, chiu2022self, chen2022wavlm}. These SSL models have shown superior capabilities in contextualizing speech, achieving state-of-the-art results on various benchmarks and challenges \citep{panayotov2015librispeech, superb, evain21_interspeech, mohamed2022self, mlsuperb, agrawal2023findings}. Moreover, they demonstrate excellent generalizability to low-resource tasks \citep{baevski2020wav2vec, hsu2021hubert, berrebbi22_interspeech, zhao2022improving}. Despite these advancements, existing speech SSL models predominantly follow a similar approach when it comes to processing speech signals. They typically extract speech frames of 20ms as their fundamental units for pre-training \citep{baevski2020wav2vec, hsu2021hubert, chung2021w2v, chiu2022self, chen2022wavlm}. This extraction can be accomplished using either a convolutional feature extractor \citep{baevski2020wav2vec, hsu2021hubert, chen2022wavlm} or traditional features like Mel filter banks \citep{lin2022melhubert, barrault2023seamlessm4t}.

Notably, this uniform frame size of 20ms may not be universally optimal across different downstream tasks. In line with conventional spectral features, existing literature suggests that multi-resolution modeling could enhance performance in various speech processing tasks, such as Automatic Speech Recognition (ASR) \citep{mallidi2016novel, mallidi2018practical, hermansky2013multistream, han2021multistream, luo2021multi, li2019multi, andrusenko2023uconv, kimsqueezeformer, burchi2021efficient}, Speaker Verification (SV) \citep{gao2022unet}, Speech Enhancement (SE) \citep{zhao2021unet++, zhang2019research}, and Voice Conversion (VC) \citep{li2022unet}. Supporting this notion, recent work by \citet{shi23h_interspeech} demonstrated the advantages of multi-resolution training by using three separate SSL models. Their findings indicate that combining these models focusing on different representations can yield superior results across various tasks, whether used in fine-tuning or as frozen feature extractors. However, the method needs to train different SSL models for each resolution, resulting in a huge computation burden from pre-training.

Despite existing efforts to utilize SSL models for speech at multiple resolutions, no work has explicitly addressed the integration of multi-resolution information during the pre-training phase. This study aims to fill that gap by focusing on multi-resolution pre-training for speech representation. We introduce a novel hierarchical framework, namely multi-resolution HuBERT (MR-HuBERT) designed to encode speech information across multiple resolutions in a single model. The model is pre-trained using objectives for multi-resolution masked unit prediction, which are integrated with HuBERT-style clustering units \citep{hsu2021hubert}. Our model shows substantial performance improvements over baseline SSL models across a variety of benchmarks. These include different subsets of the LibriSpeech dataset, the Speech Universal PERformance Benchmark (SUPERB), and the Multilingual SUPERB (ML-SUPERB) \citep{panayotov2015librispeech, superb, mlsuperb}. Another of the key advantages of our approach is efficiency; the reduced sequence length resulting from multi-resolution processing enables faster inference to 9-13\% computation reduction. We have made the implementation of MR-HuBERT, along with the pre-trained models, available as open-source resources on Fairseq and S3PRL~\citep{ott2019fairseq, superb}.\footnote{\scriptsize{Fairseq: \url{https://github.com/facebookresearch/fairseq/tree/main/examples/mr\_hubert}; S3PRL: \url{https://s3prl.github.io/s3prl/tutorial/upstream\_collection.html\#multiresolution-hubert-mr-hubert}.}}

\section{Background}
\label{sec: related works}


Self-supervised learning has achieved remarkable success in a wide array of domains, such as computer vision and natural language processing. As detailed in Section~\ref{sec: intro}, similar advancements have been made in the speech processing community. According to the classification scheme by \citet{mohamed2022self}, current speech SSL models can be categorized into generative, contrastive, and predictive approaches. Among these, predictive models have shown particularly promising results in recent benchmarks for SSL representation \citep{superb, feng2023superb, wang2021fine, masuyama2023end, hsu2021hubert, chen2022wavlm}.

As introduced in Section~\ref{sec: intro}, speech SSL models can be applied to various downstream tasks through either fine-tuning or as frozen feature extractors. The architecture of the downstream models can vary widely, including a simple linear probing layer, recurrent neural network layers, Transformer layers, or more complex encoder-decoder frameworks \citep{baevski2020wav2vec, hsu2021hubert, chung2021w2v, superb, chang2021exploration,  mlsuperb, inaguma-etal-2023-unity, barrault2023seamlessm4t}. In all these applications, SSL models generate a sequence of hidden representations with a fixed frameshift, usually around 20ms, which serve as inputs to the downstream tasks.

Two models that have notably excelled in recent benchmarks are HuBERT and WavLM \citep{hsu2021hubert, chen2022wavlm}. HuBERT employs quantized features for masked unit prediction in the context of masked speech signals \citep{hsu2021hubert}. Specifically, the model uses the classic $K$-means algorithm with a fixed cluster size $K$ to perform quantization, where cluster centroid IDs represent the target for each 20ms frame. A noteworthy aspect of HuBERT's pre-training strategy is its iterative training concept. Initially, clustering is performed on Mel Filter-bank Cepstral Coefficients (MFCC), termed as the first iteration. Subsequently, a hidden layer from the first iteration model is extracted and clustered to improve performance. Through this two-stage iterative approach, HuBERT has been shown to either match or exceed the performance of prior state-of-the-art models across various tasks \citep{hsu2021hubert, superb}. With a similar training scheme as HuBERT, WavLM differentiates itself by employing modified self-attention mechanisms and incorporating utterance mixing as a data augmentation technique. As these modifications are not the focus of this paper, our work mainly focuses on the framework of HuBERT and extends over that.




\section{MR-HuBERT}
\label{sec: method}


\subsection{HuBERT}
\label{ssec: hubert}
Consider a sequence of single-channel speech signal $ \mS \in \R^{1 \times L_s}$, where $L_s$ represents the length of the speech signal. For a given iteration $q$, the speech signal $\mS$ is initially quantized by a pre-trained $K$-means clustering model $g^q(\cdot)$, which is trained on the hidden states from the $q-1$ iteration.\footnote{The initial iteration ($q=0$) employs representations derived from MFCC features.} 

As detailed in Sections~\ref{sec: intro} and ~\ref{sec: related works}, HuBERT employs a convolutional feature extractor $f^q_0(\cdot)$ to first transform the speech signal $\mS$ into hidden representations at a frame size of 20ms. Following the masking strategies of wav2vec 2.0 and SpanBERT \citep{baevski2020wav2vec, joshi2020spanbert}, 
$\alpha\%$ of the frames are chosen randomly as starting indices, and $l$ subsequent frames are masked. The set of masked indices is denoted by $\mathbb{M}$.

A Transformer encoder $f^q_1(\cdot)$ is then tasked with predicting the quantized clusters of the masked regions, utilizing cross-entropy loss. The loss function at iteration $q$ is given by:
\begin{equation}
    \label{eq: hubert}
    \mathcal{L}^q_{m}(\theta; \mS, \mathbb{M}, g^q) = \sum_{t \in \mathbb{M}} \log p_{\theta} (g^q(\mS) \mid \Tilde{H_0^q}, t),
\end{equation}
where $\theta$ is the model parameters, $\Tilde{H_0^q}$ denotes the masked speech frames from the convolutional feature extractor and $t$ is the time step. It is worth noting that while one could define an unmasked loss $\mathcal{L}_u$, previous experiments have shown that this does not yield significant improvements in the quality of HuBERT's pre-training \citep{hsu2021hubert}.

\begin{figure}[t]
\begin{center}
\includegraphics[width=1.0\linewidth]{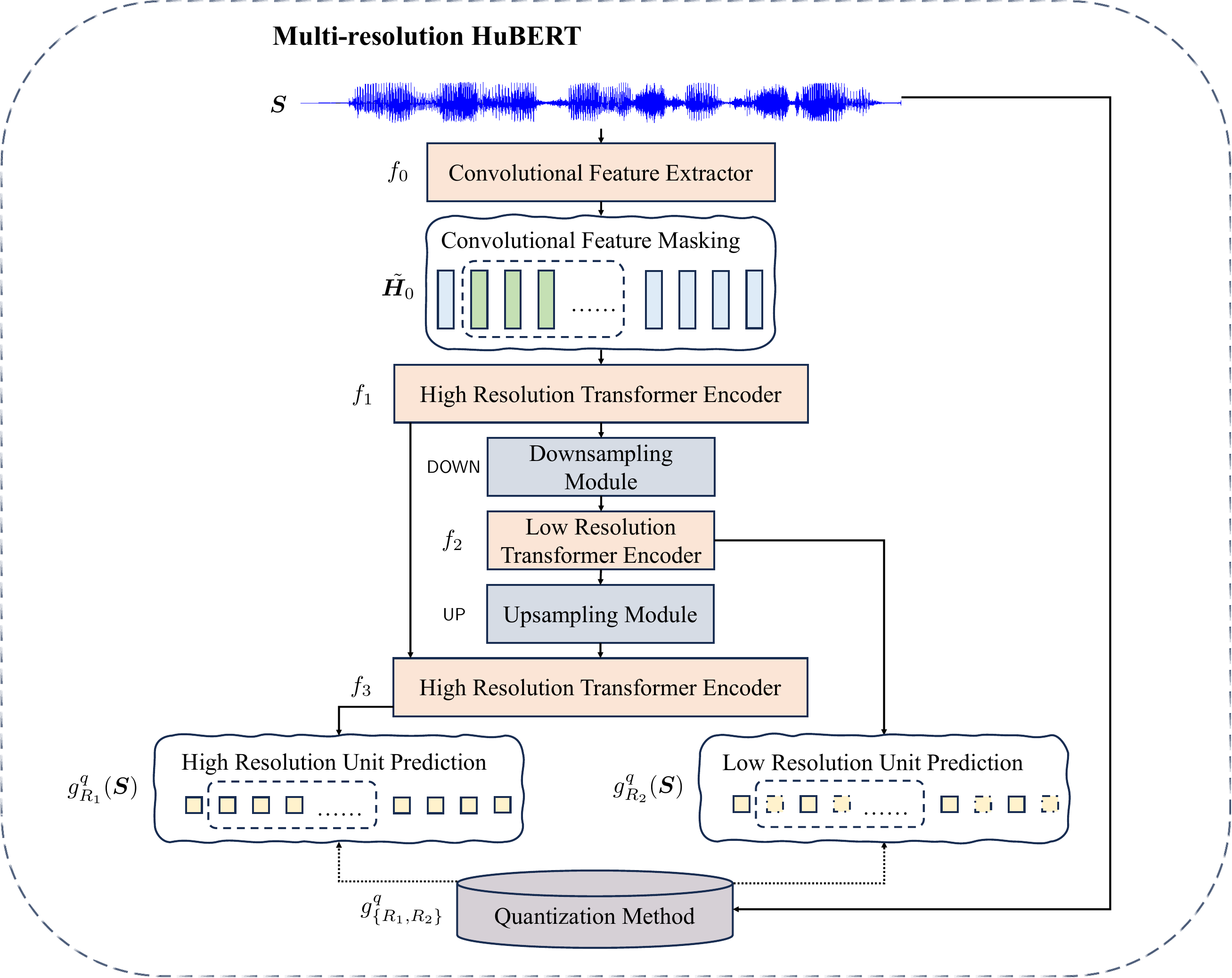}
\end{center}
\vspace{-10pt}
\caption{MR-HuBERT pre-training framework. The framework utilizes multi-resolution masked units prediction. The details of each module are discussed in Section~\ref{sec: method}}
\label{fig: multiresolution hubert}
\vspace{-10pt}
\end{figure}

\subsection{Architecture}

\label{ssec: architecture}

The proposed architecture for MR-HuBERT is schematically shown in Figure~\ref{fig: multiresolution hubert}. For this explanation, we exemplify a model with two resolutions. This architecture employs a hierarchical Transformer to explicitly encode hidden representations at multiple resolutions while retaining the iterative strategy found in the original HuBERT. The components of the framework are as follows:

Given an speech signal $\mS$, the convolutional feature extractor $f_0^q$ yields frame-wise feature $\mH_0 \in \R^{L_{R_1} \times D}$ at a high resolution $R_1$. $L_{R_1}$ is the frame length and $D$ is the feature dimension, which corresponds to the size of the convolutional channels. As outlined in Section~\ref{ssec: hubert}, a masking function $m(\cdot, \mathbb{M})$ is applied to $\mH_0$ to generate a sequence of masked features $\Tilde{\mH_0} \in \R^{L_{R_1} \times D}$. This function replaces the feature frames corresponding to the indices in $\mathbb{M}$ with zero vectors.

Next, the masked features $\Tilde{\mH_0}$ are processed by a HuBERT-style Transformer encoder $f_1^q$, noted as High Resolution Transformer Encoder in Figure~\ref{fig: multiresolution hubert} to produce $\Tilde{\mH}_1^q$. The encoder consists of a pre-convolutional module as well as a stack of transformer layers. The pre-convolutional module includes a 1D-convolutional layer, followed by Layer Normalization and a GELU activation function.

After the high-resolution encoding, the output $\Tilde{\mH}_1^q \in  \R^{L_{R_1} \times D}$ is subjected to a  downsampling module $\mathsf{DOWN}(\cdot)$ to produce a downsampled representation $\Tilde{\mH}_2^q \in \R^{L_{R_2} \times D}$. Here, $R_2$ denotes the lower resolution, and $L_{R_2}$ is the corresponding length of the downsampled hidden representation. The downsampled $\Tilde{\mH}_2^q$ serves as the input for a Low Resolution Transformer Encoder $f_2^q$, as illustrated in Figure~\ref{fig: multiresolution hubert}. Unlike $f_1^q$, $f_2^q$ does not include a pre-convolutional module. Its output $\Tilde{\mH}_3^q$, when coupled with a linear projection, is utilized to predict low-resolution units $g^q_{R_2}(\mS) \in \mathbb{N}^{+L_{R_2}}$ based on the quantization method $g^q_{R_2}(\cdot)$, detailed in Section~\ref{ssec: objective}. The whole process of generating $\Tilde{\mH}_3^q$ can be summarized into:
\begin{equation}
    \label{eq: low res output}
    \Tilde{\mH}_3^q = f_2^q \circ \mathsf{DOWN} \circ f_1^q(m(f_0^q(\mS), \mathbb{M})).
\end{equation}

Finally, an upsampling module $\mathsf{UP}(\cdot)$ expands $\Tilde{\mH}_3^q$ back to high resolution $R_1$, resulting in $\Tilde{\mH}_4^q \in \R^{L_{R_1} \times D}$. This output, when summed with $\Tilde{\mH}_1^q$, is fed into another High Resolution Transformer Encoder $f_3^q(\cdot)$. The ultimate output $\Tilde{\mH}_5^q \in \R^{L_{R_1} \times D}$ is then employed to predict high-resolution units obtained via the quantization method $g^q_{R_1}(\cdot)$. Given $\Tilde{\mH}_3^q$, the process of generating $\Tilde{\mH}_5^q$ can be summarized into:
\begin{equation}
    \label{eq: high res output}
    \Tilde{\mH}_5^q = f_3^q(\mathsf{UP}(\Tilde{\mH}_3^q) + f_1^q(m(f_0^q(\mS), \mathbb{M}))).
\end{equation}

\subsection{Sampling Modules}
\label{ssec: sampling module}

\begin{figure}[t]
\begin{center}
\includegraphics[width=0.8\linewidth]{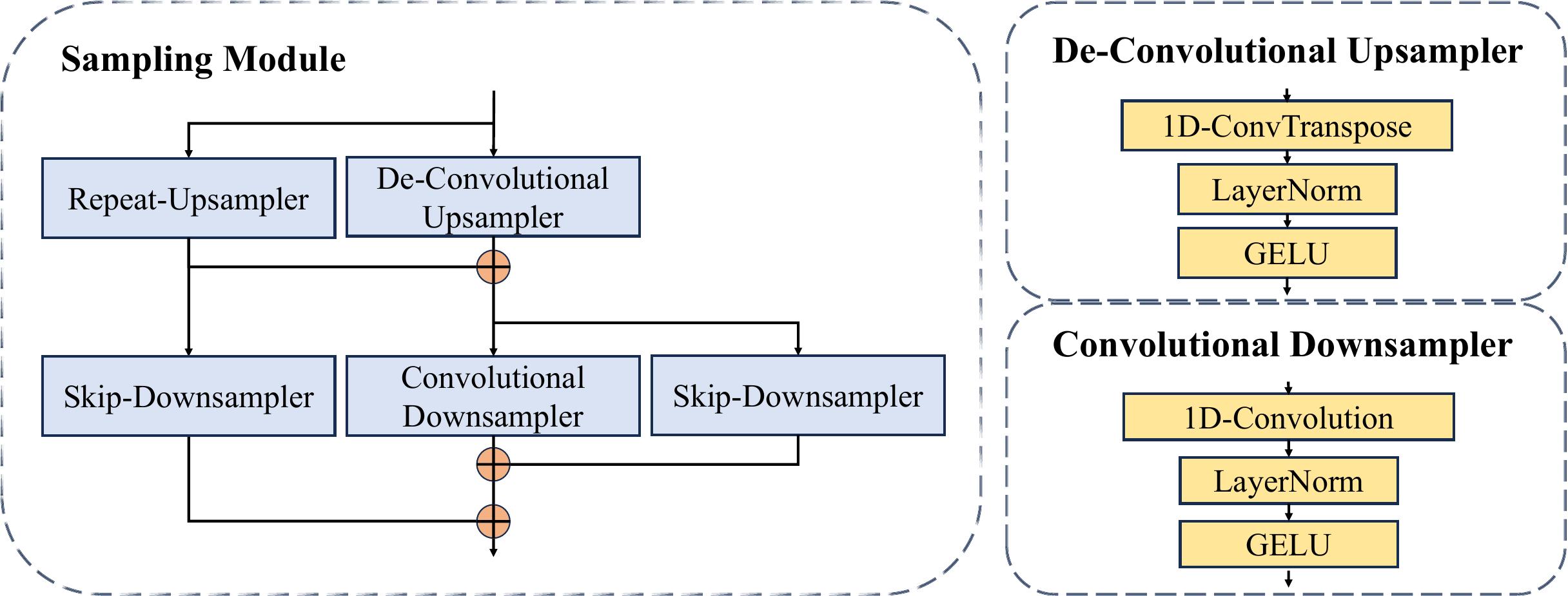}
\end{center}
\vspace{-15pt}
\caption{Sampling modules. The proposed sampling modules utilize a residual-based learning framework in either upsampling or downsampling. Details of the module are discussed in Section~\ref{ssec: sampling module}.}
\label{fig: sampler}
\end{figure}

As introduced in Section~\ref{ssec: architecture}, the proposed architecture utilizes an upsampling module $\mathsf{UP}(\cdot)$ and a downsampling module $\mathsf{DOWN}(\cdot)$.  The two sampling modules share the same design, as illustrated in Figure~\ref{fig: sampler}. The architecture is adapted from the multi-resolution fusion module in \citet{shi23h_interspeech}. 

To exemplify, we consider the downsampling module. 
The module first rescale $\Tilde{\mH}_1^q$ into a higher resolution $R_1 \cdot R_1'$ through De-Convolutional Upsampler $\mathrm{DeConv}(\cdot)$ and Repeat-Upsampler $\mathrm{Repeat}(\cdot)$, respectively.\footnote{Given $\Tilde{\mH}_1^q \in  \R^{L_{R_1} \times D}$ and the target resolution $R_2$, $R_1'$ and $R_2'$ are the numerator and denominator of the reduced fraction between $R_1$ and $R_2$. They are used as the upsampling factor and the downsampling factor, respectively.} 
The output, $\Tilde{\mH}_1^{q-\text{up}}  \in \R^{(L_{R_1 }\cdot R_1') \times D}$ is fed into a Convolutional Downsampler $\mathrm{Conv}(\cdot)$ and a Skip-Downsampler $\mathrm{Skip}(\cdot)$, respectively. 
The final output of the downsampling module, denoted as $\Tilde{\mH}_2^q$ in Section~\ref{ssec: architecture}, is the defined as:
\begin{equation}
    \Tilde{\mH}_2^q = \phi \cdot [\mathrm{Skip}(\mathrm{Repeat}(\Tilde{\mH}_1^q)) + \phi \cdot (\mathrm{Conv}(\Tilde{\mH}_1^{q-\text{up}}) + \mathrm{Skip}(\Tilde{\mH}_1^{q-\text{up}}))]
\end{equation}

\subsection{Objectives}
\label{ssec: objective}

Similar to HuBERT discussed in Section~\ref{ssec: hubert}, the objectives of MR-HuBERT focus on masked unit prediction. The major design question for MR-HuBERT, however, is how to construct units for different resolutions. In our experiments discussed in Section~\ref{sec: exp}, we compare different settings in multi-resolution units preparation. The default and most effective approach is simply start from high resolution units extraction and then subsample the low resolution units to match the low resolution sequence from the the Low Resolution Transformer Encoder $f_2^q$. The high resolution units extraction process is similar to HuBERT, by applying  $K$-means over hidden representations from $q-1$ iteration. To be specific, $g^q_{R_1}(\cdot)$ is the $K$-means model, where $g^q_{R_2}$ is $g^q_{R_1} \circ d(\cdot)$, where $d$ is a subsampling function. 

The pre-training involves two losses: one for high-resolution and another for low-resolution masked unit prediction:
\begin{equation}
    \mathcal{L}_m^{q-\text{\{high, low\}}} (\theta_{\text{\{high, low\}}}; \mS, \mathbb{M}, g^q_{\{R_1, R_2\}}) = \sum_{t \in \mathbb{M}} \log p_{\theta_{\text{\{high, low\}}}}(g^q_{\{R_1, R_2\}}(\mS) | \Tilde{\mH_0^q}, t),
\end{equation}
where $\theta_\text{high}$ are the model parameters of the MR-HuBERT, while $\theta_{\text{low}}$ are partial model parameters that exclude $\mathsf{UP}(\cdot)$ and $f_3^q(\cdot)$. The final objective combines these losses:
\begin{equation}
    \mathcal{L}_m^q = \beta \cdot \mathcal{L}_m^{q-\text{high}} + \gamma \cdot \mathcal{L}_m^{q-\text{low}},
\end{equation}
where $\beta$ and $\gamma$ are hyperparameters.

\section{Experiments}
\label{sec: exp}

We evaluate the proposed methods using a variety of speech processing tasks, segmented into four key categories: speech recognition on the LibriSpeech benchmarks \citep{panayotov2015librispeech}, SUPERB benchmark evaluation \citep{superb} and multilingual SUPERB (ML-SUPERB) benchmark evaluation \citep{mlsuperb, shi2023findings}.

\subsection{Pre-training}
\label{ssec: pretraining}

\noindent \textbf{Datasets}: We perform pre-training on three corpora: LibriSpeech \citep{panayotov2015librispeech}, LibriLight \citep{kahn2020libri}, and Voxpopuli \citep{wang2021voxpopuli}. LibriSpeech and LibriLight focus exclusively on English, while Voxpopuli is a multilingual dataset encompassing 23 European languages. The total dataset sizes amount to 960 hours for LibriSpeech, 60,000 hours for LibriLight, and 100,000 hours for Voxpopuli.\footnote{We use the same 100,000 hours split as \citet{wang2021voxpopuli}.}

\noindent \textbf{Model Configuration}: Following previous work in self-supervised speech learning \citep{baevski2020wav2vec, hsu2021hubert, chen2022wavlm}, we employ two model sizes for pre-training: \textit{base} and \textit{large}. As outlined in Section~\ref{sec: method}, we evaluate a two resolution variant of MR-HuBERT with 40ms and the commonly used 20ms. Ablation studies concerning resolutions are elaborated in Appendix~\ref{appendix: three-resolution}.

For both the \textit{base} and \textit{large} models, we adhere to the configurations used in the original HuBERT model \citep{hsu2021hubert}. Each encoder (i.e., $f_1^q(\cdot)$, $f_2^q(\cdot)$, and $f_3^q(\cdot)$) as detailed in Section~\ref{ssec: architecture}, has an evenly assigned number of Transformer layers. Specifically, the \textit{base} model uses a four-layer Transformer for each encoder, whereas the \textit{large} model deploys an eight-layer Transformer for each encoder. For an in-depth discussion on the effects of layer allocation, please refer to Appendix~\ref{appendix: layer size}.

\noindent \textbf{Unit Preparation}: To enhance efficiency of pre-training, we directly extract units from the publicly available \texttt{HuBERT-base}\footnote{\scriptsize{\url{https://dl.fbaipublicfiles.com/hubert/hubert_base_ls960.pt}}}. We first train a $K$-means model on 50\% of the LibriSpeech training set, with $K = 1,000$. Subsequently, the pre-trained $K$-means model is employed to extract target units from LibriSpeech, LibriLight, and Voxpopuli datasets. For multi-resolution scenarios, we perform subsampling of target units by skipping every second unit. Further experiments on unit extraction variants are available in Appendix~\ref{appendix: units}.

\noindent \textbf{Pre-trained Models}: We pre-train monolingual and multilingual models for both \textit{base} and \textit{large} settings. Specifically, \textbf{\texttt{mono-base}} and \textbf{\texttt{mono-large}} are trained on LibriSpeech (960 hours) and LibriLight (60,000 hours) respectively for 400,000 steps. The \textbf{\texttt{multi-base}} model is trained on Voxpopuli (384,000 hours) for 800,000 steps. More training details are available in Appendix~\ref{appendix: pre-train config}.

\noindent \textbf{Baselines}: Our primary comparisons are made with HuBERT models of matching sizes, specifically \texttt{HuBERT-base} and \texttt{HuBERT-large}. As noted in the Unit Preparation part, units are consistently extracted from \texttt{HuBERT-base}. To account for this, we include an additional iteration trained on this \textit{base} architecture, referred to as \texttt{HuBERT-base}\textsuperscript{+}. Furthermore, recognizing that our $K$-means model may not be identical to the one used in \texttt{HuBERT-large}, we introduce another setting that uses the same \textit{large} configuration but with our extracted units; we label this as \texttt{HuBERT-large}\textsuperscript{*}. For multilingual experiments, we include the public multilingual \texttt{mHuBERT-base}, introduced in \citet{lee2022textless} as well as a multilingual \texttt{HuBERT-base}\textsuperscript{*} that is trained with the same training configuration of \textbf{\texttt{multi-base}}.

To isolate the effects of individual components in our MR-HuBERT, we perform additional ablation studies detailed in Appendix~\ref{appendix: ablation}. These studies encompass mono-resolution models, models using a single high-resolution pre-training target, models with simplified sampling modules, models with less complex settings, etc.

\subsection{Speech Recognition}
\label{ssec: asr}

\begin{table}[t]
    \centering
    \caption{Word error rate for speech recognition on LibriSpeech benchmark, evaluated on 1-hour, 10-hour and 100-hour labeled data. Results with a 4-gram language model joint decoding are in parentheses. Model settings are discussed in Section~\ref{ssec: pretraining}.}
    \resizebox {0.8\linewidth} {!} {
\begin{tabular}{lccccc}
\toprule
Model &  Unlabeled Data (h) & dev-clean & dev-other & test-clean & test-other\\
\midrule \midrule
\multicolumn{6}{c}{\textit{\textbf{1-hour labeled}}} \\
\texttt{HuBERT-base} & 960 & 20.17 (8.75) & 28.11 (16.09) & 20.64 (8.88) & 28.87 (16.71)  \\
\texttt{HuBERT-base}\textsuperscript{+} & 960 & 19.64 (8.14) & 25.08 (12.36)  & 20.15 (8.31) & 25.63 (12.82) \\
\texttt{HuBERT-large} & 60,000 & 14.42 (5.84) & 18.80 (9.53) & 14.40 (5.81) & 19.29 (9.91) \\
\texttt{HuBERT-large}\textsuperscript{*} & 60,000 & 15.09 (4.30) & 18.20  (\textbf{6.84}) & 14.90 (4.30) & 18.05 (\textbf{7.23})  \\
\midrule
\textbf{\texttt{mono-base}} & 960 & 18.78 (7.33) & 23.72 (11.53) & 19.26 (7.41) & 24.46 (12.14) \\
\textbf{\texttt{mono-large}} & 60,000 & \textbf{6.44} (\textbf{3.64}) & \textbf{10.94} (6.85) & \textbf{6.37} (\textbf{3.75}) & \textbf{11.41} (\textbf{7.23}) \\
\midrule \midrule

\multicolumn{6}{c}{\textit{\textbf{10-hour labeled}}} \\
\texttt{HuBERT-base} & 960 & 9.62 (4.88) & 16.60 (8.51) & 9.71 (4.97) & 17.00 (9.15) \\
\texttt{HuBERT-base}\textsuperscript{+} & 960 & 9.51 (4.85) & 14.27 (8.37) & 9.72 (4.88) & 14.89 (8.94) \\
\texttt{HuBERT-large} & 60,000 & 5.68 (3.27) & 8.67 (5.51) & 5.75 (3.50) & 8.96 (5.93) \\
\texttt{HuBERT-large}\textsuperscript{*} & 60,000 & 5.61 (3.24) & 8.68 (5.55) & 5.57 (3.25) & 9.02 (6.00) \\
\midrule
\textbf{\texttt{mono-base}} & 960 & 8.51 (4.80) & 13.18 (8.29) & 8.46 (4.91) & 13.51 (8.33) \\
\textbf{\texttt{mono-large}} & 60,000 & \textbf{5.58} (\textbf{3.12}) & \textbf{8.57} (\textbf{5.44}) & \textbf{5.52} (\textbf{3.15}) & \textbf{8.74} (\textbf{5.86}) \\
\midrule \midrule

\multicolumn{6}{c}{\textit{\textbf{100-hour labeled}}} \\
\texttt{HuBERT-base} & 960 & 5.76 (3.66) & 12.90 (8.45) & 5.81 (3.84) & 12.76 (8.48) \\
\texttt{HuBERT-base}\textsuperscript{+} & 960 & 5.71 (3.33) & 10.66 (6.51) & 5.97 (3.55) & 10.87 (7.09) \\
\texttt{HuBERT-large} & 60,000 & 3.11 (2.37) & \textbf{6.01} (\textbf{4.22}) & 3.14 (2.48) & 6.15 (4.67) \\
\texttt{HuBERT-large}\textsuperscript{*} & 60,000 & \textbf{3.03} (2.44) & 6.30 (4.61) & 3.12 (2.62) & 6.14 (4.69) \\
\midrule
\textbf{\texttt{mono-base}} & 960 & 4.89 (3.21) & 9.04 (6.47) & 4.92 (3.57) & 9.17 (6.81) \\
\textbf{\texttt{mono-large}} & 60,000 & 3.06 (\textbf{2.33}) & 6.04 (4.54) & \textbf{3.01} (\textbf{2.44}) & \textbf{5.98} (\textbf{4.61}) \\

\bottomrule
\end{tabular}
}

    \label{tab: librispeech}
    \vspace{-15pt}
\end{table}

\noindent \textbf{Experimental Settings}: We conduct speech recognition experiments using various subsets of the LibriSpeech corpus for training. Specifically, we fine-tune the SSL models as a whole encoder using 1-hour, 10-hour, and 100-hour training subsets. Subsequently, we evaluate each fine-tuned model on four evaluation sets, namely dev-clean, test-clean, dev-other, and test-other. For training configurations, we adhere to the established settings with Connectionist Temporal Classification (CTC) used in wav2vec 2.0 and HuBERT, as outlined in the Fairseq framework~\citep{ott2019fairseq}.\footnote{\scriptsize{\url{https://github.com/facebookresearch/fairseq}}} Beyond decoding via beam search directly from the fine-tuned acoustic model, we also incorporate language model shallow fusion for enhanced performance \citep{karita19_interspeech}. To ensure result reproducibility, we employ an open-source four-gram language model pre-trained on LibriSpeech textual data, along with its associated lexicon \citep{panayotov2015librispeech}.\footnote{\scriptsize{\url{https://www.openslr.org/11/}}} Our chosen evaluation metric is the Word Error Rate~(WER).

\noindent \textbf{Results}: Our findings, illustrated in Table~\ref{tab: librispeech}, provide compelling evidence of the efficacy of our introduced methods. When subjected to a range of training durations—namely, 1-hour, 10-hour, and 100-hour—the techniques we have implemented consistently surpass the Word Error Rate (WER) results of the four reference baseline models. In the \textit{base} model variant, the \textbf{\texttt{mono-base}} model we introduce consistently showcases a marked 1\%-2\% WER improvement across the board, when measured against all four evaluation datasets. For the \textit{large} model configuration, the results become even more compelling. The \textbf{\texttt{mono-large}} model, in particular, stands out: when trained on the 1-hour dataset, it achieves a WER reduction oscillating between 40\% and 50\%. For the 10-hour training set, the dev-other and test-other evaluation datasets reflect the most pronounced improvements. Shifting to the 100-hour training set, the test-clean and test-other sets emerge as the beneficiaries of the largest boosts in performance. Furthermore, when a joint-decoding strategy with the language model is in place, while the performance differential becomes less pronounced, the proposed MR-HuBERT still maintains a performance edge, always matching or outperforming the baseline HuBERT models. A salient takeaway is that our proposed models consistently rival or outstrip the baseline models, underscoring the robustness and superiority of the methodologies we've employed.

\subsection{SUPERB Evaluation}
\label{ssec: superb evaluation}

\noindent \textbf{Experimental Settings}: Our evaluation within the SUPERB framework aims to provide a holistic assessment of the quality of SSL representations across a broad array of speech processing tasks \citep{superb, tsai-etal-2022-superb, feng2023superb}. Specifically, we assess our proposed models on tasks including Phone Recognition (PR), Automatic Speech Recognition (ASR), Intent Classification (IC), Keyword Spotting (KS), Slot Filling (SF), Speech Translation (ST), Speech Enhancement (SE), and Speech Separation (SS).\footnote{Besides the SUPERB public benchmark tasks, we also explore Voice Conversion (VC) as outlined in \citet{huang2021s3prl, s3prl-vc-journal}. For more details, see Appendix~\ref{appendix: superb details}.}

\begin{wraptable}{r}{8cm}
    \centering
    \small
    \vspace{-15pt}
    \caption{Categorical SUPERB score. Category information and SUPERB score definition are discussed in Section~\ref{ssec: superb evaluation}.}
    \resizebox {\linewidth} {!} {
\begin{tabular}{l|cc|c}
\toprule
Model & Understanding  &  Enhancement & General \\
\midrule 
\texttt{HuBERT-base} & 861.2 & 98.20 & 670.4 \\
\texttt{HuBERT-base}\textsuperscript{+} & 876.9  & 150.2 & 695.2 \\
\texttt{HuBERT-large} & 932.6  & 456.0 & 813.4 \\
\texttt{HuBERT-large}\textsuperscript{*} & 936.2  & 501.5 & 827.5 \\
\midrule
\textbf{\texttt{mono-base}} & 885.8  & 195.0 & 708.7 \\
\textbf{\texttt{mono-large}} & \textbf{949.7}  & \textbf{609.5} & \textbf{864.6} \\ 
\bottomrule
\end{tabular}
}
    \label{tab: superb-cate}
    \vspace{-15pt}
\end{wraptable}
To ensure consistent evaluations, we adopt metrics outlined in \citet{superb}: Phone Error Rate (PER) for PR, WER for ASR, Accuracy (ACC) for IC and KS, F-1 measure and Character Error Rate (CER) for SF, BLEU for ST, Short-Time Objective Intelligibility (STOI) and Perceptual Evaluation of Speech Quality (PESQ) for SE, and Scale-Invariant Signal-to-Distortion Ratio improvement (SI-SDRi) for SS.

We adhere to the SUPERB policy for downstream model training. In particular, we keep the SSL upstream models fixed and only adjust the learning rate. To address reproducibility, we perform a simple grid search for the learning rate, considering only the default rate in S3PRL along with its 0.1x and 10x variations. We also use the weighted summation strategy for the frozen SSL representation. To mitigate the resolution differences across layers, we conduct simple repeat upsampling or skip downsampling as outlined in \citep{shi23h_interspeech}.

To gauge the performance of SSL representations across tasks, we categorize SUPERB tasks into two main clusters: Understanding and Enhancement (Generation). We calculate the SUPERB score (denoted as SUPERB$_s$), as defined in the SLT 2022 SUPERB challenge \citep{feng2023superb}, which employs linear scaling between conventional spectral features and state-of-the-art upstream representations in the corresponding tasks. Comprehensive performance metrics that take into account all evaluated tasks are also calculated. More information on the SUPERB is available in Appendix~\ref{appendix: superb details}.

\begin{table}[t]
    \centering
    \caption{Detailed SUPERB evaluation. Detailed metrics and settings are detailed in Section~\ref{ssec: superb evaluation}.}
    \resizebox {\linewidth} {!} {
\begin{tabular}{l|ccccccc|ccc}
\toprule
\multirow{2}{*}{Model} & \multicolumn{7}{c}{Understanding}  &  \multicolumn{3}{|c}{Enhancement} \\
 & PR($\downarrow$) & ASR($\downarrow$) & IC($\uparrow$)  & KS($\uparrow$) & SF-F1($\uparrow$) & SF-CER($\downarrow$) & ST($\uparrow$) & SE-STOI($\uparrow$) & SE-PESQ($\uparrow$) & SS($\uparrow$) \\
\midrule 
\texttt{HuBERT-base} & 5.40 & 6.42 & 98.34  & 96.30 & 88.53 & 25.20 & 15.53  & \textbf{0.94} & 2.58 & 9.36 \\
\texttt{HuBERT-base}\textsuperscript{+} & 4.56 & 6.34 & 98.39  & 96.46 & 89.12 & 23.10 & 16.33  & 0.93 & 2.55 & 9.72 \\
\texttt{HuBERT-large} & 3.54 & 3.62 & 98.76  & 95.29 & 89.81 & 21.76 & 20.01 &  \textbf{0.94} & 2.64 & 10.45 \\
\texttt{HuBERT-large}\textsuperscript{*} & 3.59 & \textbf{3.53} & 98.73  & 97.70 & 89.88 & 22.51 & 20.02  & \textbf{0.94} & 2.65 & 10.61 \\
\midrule
\textbf{\texttt{mono-base}} & 4.16 & 5.76 & 98.68  & 96.49 & 88.96 & 23.59 & 16.94  & \textbf{0.94} & 2.55 & 9.92 \\
\textbf{\texttt{mono-large}} & \textbf{3.15} & 3.78 & \textbf{98.76}  & \textbf{97.76} & \textbf{90.57} & \textbf{20.60} & \textbf{21.05} & \textbf{0.94} & \textbf{2.67} & \textbf{10.97} \\ 
\bottomrule
\end{tabular}
}
    \label{tab: superb}
\end{table}

\noindent \textbf{Results}: The comprehensive results, divided by task category, are presented in Table~\ref{tab: superb-cate} and Table~\ref{tab: superb}. Our proposed MR-HuBERT demonstrates marked improvements over a variety of understanding and enhancement tasks in both \textit{base} and \textit{large} configurations.

\subsection{ML-SUPERB Evaluation}
\label{ssec: ml_superb}

\noindent \textbf{Experimental Settings}: We evaluate the performance of our proposed multilingual speech processing method using the ML-SUPERB benchmark \citep{mlsuperb}. This benchmark, which is supported by 143 languages, has been implemented as a recipe within the ESPnet framework \citep{watanabe18_interspeech}\footnote{\scriptsize{\url{https://github.com/espnet/espnet/tree/master/egs2/ml_superb/asr1}}}. The ML-SUPERB benchmark comprises two sets of general benchmarks—specifically, a 10-minute set and a 1-hour set—across four tasks: Monolingual ASR, Multilingual ASR, Language Identification (LID), and a joint task of Multilingual ASR+LID. To maintain the integrity of the experimental comparison, we adhere to the ML-SUPERB guidelines for downstream architectures and training configurations, including the use of frozen SSL representations \citep{mlsuperb}. For the evaluation, we employ the standard metrics: Character Error Rate (CER) or PER for ASR tasks, and ACC for LID tasks. Furthermore, we calculate a composite ML-SUPERB score as defined by \citet{mlsuperb} to provide an overall measure of performance. Additional information on the SUPERB evaluation is available in Appendix~\ref{appendix: ml-superb details}.

\begin{table}[t]
    \centering
    \caption{Results on ML-SUPERB \{10-minute/1-hour\} settings. Detailed metrics and settings are detailed in Section~\ref{ssec: ml_superb}.}
    \resizebox {\linewidth} {!} {
\begin{tabular}{l|c|cc|c|ccc|c}
\toprule
\multirow{3}{*}{SSL} & Monolingual ASR & \multicolumn{2}{c|}{Multilingual ASR} & \multicolumn{1}{c|}{LID} & \multicolumn{3}{c|}{Multilingual ASR + LID} & \multirow{3}{*}{SUPERB$_{s}$($\uparrow$)} \\
&          &             Normal & Few-shot & Normal & \multicolumn{2}{c}{Normal} & \multicolumn{1}{c|}{Few-shot} \\
& CER/PER($\downarrow$) & CER($\downarrow$) & CER($\downarrow$) & ACC($\uparrow$) & ACC($\uparrow$) & CER($\downarrow$) & \multicolumn{1}{c|}{CER($\downarrow$)} &  \\
\midrule
\texttt{HuBERT-base} & 42.8 / 35.3 & 39.8 / 31.4 & 44.5 / 42.7 & 61.2 / 86.1 & \textbf{71.5} / 86.0 &  39.2 / 30.9 & 43.8 / 41.8 & 831.9 / 884.9 \\
\texttt{HuBERT-base}\textsuperscript{+} & 42.9 / 35.3 & 41.5 / 31.2 & 45.8 / 42.8 & 63.8 / 81.9 & 70.1 / 85.8 & 39.6 / 31.3 & 44.6 / 40.7 & 819.1 / 875.8 \\
\texttt{HuBERT-large} & \textbf{38.2} / 32.2 & 44.4 / 37.7 & 48.2 / 43.5 &  46.5 / 64.1 & 55.4 / 77.7 & 45.6 / 35.1 & 49.3 / 42.2 & 678.7 / 783.6  \\
\texttt{HuBERT-large}\textsuperscript{*} & 41.2 / 32.6 & 42.8 / 32.8 & 45.6 / 42.5 & 42.3 / 58.9 & 59.2 / 84.7 & 42.3 / 29.8 & 44.1 / 41.4 & 704.5 / 817.6 \\
\texttt{mHuBERT-base} & 41.0 / 33.0 & 40.5 / 33.4 & 45.6 / 43.6 & 52.4 / 72.5  & 46.6 / 70.9 & 36.8 / 29.7 & 44.2 / 43.1 &  746.2 / 812.7 \\
\texttt{mHuBERT-base}\textsuperscript{*} & 40.1 / 32.3 & 36.3 / \textbf{27.3} & \textbf{38.6} / 39.0 & \textbf{64.0} / 82.0 & 70.4 / 84.6 & 35.4 / \textbf{27.1} & \textbf{39.0} / 37.0 & 950.8 / 964.5 \\
\midrule
\textbf{\texttt{mono-base}} & 42.8 / 34.6 & 40.2 / 30.6 & 45.0 / 42.2 & 67.2 / \textbf{86.3} & 68.7 / \textbf{86.9} & 40.3 / 30.6 & 44.1 / 41.6 & 843.5 / 899.9 \\
\textbf{\texttt{mono-large}} & 40.5 / 32.0  & 38.9 / 29.4 & 42.7 / 40.5 & 45.1 / 75.4 & 67.6 / 85.9 & 39.0 / 29.7 & 43.8 / 40.8 & 785.2 / 905.4  \\
\textbf{\texttt{multi-base}} & 38.3 / \textbf{30.6}  & \textbf{34.1} / 27.5  & 39.6 / \textbf{38.9} & \textbf{64.0} / 85.1 & 69.9 / 84.4 & \textbf{34.4} / 28.0 & 40.9 / \textbf{36.6} & \textbf{957.2} / \textbf{986.8}  \\
\bottomrule
\end{tabular}
}

    \vspace{-15pt}
    \label{tab: ml_superb}
\end{table}

\noindent \textbf{Results}: Our evaluations on the ML-SUPERB benchmark are summarized in Table~\ref{tab: ml_superb}. The data reveals that our proposed multilingual model, \textbf{\texttt{multi-base}}, stands out with the topmost performance. Notably, even our monolingual pre-trained models, \textbf{\texttt{mono-base}} and \textbf{\texttt{mono-large}}, surpass the overall monolingual baselines. Furthermore, they outperform the multilingual model \texttt{mHuBERT-base} and \texttt{mHuBERT-base}\textsuperscript{*} in the overall ML-SUPERB score.



\subsection{Discussion: Inference Speed}
\label{ssec: speed}

In addition to achieving notable gains in performance across various test scenarios, the proposed method also offers advantages in terms of computational efficiency, particularly during the inference stage. This efficiency is primarily attributable to the reduced sequence length required for self-attention computations. To quantitatively evaluate this improvement, we employ Multiply-Add Cumulations (MACs) as our metric of comparison between the baseline models and our proposed method. We utilize the TorchProfile toolkit to calculate MACs\footnote{\scriptsize{\url{https://github.com/zhijian-liu/torchprofile}}}. Specifically, we analyze audio samples of varying lengths—2s, 4s, 8s, 16s, and 32s—to calculate the total MACs for each method. The results indicate a clear computational advantage for the proposed method: in the \textit{base} model configuration, the total MACs were reduced from 431G to 394G, representing an improvement of 9\%. In the \textit{large} model configuration, the MACs decreased from 1116G to 971G, corresponding to a 13\% improvement.

\section{Relation to Similar Approaches in Other Contexts}

The idea of leveraging multiple resolutions has been explored in various other contexts. In speech understanding, downsampled spoken feature sequences are commonly employed to extract high-level linguistic or semantic features for efficiency \citep{chen2019big, meng2023compressing, chen2023once} or to better integrate pre-trained language models \citep{gaido2021ctc, shi2023bridging, wu2023decoder, li23i_interspeech}. In speech synthesis, multi-resolution discriminators have been instrumental in recent adversarial-based vocoders \citep{yamamoto2020parallel, kong2020hifi, yoneyama2023source}. Additionally, multi-resolution or multi-scale networks have shown robust performance in speech enhancement \citep{zhang20x_interspeech,zhang2022multi, xiang2021convolutional, xu2020spex, shi19f_interspeech}. While prior work exists, our paper stands out for its focus on a novel hierarchical architecture for speech pre-training. The resulting models offer not only substantial performance gains across downstream tasks but also computational efficiencies during inference.

Similar multi-resolution strategies have also found applications in other domains. In computer vision, multi-scale convolutional networks are employed for various tasks such as object detection and human pose estimation \citep{yang2015multi, cai2016unified, ghiasi2019fpn, mathieu2016deep}. Among these, Hourglass networks stand out for their hierarchical multi-resolution processing, which has resulted in significant performance gains \citep{newell2016stacked, melekhov2017image, yang2017stacked}. This concept has been extended to the text domain as the Hourglass transformer, which has proven effective for sequence processing \citep{ijcai2023p585, guo2022u, nawrot-etal-2023-efficient, nawrot2022hierarchical}. Our work has a similar architecture to the Hourglass transformer in speech pre-training with specific features like masked unit prediction, multi-resolution targets, and other speech-related architectural nuances.

\section{Conclusion}

This paper introduces MR-HuBERT, a self-supervised speech learning model that extends HuBERT by employing multi-resolution masked unit prediction in conjunction with a hierarchical transformer architecture. Comprehensive evaluations across various benchmarks reveal that MR-HuBERT substantially outperforms the original HuBERT model across a broad spectrum of speech processing tasks. These include, but are not limited to, speech recognition, spoken language understanding, multilingual speech recognition, and speech enhancement. Beyond these performance gains, the model also exhibits computational efficiencies, specifically a 9-13\% reduction in computational complexity, addressing efficiency concerns.\footnote{Limitations of the work are discussed in Appendix~\ref{appendix: limitation}, while some future directions are discussed in Appendix~\ref{appendix: future works}.}

\section{Ethics Statement}
The development and implementation of MR-HuBERT represent a significant step forward in self-supervised pre-training for speech models. While this model demonstrates substantial potential and effectiveness across various tasks, it's crucial to approach its adoption and application ethically:

\begin{itemize}
    \item Openness and Transparency: We remain committed to the principles of open research. By releasing the complete codebase and associated checkpoints of our MR-HuBERT model, we aim to foster an environment of transparency and reproducibility. This initiative encourages peer reviews and allows researchers to independently validate our findings.
    \item Potential Misuse: Like any advanced technology, MR-HuBERT's capabilities could be misappropriated for malicious purposes. While the model offers enhanced performance across various speech tasks, users must employ it responsibly, respecting individual privacy and avoiding potential misuse in surveillance or unauthorized information extraction. MR-HuBERT presents an unforeseen avenue for speech disentanglement, especially in its large configurations, as detailed in Appendix~\ref{appendix: superb details}. As the model evolves, ensuring that it doesn't unintentionally disentangle or misinterpret cultural nuances, accents, or dialects becomes paramount. This concern is essential for avoiding potential biases or misrepresentations.
\end{itemize}

While MR-HuBERT represents a promising stride in speech model advancement, its ethical implications are at the forefront of our considerations. We urge the community to employ this technology with caution, respect, and a commitment to the broader good.

\section{Reproducibility Statement}
In the spirit of open research and fostering further advancements in the field, we will be releasing the complete codebase associated with our MR-HuBERT model. This encompasses the entire spectrum of models discussed in our work, including models presented in Appendices. Researchers, academicians, and enthusiasts can access, reproduce, and potentially build upon our findings. We believe that this transparent sharing will not only validate our findings but also inspire innovative research directions anchored around MR-HuBERT. Details regarding access and implementation will be updated after the double-blind review. We eagerly anticipate the community's engagement and are open to collaborations, feedback, and further enhancements to the model.

\section{Acknowledgement}
We extend our heartfelt gratitude to Juan Pino, Paden Tomasello, Changhan Wang, Andy Chung, Ning Dong, Hongyu Gong, and Maha Elbayad for their invaluable advice and unwavering support throughout this project. Their insights and expertise have been indispensable to this work.

Special recognition is owed to Yun Tang and Shinji Watanabe. Their contributions, particularly in the formative stages of our research, have been instrumental. Their guidance in shaping our initial research idea has set a strong foundation for the entirety of this work.

\bibliography{iclr2024_conference}

\begin{thebibliography}{125}
\providecommand{\natexlab}[1]{#1}
\providecommand{\url}[1]{\texttt{#1}}
\expandafter\ifx\csname urlstyle\endcsname\relax
  \providecommand{\doi}[1]{doi: #1}\else
  \providecommand{\doi}{doi: \begingroup \urlstyle{rm}\Url}\fi

\bibitem[Agrawal et~al.(2023)Agrawal, Anastasopoulos, Bentivogli, Bojar, Borg,
  Carpuat, Cattoni, Cettolo, Chen, Chen, et~al.]{agrawal2023findings}
Sweta Agrawal, Antonios Anastasopoulos, Luisa Bentivogli, Ond{\v{r}}ej Bojar,
  Claudia Borg, Marine Carpuat, Roldano Cattoni, Mauro Cettolo, Mingda Chen,
  William Chen, et~al.
\newblock Findings of the {IWSLT} 2023 evaluation campaign.
\newblock In \emph{Proceedings of the 20th International Conference on Spoken
  Language Translation (IWSLT 2023)}, pp.\  1--61, 2023.

\bibitem[Andrusenko et~al.(2023)Andrusenko, Nasretdinov, and
  Romanenko]{andrusenko2023uconv}
Andrei Andrusenko, Rauf Nasretdinov, and Aleksei Romanenko.
\newblock {UCONV-Conformer}: High reduction of input sequence length for
  end-to-end speech recognition.
\newblock In \emph{ICASSP 2023-2023 IEEE International Conference on Acoustics,
  Speech and Signal Processing (ICASSP)}, pp.\  1--5. IEEE, 2023.

\bibitem[Babu et~al.(2021)Babu, Wang, Tjandra, Lakhotia, Xu, Goyal, Singh, von
  Platen, Saraf, Pino, et~al.]{babu2021xls}
Arun Babu, Changhan Wang, Andros Tjandra, Kushal Lakhotia, Qiantong Xu, Naman
  Goyal, Kritika Singh, Patrick von Platen, Yatharth Saraf, Juan Pino, et~al.
\newblock {XLS-R}: Self-supervised cross-lingual speech representation learning
  at scale.
\newblock \emph{arXiv preprint arXiv:2111.09296}, 2021.

\bibitem[Baevski et~al.(2020)Baevski, Zhou, Mohamed, and
  Auli]{baevski2020wav2vec}
Alexei Baevski, Yuhao Zhou, Abdelrahman Mohamed, and Michael Auli.
\newblock wav2vec 2.0: A framework for self-supervised learning of speech
  representations.
\newblock \emph{Advances in neural information processing systems},
  33:\penalty0 12449--12460, 2020.

\bibitem[Baevski et~al.(2022)Baevski, Hsu, Xu, Babu, Gu, and
  Auli]{baevski2022data2vec}
Alexei Baevski, Wei-Ning Hsu, Qiantong Xu, Arun Babu, Jiatao Gu, and Michael
  Auli.
\newblock Data2vec: A general framework for self-supervised learning in speech,
  vision and language.
\newblock In \emph{International Conference on Machine Learning}, pp.\
  1298--1312. PMLR, 2022.

\bibitem[Barrault et~al.(2023)Barrault, Chung, Meglioli, Dale, Dong, Duquenne,
  Elsahar, Gong, Heffernan, Hoffman, et~al.]{barrault2023seamlessm4t}
Lo{\"\i}c Barrault, Yu-An Chung, Mariano~Cora Meglioli, David Dale, Ning Dong,
  Paul-Ambroise Duquenne, Hady Elsahar, Hongyu Gong, Kevin Heffernan, John
  Hoffman, et~al.
\newblock Seamlessm4t-massively multilingual \& multimodal machine translation.
\newblock \emph{arXiv preprint arXiv:2308.11596}, 2023.

\bibitem[Bastianelli et~al.(2020)Bastianelli, Vanzo, Swietojanski, and
  Rieser]{bastianelli-etal-2020-slurp}
Emanuele Bastianelli, Andrea Vanzo, Pawel Swietojanski, and Verena Rieser.
\newblock {SLURP}: A spoken language understanding resource package.
\newblock In \emph{Proceedings of the 2020 Conference on Empirical Methods in
  Natural Language Processing (EMNLP)}, pp.\  7252--7262, Online, November
  2020. Association for Computational Linguistics.
\newblock \doi{10.18653/v1/2020.emnlp-main.588}.

\bibitem[Berrebbi et~al.(2022)Berrebbi, Shi, Yan, López-Francisco, Amith, and
  Watanabe]{berrebbi22_interspeech}
Dan Berrebbi, Jiatong Shi, Brian Yan, Osbel López-Francisco, Jonathan Amith,
  and Shinji Watanabe.
\newblock {Combining Spectral and Self-Supervised Features for Low Resource
  Speech Recognition and Translation}.
\newblock In \emph{Proc. Interspeech 2022}, pp.\  3533--3537, 2022.
\newblock \doi{10.21437/Interspeech.2022-10796}.

\bibitem[Burchi \& Vielzeuf(2021)Burchi and Vielzeuf]{burchi2021efficient}
Maxime Burchi and Valentin Vielzeuf.
\newblock Efficient conformer: Progressive downsampling and grouped attention
  for automatic speech recognition.
\newblock In \emph{2021 IEEE Automatic Speech Recognition and Understanding
  Workshop (ASRU)}, pp.\  8--15. IEEE, 2021.

\bibitem[Cai et~al.(2016)Cai, Fan, Feris, and Vasconcelos]{cai2016unified}
Zhaowei Cai, Quanfu Fan, Rogerio~S Feris, and Nuno Vasconcelos.
\newblock A unified multi-scale deep convolutional neural network for fast
  object detection.
\newblock In \emph{Computer Vision--ECCV 2016: 14th European Conference,
  Amsterdam, The Netherlands, October 11--14, 2016, Proceedings, Part IV 14},
  pp.\  354--370. Springer, 2016.

\bibitem[Chang et~al.(2023)Chang, Liu, and Glass]{chang23_interspeech}
Heng-Jui Chang, Alexander~H. Liu, and James Glass.
\newblock {Self-supervised Fine-tuning for Improved Content Representations by
  Speaker-invariant Clustering}.
\newblock In \emph{Proc. INTERSPEECH 2023}, pp.\  2983--2987, 2023.
\newblock \doi{10.21437/Interspeech.2023-847}.

\bibitem[Chang et~al.(2021)Chang, Maekaku, Guo, Shi, Lu, Subramanian, Wang,
  Yang, Tsao, Lee, et~al.]{chang2021exploration}
Xuankai Chang, Takashi Maekaku, Pengcheng Guo, Jing Shi, Yen-Ju Lu,
  Aswin~Shanmugam Subramanian, Tianzi Wang, Shu-wen Yang, Yu~Tsao, Hung-yi Lee,
  et~al.
\newblock An exploration of self-supervised pretrained representations for
  end-to-end speech recognition.
\newblock In \emph{2021 IEEE Automatic Speech Recognition and Understanding
  Workshop (ASRU)}, pp.\  228--235. IEEE, 2021.

\bibitem[Chen et~al.(2019)Chen, Fan, Mallinar, Sercu, and Feris]{chen2019big}
Chun~Fu Chen, Quanfu Fan, Neil Mallinar, Tom Sercu, and Rogerio Feris.
\newblock Big-little net: An efficient multi-scale feature representation for
  visual and speech recognition.
\newblock In \emph{International Conference on Learning Representations}.
  International Conference on Learning Representations, ICLR, 2019.

\bibitem[Chen et~al.(2023{\natexlab{a}})Chen, Meng, and Lee]{chen2023once}
Hsuan-Jui Chen, Yen Meng, and Hung-yi Lee.
\newblock Once-for-all sequence compression for self-supervised speech models.
\newblock In \emph{ICASSP 2023-2023 IEEE International Conference on Acoustics,
  Speech and Signal Processing (ICASSP)}, pp.\  1--5. IEEE, 2023{\natexlab{a}}.

\bibitem[Chen et~al.(2022{\natexlab{a}})Chen, Wang, Chen, Wu, Liu, Chen, Li,
  Kanda, Yoshioka, Xiao, et~al.]{chen2022wavlm}
Sanyuan Chen, Chengyi Wang, Zhengyang Chen, Yu~Wu, Shujie Liu, Zhuo Chen, Jinyu
  Li, Naoyuki Kanda, Takuya Yoshioka, Xiong Xiao, et~al.
\newblock {WavLM}: Large-scale self-supervised pre-training for full stack
  speech processing.
\newblock \emph{IEEE Journal of Selected Topics in Signal Processing},
  16\penalty0 (6):\penalty0 1505--1518, 2022{\natexlab{a}}.

\bibitem[Chen et~al.(2022{\natexlab{b}})Chen, Wu, Wang, Chen, Chen, Liu, Wu,
  Qian, Wei, Li, et~al.]{chen2022unispeech}
Sanyuan Chen, Yu~Wu, Chengyi Wang, Zhengyang Chen, Zhuo Chen, Shujie Liu, Jian
  Wu, Yao Qian, Furu Wei, Jinyu Li, et~al.
\newblock Unispeech-sat: Universal speech representation learning with speaker
  aware pre-training.
\newblock In \emph{ICASSP 2022-2022 IEEE International Conference on Acoustics,
  Speech and Signal Processing (ICASSP)}, pp.\  6152--6156. IEEE,
  2022{\natexlab{b}}.

\bibitem[Chen et~al.(2022{\natexlab{c}})Chen, Wu, Wang, Liu, Chen, Wang, Liu,
  Li, Wu, Yu, and Wei]{chen22g_interspeech}
Sanyuan Chen, Yu~Wu, Chengyi Wang, Shujie Liu, Zhuo Chen, Peidong Wang, Gang
  Liu, Jinyu Li, Jian Wu, Xiangzhan Yu, and Furu Wei.
\newblock {Why does Self-Supervised Learning for Speech Recognition Benefit
  Speaker Recognition?}
\newblock In \emph{Proc. Interspeech 2022}, pp.\  3699--3703,
  2022{\natexlab{c}}.
\newblock \doi{10.21437/Interspeech.2022-10019}.

\bibitem[Chen et~al.(2023{\natexlab{b}})Chen, Yan, Shi, Peng, Maiti, and
  Watanabe]{chen2023improving}
William Chen, Brian Yan, Jiatong Shi, Yifan Peng, Soumi Maiti, and Shinji
  Watanabe.
\newblock Improving massively multilingual {ASR} with auxiliary {CTC}
  objectives.
\newblock In \emph{ICASSP 2023-2023 IEEE International Conference on Acoustics,
  Speech and Signal Processing (ICASSP)}, pp.\  1--5. IEEE, 2023{\natexlab{b}}.

\bibitem[Chiu et~al.(2022)Chiu, Qin, Zhang, Yu, and Wu]{chiu2022self}
Chung-Cheng Chiu, James Qin, Yu~Zhang, Jiahui Yu, and Yonghui Wu.
\newblock Self-supervised learning with random-projection quantizer for speech
  recognition.
\newblock In \emph{International Conference on Machine Learning}, pp.\
  3915--3924. PMLR, 2022.

\bibitem[Choi et~al.(2021)Choi, Lee, Kim, Lee, Heo, and Lee]{choi2021neural}
Hyeong-Seok Choi, Juheon Lee, Wansoo Kim, Jie Lee, Hoon Heo, and Kyogu Lee.
\newblock Neural analysis and synthesis: Reconstructing speech from
  self-supervised representations.
\newblock \emph{Advances in Neural Information Processing Systems},
  34:\penalty0 16251--16265, 2021.

\bibitem[Choi et~al.(2023)Choi, Kim, and Ro]{choi2023intelligible}
Jeongsoo Choi, Minsu Kim, and Yong~Man Ro.
\newblock Intelligible lip-to-speech synthesis with speech units.
\newblock \emph{arXiv preprint arXiv:2305.19603}, 2023.

\bibitem[Chung et~al.(2021)Chung, Zhang, Han, Chiu, Qin, Pang, and
  Wu]{chung2021w2v}
Yu-An Chung, Yu~Zhang, Wei Han, Chung-Cheng Chiu, James Qin, Ruoming Pang, and
  Yonghui Wu.
\newblock {W2V-BERT}: Combining contrastive learning and masked language
  modeling for self-supervised speech pre-training.
\newblock In \emph{2021 IEEE Automatic Speech Recognition and Understanding
  Workshop (ASRU)}, pp.\  244--250. IEEE, 2021.

\bibitem[Conneau et~al.(2020)Conneau, Baevski, Collobert, Mohamed, and
  Auli]{conneau2020unsupervised}
Alexis Conneau, Alexei Baevski, Ronan Collobert, Abdelrahman Mohamed, and
  Michael Auli.
\newblock Unsupervised cross-lingual representation learning for speech
  recognition.
\newblock \emph{arXiv preprint arXiv:2006.13979}, 2020.

\bibitem[Davis \& Mermelstein(1980)Davis and Mermelstein]{davis1980comparison}
Steven Davis and Paul Mermelstein.
\newblock Comparison of parametric representations for monosyllabic word
  recognition in continuously spoken sentences.
\newblock \emph{IEEE transactions on acoustics, speech, and signal processing},
  28\penalty0 (4):\penalty0 357--366, 1980.

\bibitem[D{\'e}fossez et~al.(2022)D{\'e}fossez, Copet, Synnaeve, and
  Adi]{defossez2022high}
Alexandre D{\'e}fossez, Jade Copet, Gabriel Synnaeve, and Yossi Adi.
\newblock High fidelity neural audio compression.
\newblock \emph{arXiv preprint arXiv:2210.13438}, 2022.

\bibitem[Evain et~al.(2021)Evain, Nguyen, Le, Boito, Mdhaffar, Alisamir, Tong,
  Tomashenko, Dinarelli, Parcollet, Allauzen, Estève, Lecouteux, Portet,
  Rossato, Ringeval, Schwab, and Besacier]{evain21_interspeech}
Solène Evain, Ha~Nguyen, Hang Le, Marcely~Zanon Boito, Salima Mdhaffar, Sina
  Alisamir, Ziyi Tong, Natalia Tomashenko, Marco Dinarelli, Titouan Parcollet,
  Alexandre Allauzen, Yannick Estève, Benjamin Lecouteux, François Portet,
  Solange Rossato, Fabien Ringeval, Didier Schwab, and Laurent Besacier.
\newblock { LeBenchmark: A Reproducible Framework for Assessing Self-Supervised
  Representation Learning from Speech}.
\newblock In \emph{Proc. Interspeech 2021}, pp.\  1439--1443, 2021.
\newblock \doi{10.21437/Interspeech.2021-556}.

\bibitem[Feng et~al.(2023)Feng, Dong, Yeh, Yang, Lin, Shi, Chang, Huang, Wu,
  Chang, et~al.]{feng2023superb}
Tzu-hsun Feng, Annie Dong, Ching-Feng Yeh, Shu-wen Yang, Tzu-Quan Lin, Jiatong
  Shi, Kai-Wei Chang, Zili Huang, Haibin Wu, Xuankai Chang, et~al.
\newblock {SUPERB @SLT} 2022: Challenge on generalization and efficiency of
  self-supervised speech representation learning.
\newblock In \emph{2022 IEEE Spoken Language Technology Workshop (SLT)}, pp.\
  1096--1103. IEEE, 2023.

\bibitem[Fitz(2007)]{fitz2007fundamentals}
Michael~P Fitz.
\newblock \emph{Fundamentals of communications systems}.
\newblock McGraw-Hill Education, 2007.

\bibitem[Gaido et~al.(2021)Gaido, Cettolo, Negri, and Turchi]{gaido2021ctc}
Marco Gaido, Mauro Cettolo, Matteo Negri, and Marco Turchi.
\newblock {CTC}-based compression for direct speech translation.
\newblock In \emph{Proceedings of the 16th Conference of the European Chapter
  of the Association for Computational Linguistics: Main Volume}, pp.\
  690--696, 2021.

\bibitem[Gao et~al.(2022)Gao, Mak, and Lin]{gao2022unet}
Zhenke Gao, Man-Wai Mak, and Weiwei Lin.
\newblock {UNet-DenseNet} for robust far-field speaker verification.
\newblock \emph{Proc. Interspeech 2022}, pp.\  3714--3718, 2022.

\bibitem[Gaur et~al.(2021)Gaur, Farris, Haghani, Leal, Moreno, Prasad,
  Ramabhadran, and Zhu]{gaur2021mixture}
Neeraj Gaur, Brian Farris, Parisa Haghani, Isabel Leal, Pedro~J Moreno, Manasa
  Prasad, Bhuvana Ramabhadran, and Yun Zhu.
\newblock Mixture of informed experts for multilingual speech recognition.
\newblock In \emph{ICASSP 2021-2021 IEEE International Conference on Acoustics,
  Speech and Signal Processing (ICASSP)}, pp.\  6234--6238. IEEE, 2021.

\bibitem[Ghiasi et~al.(2019)Ghiasi, Lin, and Le]{ghiasi2019fpn}
Golnaz Ghiasi, Tsung-Yi Lin, and Quoc~V Le.
\newblock Nas-fpn: Learning scalable feature pyramid architecture for object
  detection.
\newblock In \emph{Proceedings of the IEEE/CVF conference on computer vision
  and pattern recognition}, pp.\  7036--7045, 2019.

\bibitem[Guo et~al.(2022)Guo, Deschaintre, Noll, and Roullier]{guo2022u}
Shouchang Guo, Valentin Deschaintre, Douglas Noll, and Arthur Roullier.
\newblock U-attention to textures: hierarchical hourglass vision transformer
  for universal texture synthesis.
\newblock In \emph{Proceedings of the 19th ACM SIGGRAPH European Conference on
  Visual Media Production}, pp.\  1--10, 2022.

\bibitem[Han et~al.(2021)Han, Pan, Tadala, Ma, and Povey]{han2021multistream}
Kyu~J Han, Jing Pan, Venkata Krishna~Naveen Tadala, Tao Ma, and Dan Povey.
\newblock Multistream cnn for robust acoustic modeling.
\newblock In \emph{ICASSP 2021-2021 IEEE International Conference on Acoustics,
  Speech and Signal Processing (ICASSP)}, pp.\  6873--6877. IEEE, 2021.

\bibitem[Hermansky(1990)]{hermansky1990perceptual}
Hynek Hermansky.
\newblock Perceptual linear predictive (plp) analysis of speech.
\newblock \emph{the Journal of the Acoustical Society of America}, 87\penalty0
  (4):\penalty0 1738--1752, 1990.

\bibitem[Hermansky(2013)]{hermansky2013multistream}
Hynek Hermansky.
\newblock Multistream recognition of speech: Dealing with unknown unknowns.
\newblock \emph{Proceedings of the IEEE}, 101\penalty0 (5):\penalty0
  1076--1088, 2013.

\bibitem[Hou et~al.(2020)Hou, Dong, Zhuang, Yang, Shi, and
  Shinozaki]{hou20_interspeech}
Wenxin Hou, Yue Dong, Bairong Zhuang, Longfei Yang, Jiatong Shi, and Takahiro
  Shinozaki.
\newblock {Large-Scale End-to-End Multilingual Speech Recognition and Language
  Identification with Multi-Task Learning}.
\newblock In \emph{Proc. Interspeech 2020}, pp.\  1037--1041, 2020.
\newblock \doi{10.21437/Interspeech.2020-2164}.

\bibitem[Hsu et~al.(2021{\natexlab{a}})Hsu, Bolte, Tsai, Lakhotia,
  Salakhutdinov, and Mohamed]{hsu2021hubert}
Wei-Ning Hsu, Benjamin Bolte, Yao-Hung~Hubert Tsai, Kushal Lakhotia, Ruslan
  Salakhutdinov, and Abdelrahman Mohamed.
\newblock {HuBERT}: Self-supervised speech representation learning by masked
  prediction of hidden units.
\newblock \emph{IEEE/ACM Transactions on Audio, Speech, and Language
  Processing}, 29:\penalty0 3451--3460, 2021{\natexlab{a}}.

\bibitem[Hsu et~al.(2021{\natexlab{b}})Hsu, Sriram, Baevski, Likhomanenko, Xu,
  Pratap, Kahn, Lee, Collobert, Synnaeve, and Auli]{hsu21_interspeech}
Wei-Ning Hsu, Anuroop Sriram, Alexei Baevski, Tatiana Likhomanenko, Qiantong
  Xu, Vineel Pratap, Jacob Kahn, Ann Lee, Ronan Collobert, Gabriel Synnaeve,
  and Michael Auli.
\newblock {Robust wav2vec 2.0: Analyzing Domain Shift in Self-Supervised
  Pre-Training}.
\newblock In \emph{Proc. Interspeech 2021}, pp.\  721--725, 2021{\natexlab{b}}.
\newblock \doi{10.21437/Interspeech.2021-236}.

\bibitem[Huang et~al.(2021)Huang, Wu, and Hayashi]{huang2021any}
Wen-Chin Huang, Yi-Chiao Wu, and Tomoki Hayashi.
\newblock Any-to-one sequence-to-sequence voice conversion using
  self-supervised discrete speech representations.
\newblock In \emph{ICASSP 2021-2021 IEEE International Conference on Acoustics,
  Speech and Signal Processing (ICASSP)}, pp.\  5944--5948. IEEE, 2021.

\bibitem[Huang et~al.(2022{\natexlab{a}})Huang, Yang, Hayashi, Lee, Watanabe,
  and Toda]{huang2021s3prl}
Wen-Chin Huang, Shu-Wen Yang, Tomoki Hayashi, Hung-Yi Lee, Shinji Watanabe, and
  Tomoki Toda.
\newblock S3prl-vc: Open-source voice conversion framework with self-supervised
  speech representations.
\newblock In \emph{ICASSP 2022-2022 IEEE International Conference on Acoustics,
  Speech and Signal Processing (ICASSP)}, pp.\  6552--6556. IEEE,
  2022{\natexlab{a}}.

\bibitem[Huang et~al.(2022{\natexlab{b}})Huang, Yang, Hayashi, and
  Toda]{s3prl-vc-journal}
Wen-Chin Huang, Shu-Wen Yang, Tomoki Hayashi, and Tomoki Toda.
\newblock {A Comparative Study of Self-Supervised Speech Representation Based
  Voice Conversion}.
\newblock \emph{IEEE Journal of Selected Topics in Signal Processing},
  16\penalty0 (6):\penalty0 1308--1318, 2022{\natexlab{b}}.

\bibitem[Huang et~al.(2023)Huang, Violeta, Liu, Shi, Yasuda, and
  Toda]{huang2023singing}
Wen-Chin Huang, Lester~Phillip Violeta, Songxiang Liu, Jiatong Shi, Yusuke
  Yasuda, and Tomoki Toda.
\newblock The singing voice conversion challenge 2023.
\newblock \emph{arXiv preprint arXiv:2306.14422}, 2023.

\bibitem[Huang et~al.(2001)Huang, Acero, Hon, and Reddy]{huang2001spoken}
Xuedong Huang, Alex Acero, Hsiao-Wuen Hon, and Raj Reddy.
\newblock \emph{Spoken language processing: A guide to theory, algorithm, and
  system development}.
\newblock Prentice hall PTR, 2001.

\bibitem[Hung et~al.(2022)Hung, wei Fu, Tseng, Chiang, Tsao, and
  Lin]{hung22_interspeech}
Kuo-Hsuan Hung, Szu wei Fu, Huan-Hsin Tseng, Hsin-Tien Chiang, Yu~Tsao, and
  Chii-Wann Lin.
\newblock {Boosting Self-Supervised Embeddings for Speech Enhancement}.
\newblock In \emph{Proc. Interspeech 2022}, pp.\  186--190, 2022.
\newblock \doi{10.21437/Interspeech.2022-10002}.

\bibitem[Inaguma et~al.(2023)Inaguma, Popuri, Kulikov, Chen, Wang, Chung, Tang,
  Lee, Watanabe, and Pino]{inaguma-etal-2023-unity}
Hirofumi Inaguma, Sravya Popuri, Ilia Kulikov, Peng-Jen Chen, Changhan Wang,
  Yu-An Chung, Yun Tang, Ann Lee, Shinji Watanabe, and Juan Pino.
\newblock {U}nit{Y}: Two-pass direct speech-to-speech translation with discrete
  units.
\newblock In \emph{Proceedings of the 61st Annual Meeting of the Association
  for Computational Linguistics (Volume 1: Long Papers)}, pp.\  15655--15680,
  Toronto, Canada, July 2023. Association for Computational Linguistics.

\bibitem[Joshi et~al.(2020)Joshi, Chen, Liu, Weld, Zettlemoyer, and
  Levy]{joshi2020spanbert}
Mandar Joshi, Danqi Chen, Yinhan Liu, Daniel~S Weld, Luke Zettlemoyer, and Omer
  Levy.
\newblock {SpanBERT}: Improving pre-training by representing and predicting
  spans.
\newblock \emph{Transactions of the association for computational linguistics},
  8:\penalty0 64--77, 2020.

\bibitem[Kahn et~al.(2020)Kahn, Rivi{\`e}re, Zheng, Kharitonov, Xu, Mazar{\'e},
  Karadayi, Liptchinsky, Collobert, Fuegen, et~al.]{kahn2020libri}
Jacob Kahn, Morgane Rivi{\`e}re, Weiyi Zheng, Evgeny Kharitonov, Qiantong Xu,
  Pierre-Emmanuel Mazar{\'e}, Julien Karadayi, Vitaliy Liptchinsky, Ronan
  Collobert, Christian Fuegen, et~al.
\newblock Libri-light: A benchmark for asr with limited or no supervision.
\newblock In \emph{ICASSP 2020-2020 IEEE International Conference on Acoustics,
  Speech and Signal Processing (ICASSP)}, pp.\  7669--7673. IEEE, 2020.

\bibitem[Karita et~al.(2019)Karita, Soplin, Watanabe, Delcroix, Ogawa, and
  Nakatani]{karita19_interspeech}
Shigeki Karita, Nelson Enrique~Yalta Soplin, Shinji Watanabe, Marc Delcroix,
  Atsunori Ogawa, and Tomohiro Nakatani.
\newblock {Improving Transformer-Based End-to-End Speech Recognition with
  Connectionist Temporal Classification and Language Model Integration}.
\newblock In \emph{Proc. Interspeech 2019}, pp.\  1408--1412, 2019.
\newblock \doi{10.21437/Interspeech.2019-1938}.

\bibitem[Kim et~al.(2022)Kim, Gholami, Shaw, Lee, Mangalam, Malik, Mahoney, and
  Keutzer]{kimsqueezeformer}
Sehoon Kim, Amir Gholami, Albert~Eaton Shaw, Nicholas Lee, Karttikeya Mangalam,
  Jitendra Malik, Michael~W Mahoney, and Kurt Keutzer.
\newblock Squeezeformer: An efficient transformer for automatic speech
  recognition.
\newblock In \emph{Advances in Neural Information Processing Systems}, 2022.

\bibitem[Kong et~al.(2020)Kong, Kim, and Bae]{kong2020hifi}
Jungil Kong, Jaehyeon Kim, and Jaekyoung Bae.
\newblock {HiFi-GAN}: Generative adversarial networks for efficient and high
  fidelity speech synthesis.
\newblock \emph{Advances in Neural Information Processing Systems},
  33:\penalty0 17022--17033, 2020.

\bibitem[Lakhotia et~al.(2021)Lakhotia, Kharitonov, Hsu, Adi, Polyak, Bolte,
  Nguyen, Copet, Baevski, Mohamed, et~al.]{lakhotia2021generative}
Kushal Lakhotia, Eugene Kharitonov, Wei-Ning Hsu, Yossi Adi, Adam Polyak,
  Benjamin Bolte, Tu-Anh Nguyen, Jade Copet, Alexei Baevski, Abdelrahman
  Mohamed, et~al.
\newblock On generative spoken language modeling from raw audio.
\newblock \emph{Transactions of the Association for Computational Linguistics},
  9:\penalty0 1336--1354, 2021.

\bibitem[Lee et~al.(2022{\natexlab{a}})Lee, Chen, Wang, Gu, Popuri, Ma, Polyak,
  Adi, He, Tang, et~al.]{lee2022direct}
Ann Lee, Peng-Jen Chen, Changhan Wang, Jiatao Gu, Sravya Popuri, Xutai Ma, Adam
  Polyak, Yossi Adi, Qing He, Yun Tang, et~al.
\newblock Direct speech-to-speech translation with discrete units.
\newblock In \emph{Proceedings of the 60th Annual Meeting of the Association
  for Computational Linguistics (Volume 1: Long Papers)}, pp.\  3327--3339,
  2022{\natexlab{a}}.

\bibitem[Lee et~al.(2022{\natexlab{b}})Lee, Gong, Duquenne, Schwenk, Chen,
  Wang, Popuri, Adi, Pino, Gu, et~al.]{lee2022textless}
Ann Lee, Hongyu Gong, Paul-Ambroise Duquenne, Holger Schwenk, Peng-Jen Chen,
  Changhan Wang, Sravya Popuri, Yossi Adi, Juan Pino, Jiatao Gu, et~al.
\newblock Textless speech-to-speech translation on real data.
\newblock In \emph{Proceedings of the 2022 Conference of the North American
  Chapter of the Association for Computational Linguistics: Human Language
  Technologies}, pp.\  860--872, 2022{\natexlab{b}}.

\bibitem[Li et~al.(2019{\natexlab{a}})Li, Zhang, Sainath, Wu, and
  Chan]{li2019bytes}
Bo~Li, Yu~Zhang, Tara Sainath, Yonghui Wu, and William Chan.
\newblock Bytes are all you need: End-to-end multilingual speech recognition
  and synthesis with bytes.
\newblock In \emph{ICASSP 2019-2019 IEEE International Conference on Acoustics,
  Speech and Signal Processing (ICASSP)}, pp.\  5621--5625. IEEE,
  2019{\natexlab{a}}.

\bibitem[Li et~al.(2022)Li, Pu, Huang, and Huang]{li2022unet}
Rui Li, Dong Pu, Minnie Huang, and Bill Huang.
\newblock {Unet-TTS}: Improving unseen speaker and style transfer in one-shot
  voice cloning.
\newblock In \emph{ICASSP 2022-2022 IEEE International Conference on Acoustics,
  Speech and Signal Processing (ICASSP)}, pp.\  8327--8331. IEEE, 2022.

\bibitem[Li et~al.(2019{\natexlab{b}})Li, Wang, Mallidi, Watanabe, Hori, and
  Hermansky]{li2019multi}
Ruizhi Li, Xiaofei Wang, Sri~Harish Mallidi, Shinji Watanabe, Takaaki Hori, and
  Hynek Hermansky.
\newblock Multi-stream end-to-end speech recognition.
\newblock \emph{IEEE/ACM Transactions on Audio, Speech, and Language
  Processing}, 28:\penalty0 646--655, 2019{\natexlab{b}}.

\bibitem[Li et~al.(2023{\natexlab{a}})Li, Jia, and Chiu]{li2023textless}
Xinjian Li, Ye~Jia, and Chung-Cheng Chiu.
\newblock Textless direct speech-to-speech translation with discrete speech
  representation.
\newblock In \emph{ICASSP 2023-2023 IEEE International Conference on Acoustics,
  Speech and Signal Processing (ICASSP)}, pp.\  1--5. IEEE, 2023{\natexlab{a}}.

\bibitem[Li et~al.(2023{\natexlab{b}})Li, Yuan, Zhang, Ma, Chen, Yin, Lin,
  Ragni, Benetos, Gyenge, et~al.]{li2023mert}
Yizhi Li, Ruibin Yuan, Ge~Zhang, Yinghao Ma, Xingran Chen, Hanzhi Yin, Chenghua
  Lin, Anton Ragni, Emmanouil Benetos, Norbert Gyenge, et~al.
\newblock {MERT}: Acoustic music understanding model with large-scale
  self-supervised training.
\newblock \emph{arXiv preprint arXiv:2306.00107}, 2023{\natexlab{b}}.

\bibitem[Li et~al.(2023{\natexlab{c}})Li, Wu, Li, and Liu]{li23i_interspeech}
Yuang Li, Yu~Wu, Jinyu Li, and Shujie Liu.
\newblock {Accelerating Transducers through Adjacent Token Merging}.
\newblock In \emph{Proc. Interspeech 2023}, pp.\  1379--1383,
  2023{\natexlab{c}}.
\newblock \doi{10.21437/Interspeech.2023-599}.

\bibitem[Lian et~al.(2022)Lian, Zhang, Anumanchipalli, and Yu]{lian2022utts}
Jiachen Lian, Chunlei Zhang, Gopala~Krishna Anumanchipalli, and Dong Yu.
\newblock Utts: Unsupervised tts with conditional disentangled sequential
  variational auto-encoder.
\newblock \emph{arXiv preprint arXiv:2206.02512}, 2022.

\bibitem[Lin et~al.(2022{\natexlab{a}})Lin, Chuang, Chung, wen Yang, Chen,
  Dong, Li, Mohamed, yi~Lee, and shan Lee]{lin22c_interspeech}
Guan-Ting Lin, Yung-Sung Chuang, Ho-Lam Chung, Shu wen Yang, Hsuan-Jui Chen,
  Shuyan~Annie Dong, Shang-Wen Li, Abdelrahman Mohamed, Hung yi~Lee, and Lin
  shan Lee.
\newblock {DUAL: Discrete Spoken Unit Adaptive Learning for Textless Spoken
  Question Answering}.
\newblock In \emph{Proc. Interspeech 2022}, pp.\  5165--5169,
  2022{\natexlab{a}}.
\newblock \doi{10.21437/Interspeech.2022-612}.

\bibitem[Lin et~al.(2023)Lin, Feng, Huang, Tseng, Lin, Li, Lee, and
  Ward]{lin2023utility}
Guan-Ting Lin, Chi-Luen Feng, Wei-Ping Huang, Yuan Tseng, Tzu-Han Lin, Chen-An
  Li, Hung-yi Lee, and Nigel~G Ward.
\newblock On the utility of self-supervised models for prosody-related tasks.
\newblock In \emph{2022 IEEE Spoken Language Technology Workshop (SLT)}, pp.\
  1104--1111. IEEE, 2023.

\bibitem[Lin et~al.(2022{\natexlab{b}})Lin, Lee, and Tang]{lin2022melhubert}
Tzu-Quan Lin, Hung-yi Lee, and Hao Tang.
\newblock Melhubert: A simplified hubert on mel spectrogram.
\newblock \emph{arXiv preprint arXiv:2211.09944}, 2022{\natexlab{b}}.

\bibitem[Liu et~al.(2020{\natexlab{a}})Liu, Yang, Chi, Hsu, and
  Lee]{liu2020mockingjay}
Andy~T Liu, Shu-wen Yang, Po-Han Chi, Po-chun Hsu, and Hung-yi Lee.
\newblock Mockingjay: Unsupervised speech representation learning with deep
  bidirectional transformer encoders.
\newblock In \emph{ICASSP 2020-2020 IEEE International Conference on Acoustics,
  Speech and Signal Processing (ICASSP)}, pp.\  6419--6423. IEEE,
  2020{\natexlab{a}}.

\bibitem[Liu et~al.(2020{\natexlab{b}})Liu, Chen, Zhang, Jiang, Hu, Ling, and
  Dai]{liu2020non}
Li-Juan Liu, Yan-Nian Chen, Jing-Xuan Zhang, Yuan Jiang, Ya-Jun Hu, Zhen-Hua
  Ling, and Li-Rong Dai.
\newblock Non-parallel voice conversion with autoregressive conversion model
  and duration adjustment.
\newblock In \emph{Proc. Joint Workshop for the Blizzard Challenge and Voice
  Conversion Challenge 2020}, pp.\  126--130, 2020{\natexlab{b}}.

\bibitem[Liu et~al.(2022)Liu, Mallol-Ragolta, Parada-Cabaleiro, Qian, Jing,
  Kathan, Hu, and Schuller]{liu2022audio}
Shuo Liu, Adria Mallol-Ragolta, Emilia Parada-Cabaleiro, Kun Qian, Xin Jing,
  Alexander Kathan, Bin Hu, and Bjoern~W Schuller.
\newblock Audio self-supervised learning: A survey.
\newblock \emph{Patterns}, 3\penalty0 (12), 2022.

\bibitem[Lugosch et~al.(2022)Lugosch, Likhomanenko, Synnaeve, and
  Collobert]{lugosch2022pseudo}
Loren Lugosch, Tatiana Likhomanenko, Gabriel Synnaeve, and Ronan Collobert.
\newblock Pseudo-labeling for massively multilingual speech recognition.
\newblock In \emph{ICASSP 2022-2022 IEEE International Conference on Acoustics,
  Speech and Signal Processing (ICASSP)}, pp.\  7687--7691. IEEE, 2022.

\bibitem[Luo et~al.(2021)Luo, Wang, Cheng, Jiang, and Xiao]{luo2021multi}
Jian Luo, Jianzong Wang, Ning Cheng, Guilin Jiang, and Jing Xiao.
\newblock Multi-quartznet: Multi-resolution convolution for speech recognition
  with multi-layer feature fusion.
\newblock In \emph{2021 IEEE Spoken Language Technology Workshop (SLT)}, pp.\
  82--88. IEEE, 2021.

\bibitem[Ma et~al.(2023)Ma, Yuan, Li, Zhang, Chen, Yin, Lin, Benetos, Ragni,
  Gyenge, et~al.]{ma2023effectiveness}
Yinghao Ma, Ruibin Yuan, Yizhi Li, Ge~Zhang, Xingran Chen, Hanzhi Yin, Chenghua
  Lin, Emmanouil Benetos, Anton Ragni, Norbert Gyenge, et~al.
\newblock On the effectiveness of speech self-supervised learning for music.
\newblock \emph{arXiv preprint arXiv:2307.05161}, 2023.

\bibitem[Mallidi \& Hermansky(2016)Mallidi and Hermansky]{mallidi2016novel}
Sri~Harish Mallidi and Hynek Hermansky.
\newblock Novel neural network based fusion for multistream {ASR}.
\newblock In \emph{2016 IEEE International Conference on Acoustics, Speech and
  Signal Processing (ICASSP)}, pp.\  5680--5684. IEEE, 2016.

\bibitem[Mallidi et~al.(2018)]{mallidi2018practical}
Sri Harish~Reddy Mallidi et~al.
\newblock \emph{A practical and efficient multistream framework for noise
  robust speech recognition}.
\newblock PhD thesis, Johns Hopkins University, 2018.

\bibitem[Masuyama et~al.(2023)Masuyama, Chang, Cornell, Watanabe, and
  Ono]{masuyama2023end}
Yoshiki Masuyama, Xuankai Chang, Samuele Cornell, Shinji Watanabe, and Nobutaka
  Ono.
\newblock End-to-end integration of speech recognition, dereverberation,
  beamforming, and self-supervised learning representation.
\newblock In \emph{2022 IEEE Spoken Language Technology Workshop (SLT)}, pp.\
  260--265. IEEE, 2023.

\bibitem[Mathieu et~al.(2016)Mathieu, Couprie, and LeCun]{mathieu2016deep}
Michael Mathieu, Camille Couprie, and Yann LeCun.
\newblock Deep multi-scale video prediction beyond mean square error.
\newblock In \emph{4th International Conference on Learning Representations,
  ICLR 2016}, 2016.

\bibitem[Melekhov et~al.(2017)Melekhov, Ylioinas, Kannala, and
  Rahtu]{melekhov2017image}
Iaroslav Melekhov, Juha Ylioinas, Juho Kannala, and Esa Rahtu.
\newblock Image-based localization using hourglass networks.
\newblock In \emph{Proceedings of the IEEE international conference on computer
  vision workshops}, pp.\  879--886, 2017.

\bibitem[Meng et~al.(2023)Meng, Chen, Shi, Watanabe, Garcia, Lee, and
  Tang]{meng2023compressing}
Yen Meng, Hsuan-Jui Chen, Jiatong Shi, Shinji Watanabe, Paola Garcia, Hung-yi
  Lee, and Hao Tang.
\newblock On compressing sequences for self-supervised speech models.
\newblock In \emph{2022 IEEE Spoken Language Technology Workshop (SLT)}, pp.\
  1128--1135. IEEE, 2023.

\bibitem[Mohamed et~al.(2022)Mohamed, Lee, Borgholt, Havtorn, Edin, Igel,
  Kirchhoff, Li, Livescu, Maal{\o}e, et~al.]{mohamed2022self}
Abdelrahman Mohamed, Hung-yi Lee, Lasse Borgholt, Jakob~D Havtorn, Joakim Edin,
  Christian Igel, Katrin Kirchhoff, Shang-Wen Li, Karen Livescu, Lars
  Maal{\o}e, et~al.
\newblock Self-supervised speech representation learning: A review.
\newblock \emph{IEEE Journal of Selected Topics in Signal Processing}, 2022.

\bibitem[Nawrot et~al.(2022)Nawrot, Tworkowski, Tyrolski, Kaiser, Wu, Szegedy,
  and Michalewski]{nawrot2022hierarchical}
Piotr Nawrot, Szymon Tworkowski, Micha{\l} Tyrolski, {\L}ukasz Kaiser, Yuhuai
  Wu, Christian Szegedy, and Henryk Michalewski.
\newblock Hierarchical transformers are more efficient language models.
\newblock In \emph{Findings of the Association for Computational Linguistics:
  NAACL 2022}, pp.\  1559--1571, 2022.

\bibitem[Nawrot et~al.(2023)Nawrot, Chorowski, Lancucki, and
  Ponti]{nawrot-etal-2023-efficient}
Piotr Nawrot, Jan Chorowski, Adrian Lancucki, and Edoardo~Maria Ponti.
\newblock Efficient transformers with dynamic token pooling.
\newblock In \emph{Proceedings of the 61st Annual Meeting of the Association
  for Computational Linguistics (Volume 1: Long Papers)}, pp.\  6403--6417,
  Toronto, Canada, July 2023. Association for Computational Linguistics.
\newblock \doi{10.18653/v1/2023.acl-long.353}.

\bibitem[Newell et~al.(2016)Newell, Yang, and Deng]{newell2016stacked}
Alejandro Newell, Kaiyu Yang, and Jia Deng.
\newblock Stacked hourglass networks for human pose estimation.
\newblock In \emph{Computer Vision--ECCV 2016: 14th European Conference,
  Amsterdam, The Netherlands, October 11-14, 2016, Proceedings, Part VIII 14},
  pp.\  483--499. Springer, 2016.

\bibitem[Nguyen et~al.(2023)Nguyen, Hsu, D'Avirro, Shi, Gat, Fazel-Zarani,
  Remez, Copet, Synnaeve, Hassid, et~al.]{nguyen2023expresso}
Tu~Anh Nguyen, Wei-Ning Hsu, Antony D'Avirro, Bowen Shi, Itai Gat, Maryam
  Fazel-Zarani, Tal Remez, Jade Copet, Gabriel Synnaeve, Michael Hassid, et~al.
\newblock Expresso: A benchmark and analysis of discrete expressive speech
  resynthesis.
\newblock \emph{arXiv preprint arXiv:2308.05725}, 2023.

\bibitem[Oord et~al.(2018)Oord, Li, and Vinyals]{oord2018representation}
Aaron van~den Oord, Yazhe Li, and Oriol Vinyals.
\newblock Representation learning with contrastive predictive coding.
\newblock \emph{arXiv preprint arXiv:1807.03748}, 2018.

\bibitem[Otake et~al.(2023)Otake, Kawakami, and Inoue]{otake2023parameter}
Shinta Otake, Rei Kawakami, and Nakamasa Inoue.
\newblock Parameter efficient transfer learning for various speech processing
  tasks.
\newblock In \emph{ICASSP 2023-2023 IEEE International Conference on Acoustics,
  Speech and Signal Processing (ICASSP)}, pp.\  1--5. IEEE, 2023.

\bibitem[Ott et~al.(2019)Ott, Edunov, Baevski, Fan, Gross, Ng, Grangier, and
  Auli]{ott2019fairseq}
Myle Ott, Sergey Edunov, Alexei Baevski, Angela Fan, Sam Gross, Nathan Ng,
  David Grangier, and Michael Auli.
\newblock fairseq: A fast, extensible toolkit for sequence modeling.
\newblock In \emph{Proceedings of the 2019 Conference of the North American
  Chapter of the Association for Computational Linguistics (Demonstrations)},
  pp.\  48--53, 2019.

\bibitem[Panayotov et~al.(2015)Panayotov, Chen, Povey, and
  Khudanpur]{panayotov2015librispeech}
Vassil Panayotov, Guoguo Chen, Daniel Povey, and Sanjeev Khudanpur.
\newblock Librispeech: an asr corpus based on public domain audio books.
\newblock In \emph{2015 IEEE international conference on acoustics, speech and
  signal processing (ICASSP)}, pp.\  5206--5210. IEEE, 2015.

\bibitem[Polyak et~al.(2021)Polyak, Adi, Copet, Kharitonov, Lakhotia, Hsu,
  Mohamed, and Dupoux]{polyak21_interspeech}
Adam Polyak, Yossi Adi, Jade Copet, Eugene Kharitonov, Kushal Lakhotia,
  Wei-Ning Hsu, Abdelrahman Mohamed, and Emmanuel Dupoux.
\newblock {Speech Resynthesis from Discrete Disentangled Self-Supervised
  Representations}.
\newblock In \emph{Proc. Interspeech 2021}, pp.\  3615--3619, 2021.
\newblock \doi{10.21437/Interspeech.2021-475}.

\bibitem[Qian et~al.(2022)Qian, Zhang, Gao, Ni, Lai, Cox, Hasegawa-Johnson, and
  Chang]{qian2022contentvec}
Kaizhi Qian, Yang Zhang, Heting Gao, Junrui Ni, Cheng-I Lai, David Cox, Mark
  Hasegawa-Johnson, and Shiyu Chang.
\newblock Contentvec: An improved self-supervised speech representation by
  disentangling speakers.
\newblock In \emph{International Conference on Machine Learning}, pp.\
  18003--18017. PMLR, 2022.

\bibitem[Roberts \& Mullis(1987)Roberts and Mullis]{roberts1987digital}
Richard~A Roberts and Clifford~T Mullis.
\newblock \emph{Digital signal processing}.
\newblock Addison-Wesley Longman Publishing Co., Inc., 1987.

\bibitem[Shi et~al.(2023{\natexlab{a}})Shi, Berrebbi, Chen, Hu, Huang, Chung,
  Chang, Li, Mohamed, yi~Lee, and Watanabe]{mlsuperb}
Jiatong Shi, Dan Berrebbi, William Chen, En-Pei Hu, Wei-Ping Huang, Ho-Lam
  Chung, Xuankai Chang, Shang-Wen Li, Abdelrahman Mohamed, Hung yi~Lee, and
  Shinji Watanabe.
\newblock {ML-SUPERB: Multilingual Speech Universal PERformance Benchmark}.
\newblock In \emph{Proc. Interspeech 2023}, pp.\  884--888, 2023{\natexlab{a}}.
\newblock \doi{10.21437/Interspeech.2023-1316}.

\bibitem[Shi et~al.(2023{\natexlab{b}})Shi, Chen, Berrebbi, Wang, Huang, Hu,
  Chuang, Chang, Tang, Li, et~al.]{shi2023findings}
Jiatong Shi, William Chen, Dan Berrebbi, Hsiu-Hsuan Wang, Wei-Ping Huang,
  En-Pei Hu, Ho-Lam Chuang, Xuankai Chang, Yuxun Tang, Shang-Wen Li, et~al.
\newblock Findings of the 2023 {ML-SUPERB} challenge: Pre-training and
  evaluation over more languages and beyond.
\newblock In \emph{2023 IEEE Automatic Speech Recognition and Understanding
  Workshop (ASRU)}, pp.\  1--8. IEEE, 2023{\natexlab{b}}.

\bibitem[Shi et~al.(2023{\natexlab{c}})Shi, Hsu, Chung, Gao, Garcia, Watanabe,
  Lee, and Lee]{shi2023bridging}
Jiatong Shi, Chan-Jan Hsu, Holam Chung, Dongji Gao, Paola Garcia, Shinji
  Watanabe, Ann Lee, and Hung-yi Lee.
\newblock Bridging speech and textual pre-trained models with unsupervised
  {ASR}.
\newblock In \emph{ICASSP 2023-2023 IEEE International Conference on Acoustics,
  Speech and Signal Processing (ICASSP)}, pp.\  1--5. IEEE, 2023{\natexlab{c}}.

\bibitem[Shi et~al.(2023{\natexlab{d}})Shi, Tang, Inaguma, Gong, Pino, and
  Watanabe]{shi23h_interspeech}
Jiatong Shi, Yun Tang, Hirofumi Inaguma, Hongyu Gong, Juan Pino, and Shinji
  Watanabe.
\newblock {Exploration on HuBERT with Multiple Resolution}.
\newblock In \emph{Proc. Interspeech 2023}, pp.\  3287--3291,
  2023{\natexlab{d}}.
\newblock \doi{10.21437/Interspeech.2023-1337}.

\bibitem[Shi et~al.(2023{\natexlab{e}})Shi, Tang, Lee, Inaguma, Wang, Pino, and
  Watanabe]{shi2023enhancing}
Jiatong Shi, Yun Tang, Ann Lee, Hirofumi Inaguma, Changhan Wang, Juan Pino, and
  Shinji Watanabe.
\newblock Enhancing speech-to-speech translation with multiple tts targets.
\newblock In \emph{ICASSP 2023-2023 IEEE International Conference on Acoustics,
  Speech and Signal Processing (ICASSP)}, pp.\  1--5. IEEE, 2023{\natexlab{e}}.

\bibitem[Shi et~al.(2021)Shi, Chang, Hayashi, Lu, Watanabe, and
  Xu]{shi2021discretization}
Jing Shi, Xuankai Chang, Tomoki Hayashi, Yen-Ju Lu, Shinji Watanabe, and Bo~Xu.
\newblock Discretization and re-synthesis: an alternative method to solve the
  cocktail party problem.
\newblock \emph{arXiv preprint arXiv:2112.09382}, 2021.

\bibitem[Shi et~al.(2019)Shi, Lin, Liu, Liu, Hayakawa, Harada, and
  Han]{shi19f_interspeech}
Ziqiang Shi, Huibin Lin, Liu Liu, Rujie Liu, Shoji Hayakawa, Shouji Harada, and
  Jiqing Han.
\newblock {End-to-End Monaural Speech Separation with Multi-Scale Dynamic
  Weighted Gated Dilated Convolutional Pyramid Network}.
\newblock In \emph{Proc. Interspeech 2019}, pp.\  4614--4618, 2019.
\newblock \doi{10.21437/Interspeech.2019-1292}.

\bibitem[Sicherman \& Adi(2023)Sicherman and Adi]{sicherman2023analysing}
Amitay Sicherman and Yossi Adi.
\newblock Analysing discrete self supervised speech representation for spoken
  language modeling.
\newblock In \emph{ICASSP 2023-2023 IEEE International Conference on Acoustics,
  Speech and Signal Processing (ICASSP)}, pp.\  1--5. IEEE, 2023.

\bibitem[Toshniwal et~al.(2018)Toshniwal, Sainath, Weiss, Li, Moreno,
  Weinstein, and Rao]{toshniwal2018multilingual}
Shubham Toshniwal, Tara~N Sainath, Ron~J Weiss, Bo~Li, Pedro Moreno, Eugene
  Weinstein, and Kanishka Rao.
\newblock Multilingual speech recognition with a single end-to-end model.
\newblock In \emph{2018 IEEE international conference on acoustics, speech and
  signal processing (ICASSP)}, pp.\  4904--4908. IEEE, 2018.

\bibitem[Tsai et~al.(2022)Tsai, Chang, Huang, Huang, Lakhotia, Yang, Dong, Liu,
  Lai, Shi, Chang, Hall, Chen, Li, Watanabe, Mohamed, and
  Lee]{tsai-etal-2022-superb}
Hsiang-Sheng Tsai, Heng-Jui Chang, Wen-Chin Huang, Zili Huang, Kushal Lakhotia,
  Shu-wen Yang, Shuyan Dong, Andy Liu, Cheng-I Lai, Jiatong Shi, Xuankai Chang,
  Phil Hall, Hsuan-Jui Chen, Shang-Wen Li, Shinji Watanabe, Abdelrahman
  Mohamed, and Hung-yi Lee.
\newblock {SUPERB}-{SG}: Enhanced speech processing universal {PER}formance
  benchmark for semantic and generative capabilities.
\newblock In \emph{Proceedings of the 60th Annual Meeting of the Association
  for Computational Linguistics (Volume 1: Long Papers)}, pp.\  8479--8492,
  Dublin, Ireland, May 2022. Association for Computational Linguistics.

\bibitem[Turian et~al.(2022)Turian, Shier, Khan, Raj, Schuller, Steinmetz,
  Malloy, Tzanetakis, Velarde, McNally, et~al.]{turian2022hear}
Joseph Turian, Jordie Shier, Humair~Raj Khan, Bhiksha Raj, Bj{\"o}rn~W
  Schuller, Christian~J Steinmetz, Colin Malloy, George Tzanetakis, Gissel
  Velarde, Kirk McNally, et~al.
\newblock Hear: Holistic evaluation of audio representations.
\newblock In \emph{NeurIPS 2021 Competitions and Demonstrations Track}, pp.\
  125--145. PMLR, 2022.

\bibitem[Wang et~al.(2021{\natexlab{a}})Wang, Riviere, Lee, Wu, Talnikar,
  Haziza, Williamson, Pino, and Dupoux]{wang2021voxpopuli}
Changhan Wang, Morgane Riviere, Ann Lee, Anne Wu, Chaitanya Talnikar, Daniel
  Haziza, Mary Williamson, Juan Pino, and Emmanuel Dupoux.
\newblock Voxpopuli: A large-scale multilingual speech corpus for
  representation learning, semi-supervised learning and interpretation.
\newblock In \emph{Proceedings of the 59th Annual Meeting of the Association
  for Computational Linguistics and the 11th International Joint Conference on
  Natural Language Processing (Volume 1: Long Papers)}, pp.\  993--1003,
  2021{\natexlab{a}}.

\bibitem[Wang et~al.(2022)Wang, Zhang, Wang, Cheng, and Xiao]{wang2022drvc}
Qiqi Wang, Xulong Zhang, Jianzong Wang, Ning Cheng, and Jing Xiao.
\newblock {DRVC}: A framework of any-to-any voice conversion with
  self-supervised learning.
\newblock In \emph{ICASSP 2022-2022 IEEE International Conference on Acoustics,
  Speech and Signal Processing (ICASSP)}, pp.\  3184--3188. IEEE, 2022.

\bibitem[Wang et~al.(2021{\natexlab{b}})Wang, Boumadane, and
  Heba]{wang2021fine}
Yingzhi Wang, Abdelmoumene Boumadane, and Abdelwahab Heba.
\newblock A fine-tuned wav2vec 2.0/{HuBERT} benchmark for speech emotion
  recognition, speaker verification and spoken language understanding.
\newblock \emph{arXiv preprint arXiv:2111.02735}, 2021{\natexlab{b}}.

\bibitem[Watanabe et~al.(2017)Watanabe, Hori, and
  Hershey]{watanabe2017language}
Shinji Watanabe, Takaaki Hori, and John~R Hershey.
\newblock Language independent end-to-end architecture for joint language
  identification and speech recognition.
\newblock In \emph{2017 IEEE Automatic Speech Recognition and Understanding
  Workshop (ASRU)}, pp.\  265--271. IEEE, 2017.

\bibitem[Watanabe et~al.(2018)Watanabe, Hori, Karita, Hayashi, Nishitoba, Unno,
  {Enrique Yalta Soplin}, Heymann, Wiesner, Chen, Renduchintala, and
  Ochiai]{watanabe18_interspeech}
Shinji Watanabe, Takaaki Hori, Shigeki Karita, Tomoki Hayashi, Jiro Nishitoba,
  Yuya Unno, Nelson {Enrique Yalta Soplin}, Jahn Heymann, Matthew Wiesner,
  Nanxin Chen, Adithya Renduchintala, and Tsubasa Ochiai.
\newblock {ESPnet: End-to-End Speech Processing Toolkit}.
\newblock In \emph{Proc. Interspeech 2018}, pp.\  2207--2211, 2018.
\newblock \doi{10.21437/Interspeech.2018-1456}.

\bibitem[Wu et~al.(2023)Wu, Gaur, Chen, Zhou, Zhu, Wang, Li, Liu, Ren, Liu,
  et~al.]{wu2023decoder}
Jian Wu, Yashesh Gaur, Zhuo Chen, Long Zhou, Yimeng Zhu, Tianrui Wang, Jinyu
  Li, Shujie Liu, Bo~Ren, Linquan Liu, et~al.
\newblock On decoder-only architecture for speech-to-text and large language
  model integration.
\newblock \emph{arXiv preprint arXiv:2307.03917}, 2023.

\bibitem[Wu et~al.(2022)Wu, Polyak, Taigman, Fong, Agrawal, and
  He]{wu2022multilingual}
Jilong Wu, Adam Polyak, Yaniv Taigman, Jason Fong, Prabhav Agrawal, and Qing
  He.
\newblock Multilingual text-to-speech training using cross language voice
  conversion and self-supervised learning of speech representations.
\newblock In \emph{ICASSP 2022-2022 IEEE International Conference on Acoustics,
  Speech and Signal Processing (ICASSP)}, pp.\  8017--8021. IEEE, 2022.

\bibitem[Xiang et~al.(2021)Xiang, Zhang, and Chen]{xiang2021convolutional}
Xiaoxiao Xiang, Xiaojuan Zhang, and Haozhe Chen.
\newblock A convolutional network with multi-scale and attention mechanisms for
  end-to-end single-channel speech enhancement.
\newblock \emph{IEEE Signal Processing Letters}, 28:\penalty0 1455--1459, 2021.

\bibitem[Xu et~al.(2020)Xu, Rao, Chng, and Li]{xu2020spex}
Chenglin Xu, Wei Rao, Eng~Siong Chng, and Haizhou Li.
\newblock Spex: Multi-scale time domain speaker extraction network.
\newblock \emph{IEEE/ACM transactions on audio, speech, and language
  processing}, 28:\penalty0 1370--1384, 2020.

\bibitem[Yamamoto et~al.(2020)Yamamoto, Song, and Kim]{yamamoto2020parallel}
Ryuichi Yamamoto, Eunwoo Song, and Jae-Min Kim.
\newblock Parallel {WaveGAN}: A fast waveform generation model based on
  generative adversarial networks with multi-resolution spectrogram.
\newblock In \emph{ICASSP 2020-2020 IEEE International Conference on Acoustics,
  Speech and Signal Processing (ICASSP)}, pp.\  6199--6203. IEEE, 2020.

\bibitem[Yan et~al.(2023)Yan, Shi, Tang, Inaguma, Peng, Dalmia, Pol{\'a}k,
  Fernandes, Berrebbi, Hayashi, Zhang, Ni, Hira, Maiti, Pino, and
  Watanabe]{yan-etal-2023-espnet}
Brian Yan, Jiatong Shi, Yun Tang, Hirofumi Inaguma, Yifan Peng, Siddharth
  Dalmia, Peter Pol{\'a}k, Patrick Fernandes, Dan Berrebbi, Tomoki Hayashi,
  Xiaohui Zhang, Zhaoheng Ni, Moto Hira, Soumi Maiti, Juan Pino, and Shinji
  Watanabe.
\newblock {ESP}net-{ST}-v2: Multipurpose spoken language translation toolkit.
\newblock In \emph{Proceedings of the 61st Annual Meeting of the Association
  for Computational Linguistics (Volume 3: System Demonstrations)}, pp.\
  400--411, Toronto, Canada, July 2023. Association for Computational
  Linguistics.
\newblock \doi{10.18653/v1/2023.acl-demo.38}.

\bibitem[Yang et~al.(2017)Yang, Liu, and Zhang]{yang2017stacked}
Jing Yang, Qingshan Liu, and Kaihua Zhang.
\newblock Stacked hourglass network for robust facial landmark localisation.
\newblock In \emph{Proceedings of the IEEE conference on computer vision and
  pattern recognition workshops}, pp.\  79--87, 2017.

\bibitem[Yang et~al.(2021)Yang, Chi, Chuang, Lai, Lakhotia, Lin, Liu, Shi,
  Chang, Lin, Huang, Tseng, tik Lee, Liu, Huang, Dong, Li, Watanabe, Mohamed,
  and yi~Lee]{superb}
Shu-Wen Yang, Po-Han Chi, Yung-Sung Chuang, Cheng-I~Jeff Lai, Kushal Lakhotia,
  Yist~Y. Lin, Andy~T. Liu, Jiatong Shi, Xuankai Chang, Guan-Ting Lin,
  Tzu-Hsien Huang, Wei-Cheng Tseng, Ko~tik Lee, Da-Rong Liu, Zili Huang, Shuyan
  Dong, Shang-Wen Li, Shinji Watanabe, Abdelrahman Mohamed, and Hung yi~Lee.
\newblock {SUPERB: Speech Processing Universal PERformance Benchmark}.
\newblock In \emph{Proc. Interspeech 2021}, pp.\  1194--1198, 2021.
\newblock \doi{10.21437/Interspeech.2021-1775}.

\bibitem[Yang \& Ramanan(2015)Yang and Ramanan]{yang2015multi}
Songfan Yang and Deva Ramanan.
\newblock Multi-scale recognition with dag-cnns.
\newblock In \emph{Proceedings of the IEEE international conference on computer
  vision}, pp.\  1215--1223, 2015.

\bibitem[Yi et~al.(2020)Yi, Huang, Tian, Yamagishi, Das, Kinnunen, Ling, and
  Toda]{yi20_vccbc}
Zhao Yi, Wen-Chin Huang, Xiaohai Tian, Junichi Yamagishi, Rohan~Kumar Das, Tomi
  Kinnunen, Zhen-Hua Ling, and Tomoki Toda.
\newblock {Voice Conversion Challenge 2020 –- Intra-lingual semi-parallel and
  cross-lingual voice conversion}.
\newblock In \emph{Proc. Joint Workshop for the Blizzard Challenge and Voice
  Conversion Challenge 2020}, pp.\  80--98, 2020.
\newblock \doi{10.21437/VCCBC.2020-14}.

\bibitem[Yoneyama et~al.(2023)Yoneyama, Wu, and Toda]{yoneyama2023source}
Reo Yoneyama, Yi-Chiao Wu, and Tomoki Toda.
\newblock Source-filter hifi-gan: Fast and pitch controllable high-fidelity
  neural vocoder.
\newblock In \emph{ICASSP 2023-2023 IEEE International Conference on Acoustics,
  Speech and Signal Processing (ICASSP)}, pp.\  1--5. IEEE, 2023.

\bibitem[Yu \& Deng(2016)Yu and Deng]{yu2016automatic}
Dong Yu and Lin Deng.
\newblock \emph{Automatic speech recognition}, volume~1.
\newblock Springer, 2016.

\bibitem[Yuan et~al.(2023)Yuan, Ma, Li, Zhang, Chen, Yin, Zhuo, Liu, Huang,
  Tian, et~al.]{yuan2023marble}
Ruibin Yuan, Yinghao Ma, Yizhi Li, Ge~Zhang, Xingran Chen, Hanzhi Yin, Le~Zhuo,
  Yiqi Liu, Jiawen Huang, Zeyue Tian, et~al.
\newblock {MARBLE}: Music audio representation benchmark for universal
  evaluation.
\newblock \emph{arXiv preprint arXiv:2306.10548}, 2023.

\bibitem[Zhai et~al.(2023)Zhai, Li, Qin, Yi, Xie, Zhang, Yao, Wu, and
  Jia]{ijcai2023p585}
Mingliang Zhai, Yulin Li, Xiameng Qin, Chen Yi, Qunyi Xie, Chengquan Zhang, Kun
  Yao, Yuwei Wu, and Yunde Jia.
\newblock Fast-struc{TexT}: An efficient {H}ourglass transformer with
  modality-guided dynamic token merge for document understanding.
\newblock In Edith Elkind (ed.), \emph{Proceedings of the Thirty-Second
  International Joint Conference on Artificial Intelligence, {IJCAI-23}}, pp.\
  5269--5277. International Joint Conferences on Artificial Intelligence
  Organization, 8 2023.
\newblock \doi{10.24963/ijcai.2023/585}.
\newblock Main Track.

\bibitem[Zhang et~al.(2022{\natexlab{a}})Zhang, Li, Sainath, Strohman,
  Mavandadi, Chang, and Haghani]{zhang22da_interspeech}
Chao Zhang, Bo~Li, Tara Sainath, Trevor Strohman, Sepand Mavandadi, Shuo-Yiin
  Chang, and Parisa Haghani.
\newblock {Streaming End-to-End Multilingual Speech Recognition with Joint
  Language Identification}.
\newblock In \emph{Proc. Interspeech 2022}, pp.\  3223--3227,
  2022{\natexlab{a}}.
\newblock \doi{10.21437/Interspeech.2022-11249}.

\bibitem[Zhang et~al.(2022{\natexlab{b}})Zhang, Yu, Wang, and
  Wei]{zhang2022multi}
Guochang Zhang, Libiao Yu, Chunliang Wang, and Jianqiang Wei.
\newblock Multi-scale temporal frequency convolutional network with axial
  attention for speech enhancement.
\newblock In \emph{ICASSP 2022-2022 IEEE International Conference on Acoustics,
  Speech and Signal Processing (ICASSP)}, pp.\  9122--9126. IEEE,
  2022{\natexlab{b}}.

\bibitem[Zhang \& Wang(2020)Zhang and Wang]{zhang20x_interspeech}
Lu~Zhang and Mingjiang Wang.
\newblock {Multi-Scale TCN: Exploring Better Temporal DNN Model for Causal
  Speech Enhancement}.
\newblock In \emph{Proc. Interspeech 2020}, pp.\  2672--2676, 2020.
\newblock \doi{10.21437/Interspeech.2020-1104}.

\bibitem[Zhang et~al.(2019)Zhang, Duan, Liao, Liu, Wu, and
  Xie]{zhang2019research}
Yi~Zhang, Qing Duan, Yun Liao, Junhui Liu, Ruiqiong Wu, and Bisen Xie.
\newblock Research on speech enhancement algorithm based on {SA-Unet}.
\newblock In \emph{2019 4th International Conference on Mechanical, Control and
  Computer Engineering (ICMCCE)}, pp.\  818--8183. IEEE, 2019.

\bibitem[Zhao \& Zhang(2022)Zhao and Zhang]{zhao2022improving}
Jing Zhao and Wei-Qiang Zhang.
\newblock Improving automatic speech recognition performance for low-resource
  languages with self-supervised models.
\newblock \emph{IEEE Journal of Selected Topics in Signal Processing},
  16\penalty0 (6):\penalty0 1227--1241, 2022.

\bibitem[Zhao et~al.(2021)Zhao, Zhao, Wang, and Han]{zhao2021unet++}
Tuo Zhao, Yunxin Zhao, Shaojun Wang, and Mei Han.
\newblock Unet++-based multi-channel speech dereverberation and distant speech
  recognition.
\newblock In \emph{2021 12th International Symposium on Chinese Spoken Language
  Processing (ISCSLP)}, pp.\  1--5. IEEE, 2021.

\bibitem[Zhu \& Alwan(2000)Zhu and Alwan]{zhu2000use}
Qifeng Zhu and Abeer Alwan.
\newblock On the use of variable frame rate analysis in speech recognition.
\newblock In \emph{2000 IEEE international conference on acoustics, speech, and
  signal processing (ICASSP)}, pp.\  1783--1786. IEEE, 2000.

\end{thebibliography}
\bibliographystyle{iclr2024_conference}

\newpage

\appendix
\section{Pre-training Settings}
\label{appendix: pre-train config}

The pre-training configurations of the models presented in the main content can be found in Table~\ref{tab: appendix-pretrain}. Generally, MR-HuBERT possesses a parameter count analogous to the original HuBERT model. We've made concerted efforts to mitigate the impact of incorporating an additional sampling module, which naturally adds more parameters. Specifically, we consistently employ a kernel size of 1 for both convolutional and de-convolutional layers in the sampling module, as elaborated in Section~\ref{ssec: sampling module}. Nonetheless, the model experiences a modest increase in parameter size, but this surge is less than 3\%. To ensure that the performance boosts highlighted in Section~\ref{sec: exp} aren't merely due to this increase, we've carried out comprehensive ablation studies, detailed in Appendix~\ref{appendix: ablation}.

In line with the insights from \citet{hsu2021hubert}, a more substantial batch size can typically augment model performance. In our research, when juxtaposing our method against the baselines, we've meticulously ensured that the batch size of our approach is either equivalent to or smaller than that of the baseline, to offset potential biases. All model training was executed on V100-32GB GPUs using the Fariseq toolkit \citep{ott2019fairseq}.

\begin{table}[t]
    \centering
    \caption{Detailed Hyper-parameters for models presented in main content.}
    \resizebox {\linewidth} {!} {
\begin{tabular}{l|c|ccc|cc|c}
\toprule
 & & \multicolumn{3}{|c}{Additional Baseline} &  \multicolumn{2}{|c}{Monolingual Models} &  \multicolumn{1}{|c}{Multilingual Models} \\
 & & \texttt{HuBERT-base}\textsuperscript{+} & \texttt{HuBERT-large}\textsuperscript{*} & \texttt{mHuBERT-base}\textsuperscript{*} &  \texttt{mono-base} & \texttt{mono-large} & \texttt{multi-base} \\
\midrule 

\multirow{10}{*}{Architecture} &  Num. Param (M) & 95 & 317 & 95 & 97 & 321 & 97 \\
& Transformer Layers & 12 & 24 & 12 & 4 * 3 & 8 * 3 & 4 * 3  \\
& - Attention Dim. & 768 & 1024 & 768 & 768 & 1024 & 768  \\
& - Linear Dim. & 3072 & 4096 & 3072 & 3072 & 4096 & 3072  \\
& - Attention Head & 12 & 16 & 12 & 12 & 16 & 12  \\
& Sampling Module & - & - & - & up+down & up+down & up+down  \\
& - Kernel size & - & - & - & 1 & 1 & 1  \\
& - Channel Size & - & - & - & 768 & 1024 & 768  \\
\cmidrule{3-8}
& Conv. Extractor & \multicolumn{6}{c}{[(512, 10, 5), (512, 3, 2) * 4, (512, 2, 2) * 2]} \\
& Mask Ratio & \multicolumn{6}{c}{0.8}  \\

\midrule
\multirow{10}{*}{Training} & Num. GPU & 32 & 128 & 32 & 32 & 128 & 32 \\
& Num. Frames & 100k  &  90k & 140k & 100k & 30k & 100k \\
& Grad. Accum. & 1 & 1 & 1 & 1 & 3 & 1  \\
& Num. Steps & 400k & 400k & 800k & 400k & 400k & 800k  \\
& Optimizer & Adamw & Adamw & Adamw & Adamw & Adamw & Adamw  \\
& Learning Rate & 0.0005 & 0.0015 & 0.0005 & 0.0005 & 0.0015 & 0.0005  \\
& Warmup Steps & 32k & 32k & 32k & 32k & 32k & 32k \\
& Dropout & 0.1 & 0.0 & 0.1 & 0.1 & 0.0 & 0.1 \\
& Loss Weights ($\beta, \gamma$) & - & - & - & (1, 1) & (1, 1) & (1, 1)  \\
& Audio Norm & true & false & true & true & false & true  \\

\bottomrule
\end{tabular}
}

    \label{tab: appendix-pretrain}
\end{table}

\section{Ablation Studies}
\label{appendix: ablation}

To garner an in-depth understanding of MR-HuBERT, we undertake extensive ablation studies. This ensures each component of MR-HuBERT is optimized and offers insight into their individual contributions to the model's superior performance. We delved into seven distinct conditions:
\begin{itemize}
    \item Encoder Layer Sizes: We explore the effect of varying the layer sizes for each encoder (Appendix~\ref{appendix: layer size}).
    \item Multi-Resolution Analysis: We evaluate the impact of utilizing multiple resolutions (Appendix~\ref{appendix: three-resolution}).
    \item Simpler Upsampling \& Downsampling Modules: A study into the implications of adopting a simplified upsampling or downsampling module is presented (Appendix~\ref{appendix: simplified updownsampling}).
    \item Single Prediction Target: Instead of multi-tasking, we scrutinize the outcome of using a singular prediction target (Appendix~\ref{appendix: single prediction target}).
    \item Single Resolution: The performance implications of deploying only one resolution are analyzed (Appendix~\ref{appendix: single resolution}).
    \item Compact Model: We test the efficacy of the model in a more compact setting (Appendix~\ref{appendix: small setting}).
    \item Target Units for Prediction: We investigate the repercussions of utilizing various target units for prediction (Appendix~\ref{appendix: units}).
\end{itemize}

The above ablations are all conducted in \textit{base} setting for efficiency, while we also conduct selected \textit{large} setting experiments in Appendix~\ref{appendix: large setting}.

As detailed in Section~\ref{ssec: asr}, we utilize the labeled LibriSpeech subsets of 1-hour, 10-hour, and 100-hour, as described in \citet{kahn2020libri}, for fine-tuning. The LibriSpeech evaluation sets serve as our testing grounds. All ASR results are presented using the word error rate. Prioritizing the quality of representation, we opt for Viterbi decoding over language model joint decoding. In addition to the ASR performance, we provide information on each model's parameter size and MACs. The calculation of MACs can be found in Section~\ref{ssec: speed}.

\subsection{Encoder Layer Sizes}
\label{appendix: layer size}

\begin{table}[t]
    \centering
    \small
    \caption{Ablation study configurations on different encoder layer sizes in the \textit{base} setting.}
\begin{tabular}{lccc}
\toprule
Model & Layers & Num. Param (M) & MACs (G) \\
\midrule 
\texttt{HuBERT-base}  & 12 & 95 & 431   \\
\texttt{HuBERT-base}\textsuperscript{+}  & 12 & 95 & 431  \\
\midrule
\textbf{\texttt{mono-base}}  & (4, 4, 4) & 97 & 394 \\
\midrule
\texttt{(B.1)-a} & (2, 4, 6) & 97 & 394 \\
\texttt{(B.1)-b} & (5, 2, 5) & 97 & 416 \\
\texttt{(B.1)-c} & (6, 4, 2) & 97 & 394 \\

\bottomrule
\end{tabular}

    \label{tab: appendix-different-layer-config}
\end{table}

\begin{table}[t]
    \centering
    \small
    \caption{Ablation study of differing encoder layer sizes for the \textit{base} setting. The experiments are conducted on ASR fine-tuning experiments over LibriSpeech subsets.}
\begin{tabular}{lccccc}
\toprule
Model & Layers & dev-clean & dev-other & test-clean & test-other\\
\midrule \midrule
\multicolumn{6}{c}{\textit{\textbf{1-hour labeled}}} \\
\texttt{HuBERT-base}  & 12  & 20.17  & 28.11  & 20.64  & 28.87 \\
\texttt{HuBERT-base}\textsuperscript{+}  & 12 &  19.64  & 25.08   & 20.15 & 25.63  \\
\midrule
\textbf{\texttt{mono-base}}  & (4, 4, 4)  &  18.78  & 23.72  & 19.26  & 24.46 \\
\midrule
\texttt{(B.1)-a} & (2, 4, 6) & 18.71 & 23.30 & 19.30 & 23.94  \\
\texttt{(B.1)-b} & (5, 2, 5) & 18.61 & \textbf{23.22} & \textbf{18.63} & \textbf{23.75} \\
\texttt{(B.1)-c} & (6, 4, 2) & \textbf{18.41} & 23.37 & 18.83 & 23.96  \\
\midrule \midrule
\multicolumn{6}{c}{\textit{\textbf{10-hour labeled}}} \\
\texttt{HuBERT-base}  & 12  &  9.62 & 16.60 & 9.71 & 17.00  \\
\texttt{HuBERT-base}\textsuperscript{+}  & 12  & 9.51 & 14.27 & 9.72 & 14.89 \\
\midrule
\textbf{\texttt{mono-base}}  & (4, 4, 4)  & 8.51  & 13.18  & 8.46  & 13.51 \\
\midrule
\texttt{(B.1)-a} & (2, 4, 6) &  8.61 & 13.33 & 8.54 & 13.64 \\
\texttt{(B.1)-b} & (5, 2, 5) & \textbf{8.30} & \textbf{12.96} & \textbf{8.38} & \textbf{13.42} \\
\texttt{(B.1)-c} & (6, 4, 2) & 8.71 & 13.24 & 8.71 & 13.72 \\
\midrule \midrule
\multicolumn{6}{c}{\textit{\textbf{100-hour labeled}}} \\
\texttt{HuBERT-base}  & 12  &  5.76 & 12.90 & 5.81 & 12.76   \\
\texttt{HuBERT-base}\textsuperscript{+}  & 12  & 5.71 & 10.66 & 5.97 & 10.87 \\
\midrule
\textbf{\texttt{mono-base}}  & (4, 4, 4)  & 4.89 & \textbf{9.04} & 4.92 & \textbf{9.17} \\
\midrule
\texttt{(B.1)-a} & (2, 4, 6) &  4.96 & 9.40 & 5.00 & 9.76 \\ 
\texttt{(B.1)-b} & (5, 2, 5) &  \textbf{4.65} & 9.22 & \textbf{4.78} & 9.44 \\
\texttt{(B.1)-c} & (6, 4, 2) &  5.11 & 9.80 & 5.10 & 9.90 \\
\bottomrule
\end{tabular}

    \label{tab: appendix-different-layer}
\end{table}
As discussed in Section~\ref{ssec: pretraining}, each encoder of MR-HuBERT maintains a consistent layer size. However, the impact of varied layer sizes for each encoder on the model's efficacy remains an open question. To address this, we explore the \textit{base} setting by altering layer counts.

The model configurations for this exploration are detailed in Table~\ref{tab: appendix-different-layer-config}. Across all new configurations, the parameter size remains consistent. Yet, in the \texttt{(B.1)-b} configuration, where low-resolution layers are minimized, the MACs rise to 416G from 394G.

The evaluation outcomes are tabulated in Table~\ref{tab: appendix-different-layer}. A key insight drawn from these results is that the \texttt{(B.1)-b} configuration excels in most LibriSpeech evaluation scenarios, especially when working with limited labeled data sets like the 1-hour and 10-hour subsets. This underscores the notion that while low-resolution modeling can effectively learn with fewer layers, the contribution of high-resolution comprehension remains pivotal to the overall model's success.

\subsection{Multi-Resolution Analysis}
\label{appendix: three-resolution}

\begin{table}[t]
    \centering
    \small
    \caption{Ablation study configurations on three-resolution MR-HuBERT in the \textit{base} setting.}
\begin{tabular}{lcccc}
\toprule
Model & Resolutions (ms) & Layers & Num. Param (M) & MACs (G) \\
\midrule 
\texttt{HuBERT-base}  & 20 & 12 & 95 & 431   \\
\texttt{HuBERT-base}\textsuperscript{+}  & 20 & 12 & 95 & 431  \\
\midrule
\textbf{\texttt{mono-base}} & (20, 40) & (4, 4, 4) & 97 & 393 \\
\midrule
\texttt{(B.2)-a} & (20, 40, 80) & (3, 2, 2, 2, 3) & 100 &  353 \\
\texttt{(B.2)-b} & (20, 40, 80) & (2, 2, 4, 2, 2) & 100 & 331 \\
\texttt{(B.2)-c} & (20, 40, 100) & (2, 2, 2, 2, 2) & 86 & 316 \\
\bottomrule
\end{tabular}

    \label{tab: appendix-multi-res-config}
\end{table}

\begin{table}[t]
    \centering
    \small
    \caption{Ablation study of three-resolution MR-HuBERT in the \textit{base} setting. The experiments are conducted on ASR fine-tuning experiments over LibriSpeech subsets.}
\begin{tabular}{lccccc}
\toprule
Model & Resolutions (ms) & dev-clean & dev-other & test-clean & test-other\\
\midrule \midrule
\multicolumn{6}{c}{\textit{\textbf{1-hour labeled}}} \\
\texttt{HuBERT-base}  & 20  & 20.17  & 28.11  & 20.64  & 28.87 \\
\texttt{HuBERT-base}\textsuperscript{+}  & 20 &  19.64  & 25.08   & 20.15 & 25.63  \\
\midrule
\textbf{\texttt{mono-base}}  & (20, 40)  &  \textbf{18.78}  & \textbf{23.72}  & \textbf{19.26}  & \textbf{24.46} \\
\midrule
\texttt{(B.2)-a} & (20, 40, 80) & 19.63 & 24.60 & 19.80 & 24.93  \\
\texttt{(B.2)-b} & (20, 40, 80) & 19.93 & 24.08 & 19.79 & 25.32 \\
\texttt{(B.2)-c} & (20, 40, 100) &  19.11 & 24.76 & 19.48 & 25.00 \\
\midrule \midrule
\multicolumn{6}{c}{\textit{\textbf{10-hour labeled}}} \\
\texttt{HuBERT-base}  & 20  &  9.62 & 16.60 & 9.71 & 17.00  \\
\texttt{HuBERT-base}\textsuperscript{+}  & 20  & 9.51 & 14.27 & 9.72 & 14.89 \\
\midrule
\textbf{\texttt{mono-base}}  & (20, 40)  & \textbf{8.51}  & \textbf{13.18}  & \textbf{8.46}  & \textbf{13.51} \\
\midrule
\texttt{(B.2)-a} & (20, 40, 80) &  8.63 & 14.19 & 8.84 & 14.31 \\
\texttt{(B.2)-b} & (20, 40, 80) &  8.81 & 14.34 & 8.90 & 14.61 \\
\texttt{(B.2)-c} & (20, 40, 100) & 9.34 & 15.08 & 9.48 & 15.15 \\
\midrule \midrule
\multicolumn{6}{c}{\textit{\textbf{100-hour labeled}}} \\
\texttt{HuBERT-base}  & 20  &  5.76 & 12.90 & 5.81 & 12.76   \\
\texttt{HuBERT-base}\textsuperscript{+}  & 20  & 5.71 & 10.66 & 5.97 & 10.87 \\
\midrule
\textbf{\texttt{mono-base}}  & (20, 40)  & 4.89 & \textbf{9.04} & 4.92 & \textbf{9.17} \\
\midrule
\texttt{(B.2)-a} & (20, 40, 80) &  \textbf{4.70} & 10.04 & \textbf{4.87} & 9.90 \\
\texttt{(B.2)-b} & (20, 40, 80) & 5.00 & 10.49 & 5.10 & 10.37 \\
\texttt{(B.2)-c} & (20, 40, 100) & 5.53 & 11.47 & 5.60 & 11.25 \\
\bottomrule
\end{tabular}

    \label{tab: appendix-multi-res}
\end{table}

While the main discussion primarily revolves around MR-HuBERT trained with two resolutions, this section explores its performance using three resolutions. This is to gauge the potential advantages or drawbacks of adopting more than two resolutions. Table~\ref{tab: appendix-multi-res-config} showcases that by adding a lower resolution, there's an increase in the parameter size to 100M, primarily due to the inclusion of extra sampling modules. However, MACs decrease further to values of 353G and 331G, contingent on layer distribution. In essence, incorporating more lower resolution components into MR-HuBERT provides the benefit of faster inference.

Table~\ref{tab: appendix-multi-res} presents the ASR results for the configurations with three resolutions. Despite showing marked improvement over baselines (i.e., \texttt{HuBERT-base} and \texttt{HuBERT-base}\textsuperscript{+}), the performance of MR-HuBERT with three resolutions isn't as robust as that of \textbf{\texttt{mono-base}}. This suggests that information from lower resolutions might not always enhance the ASR task. Given the efficiency gains observed, the inclusion of lower resolutions could be perceived as balancing efficiency against performance efficacy. It's worth noting that the performance dip observed in the three-resolution MR-HuBERT appears inconsistent with findings in \citep{shi23h_interspeech}. The latter study revealed that features fused from multi-resolution HuBERTs across varying resolutions can bolster ASR tasks. Our hypothesis is that this performance discrepancy might stem from each resolution's constrained model capacity. A deeper dive into this is required to determine if lower resolutions can indeed boost performance.

\subsection{Simpler Upsampling \& Downsampling Modules}
\label{appendix: simplified updownsampling}

\begin{table}[t]
    \centering
    \small
    \caption{Ablation study on simplified upsampling \& downsampling modules along with a singular prediction target in the \textit{base} setting. The experiments are conducted on ASR fine-tuning experiments over LibriSpeech subsets.}
\begin{tabular}{lccccc}
\toprule
Model & Note & dev-clean & dev-other & test-clean & test-other\\
\midrule \midrule
\multicolumn{6}{c}{\textit{\textbf{1-hour labeled}}} \\
\texttt{HuBERT-base}  & -  & 20.17  & 28.11  & 20.64  & 28.87 \\
\texttt{HuBERT-base}\textsuperscript{+}  & - &  19.64  & 25.08   & 20.15 & 25.63  \\
\midrule
\textbf{\texttt{mono-base}}  & -  &  18.78  & 23.72  & 19.26  & 24.46 \\
\midrule
\texttt{(B.3)-a} & Simple sampling &   \textbf{18.06} & \textbf{22.61} & \textbf{18.33} & \textbf{23.37} \\
\texttt{(B.4)-a} & Single target & 19.74 & 25.12 & 20.04 & 25.87 \\
\texttt{(B.4)-b} & Simple sampling + Single target & 19.02 & 24.30 & 19.40 & 24.94 \\
\midrule \midrule
\multicolumn{6}{c}{\textit{\textbf{10-hour labeled}}} \\
\texttt{HuBERT-base}  & -  &  9.62 & 16.60 & 9.71 & 17.00  \\
\texttt{HuBERT-base}\textsuperscript{+}  & -  & 9.51 & 14.27 & 9.72 & 14.89 \\
\midrule
\textbf{\texttt{mono-base}}  & - & 8.51  & 13.18  & \textbf{8.46}  & 13.51 \\
\midrule
\texttt{(B.3)-a} & Simple sampling &  \textbf{8.30} & \textbf{12.88} & 8.49 & \textbf{13.35}  \\
\texttt{(B.4)-a} & Single target & 9.43 & 14.49 & 9.52 & 14.99 \\
\texttt{(B.4)-b} & Simple sampling + Single target & 9.15 & 13.78 & 9.22 & 14.42 \\
\midrule \midrule
\multicolumn{6}{c}{\textit{\textbf{100-hour labeled}}} \\
\texttt{HuBERT-base}  & -  &  5.76 & 12.90 & 5.81 & 12.76   \\
\texttt{HuBERT-base}\textsuperscript{+}  & -  & 5.71 & 10.66 & 5.97 & 10.87 \\
\midrule
\textbf{\texttt{mono-base}}  & -  & \textbf{4.89} & \textbf{9.04} & \textbf{4.92} & \textbf{9.17} \\
\midrule
\texttt{(B.3)-a} & Simple sampling & 4.91 & 9.66 & 5.10 & 9.73 \\
\texttt{(B.4)-a} & Single target & 5.51 & 10.62 & 5.71 & 10.81 \\
\texttt{(B.4)-b} & Simple sampling + Single target & 5.21 & 10.00 & 5.46 & 10.34 \\
\bottomrule
\end{tabular}

    \label{tab: appendix-simple-sample-single-target}
\end{table}

As detailed in Section~\ref{ssec: sampling module}, our proposed architecture's sampling module employs a blend of upsampling and downsampling to achieve a flexible ratio between any two resolutions. However, when dealing with low resolutions that are evenly divisible by their corresponding high resolutions, there's no need to simultaneously deploy both the upsample and downsample modules. This simultaneous use introduces an unnecessary computational overhead. Given this, we delve into a more streamlined setting in this section: the upsampling module is dedicated solely to upsampling, and the downsampling module focuses only on downsampling. While this streamlined approach slightly curtails the computational load (reducing MACs from 394G to 390G) and marginally shrinks the parameter size (from 97M to 96M), it lacks the flexibility to handle unconventional ratios, such as 3:4, between resolutions.

The derived model, dubbed \texttt{(B.3)-a}, is subsequently fine-tuned for the ASR task, with outcomes presented in Table~\ref{tab: appendix-simple-sample-single-target}. From the results, it is evident that the MR-HuBERT equipped with the simplified sampling modules outperforms in low-resource situations, specifically the 1-hour and 10-hour ASR training scenarios. However, its performance isn't as consistent in the more extensive 100-hour experiment, particularly when juxtaposed against \textbf{\texttt{mono-base}}.

\subsection{Single Prediction Target}
\label{appendix: single prediction target}

As delineated in Section~\ref{ssec: objective}, our model incorporates a summation of masked unit prediction losses derived from all resolutions. In this subsection, we pivot to gauge the efficacy of deploying a singular masked unit prediction, sidelining the amalgamation of intermediate losses. Originating from \textbf{\texttt{mono-base}}, the resultant model, designated as \texttt{(B.4)-a}, benefits from an approximate reduction of 1M in parameter size. This reduction is achieved by discarding prediction heads assigned for the supplemental low-resolution masked unit prediction loss. Concurrently, we assess \texttt{(B.4)-b}, which melds the single prediction feature with the streamlined sampling module, as expounded upon in Appendix~\ref{appendix: simplified updownsampling}.

Both models, \texttt{(B.4)-a} and \texttt{(B.4)-b}, have their performance metrics tabulated in Table~\ref{tab: appendix-simple-sample-single-target}. Overall, a distinct performance hierarchy emerges: \texttt{(B.3)-a} outstrips \texttt{(B.4)-b}, which in turn surpasses \texttt{(B.4)-a}. This sequence underscores the indispensability of the multi-task objective spanning multiple resolutions for MR-HuBERT. Moreover, when navigating models fixated on a solitary prediction target, the elementary sampling modules exhibit more potency compared to their flexible counterparts.

\subsection{Single Resolution}
\label{appendix: single resolution}

A salient feature of MR-HuBERT is its concurrent utilization of diverse resolutions. In this subsection, we distill this multifaceted design down to a singular resolution. The intention behind this simplification is to probe the contributory essence of the multi-resolution concept to the model's efficacy. We harness the architectural blueprint delineated in Section~\ref{ssec: architecture}, albeit employing a consistent resolution across intermediate components. Consequently, this model forsakes the computational advantages derived from sequence reduction in self-attention calculations, culminating in a heightened computational overhead as reflected in the MACs of 439G. Intriguingly, this computational cost surpasses that of the native HuBERT, clocking in at 431G, as evidenced in Table~\ref{tab: appendix-single-resolution-small-model-config}.

\begin{table}[t]
\centering
\small
\caption{Ablation study configurations focusing on singular resolution and svelte model dimensions in the \textit{base} setting.}
\begin{tabular}{lcccc}
\toprule
Model & Layers & Resolutions (ms) & Num. Param (M) & MACs (G) \\
\midrule 
\texttt{HuBERT-base}  & 12 & 20 & 95 & 431   \\
\texttt{HuBERT-base}\textsuperscript{+}  & 12 & 20 & 95 & 431  \\
\midrule
\textbf{\texttt{mono-base}}  & (4, 4, 4) & (20, 40) & 97 & 394 \\
\midrule
\texttt{(B.5)-a} & (4, 4, 4) & (20, 20) & 97 & 439 \\
\texttt{(B.6)-a} & (3, 3, 3) & (20, 40) & 76 & 339 \\
\texttt{(B.6)-b} & (3, 3, 3) & (20, 20) & 76 & 373 \\

\bottomrule
\end{tabular}
\label{tab: appendix-single-resolution-small-model-config}
\end{table}

\begin{table}[t]
\centering
\small
\caption{Ablation study for singular resolution and svelte models within the \textit{base} context. The experiments are conducted on ASR fine-tuning experiments over LibriSpeech subsets.}
\begin{tabular}{lccccc}
\toprule
Model & MACs & dev-clean & dev-other & test-clean & test-other\\
\midrule \midrule
\multicolumn{6}{c}{\textit{\textbf{1-hour labeled}}} \\
\texttt{HuBERT-base}  & 431  & 20.17  & 28.11  & 20.64  & 28.87 \\
\texttt{HuBERT-base}\textsuperscript{+}  & 431 &  19.64  & 25.08   & 20.15 & 25.63  \\
\midrule
\textbf{\texttt{mono-base}}  & 394  &  \textbf{18.78}  & 23.72  & \textbf{19.26}  & 24.46 \\
\midrule
\texttt{(B.5)-a} & 439 & 18.87 & \textbf{23.37} &  19.69 & \textbf{24.05} \\
\texttt{(B.6)-a} & 339 & 18.73 & 24.40 & 19.37 & 24.78 \\
\texttt{(B.6)-b} & 373 & 19.41 & 25.32 & 19.67 & 26.00 \\
\midrule \midrule
\multicolumn{6}{c}{\textit{\textbf{10-hour labeled}}} \\
\texttt{HuBERT-base}  & 431  &  9.62 & 16.60 & 9.71 & 17.00  \\
\texttt{HuBERT-base}\textsuperscript{+}  & 431  & 9.51 & 14.27 & 9.72 & 14.89 \\
\midrule
\textbf{\texttt{mono-base}}  & 394 & \textbf{8.51}  & 13.18  & \textbf{8.46}  & 13.51 \\
\midrule
\texttt{(B.5)-a} & 439 & 8.56 & \textbf{12.73} & 8.69 & \textbf{12.89} \\
\texttt{(B.6)-a} & 339 & 9.13 & 14.87 & 9.36 & 15.22 \\
\texttt{(B.6)-b} & 373 & 9.13 & 14.43 & 9.38 & 14.92 \\
\midrule \midrule
\multicolumn{6}{c}{\textit{\textbf{100-hour labeled}}} \\
\texttt{HuBERT-base}  & 431  &  5.76 & 12.90 & 5.81 & 12.76   \\
\texttt{HuBERT-base}\textsuperscript{+}  & 431  & 5.71 & 10.66 & 5.97 & 10.87 \\
\midrule
\textbf{\texttt{mono-base}}  & 394  & \textbf{4.89} & \textbf{9.04} & \textbf{4.92} & \textbf{9.17} \\
\midrule
\texttt{(B.5)-a} & 439 & \textbf{4.89} & 9.46  &  4.93 & 9.59\\
\texttt{(B.6)-a} & 339  & 5.31 & 11.07 & 5.55 & 11.19 \\
\texttt{(B.6)-b} & 373 & 5.31 & 11.11 & 5.47 & 11.20 \\
\bottomrule
\end{tabular}
\label{tab: appendix-single-resolution-small-model}
\end{table}

The experimental results are cataloged in Table~\ref{tab: appendix-single-resolution-small-model}. Across the 100-hour ASR dataset, the proposed \textbf{\texttt{mono-base}} unambiguously outperforms its singular resolution counterpart, \texttt{(B.5)-a}. However, when venturing into the 1-hour and 10-hour ASR realms, the outcomes are more equivocal. Bearing both efficiency and performance in mind, these findings underscore the pivotal influence of multi-resolution strategies in bolstering MR-HuBERT's impressive performance benchmarks. Please also refer to Appendix~\ref{appendix: superb details}, where we identify more benefits from introducing multiple resolutions.

\subsection{Compact Model}
\label{appendix: small setting}

Motivated by the conspicuous performance advantage of MR-HuBERT over traditional HuBERT, we pivot our efforts towards crafting a more svelte version of MR-HuBERT, prioritizing computational economy. Eschewing the convention of a four-layer encoder, our pared-down MR-HuBERT, christened \texttt{(B.6)-a}, adopts a three-layer encoder scheme. This strategic recalibration augments inferential speed without significantly compromising on performance standards. The architectural nuances are delineated in Table~\ref{tab: appendix-single-resolution-small-model-config}. It's worth noting that our investigative purview extends to another optimized model, \texttt{(B.6)-b}, which amalgamates the principles of the single-resolution approach detailed in Section~\ref{appendix: single resolution}.

As revealed in Table~\ref{tab: appendix-single-resolution-small-model}, the compact iteration understandably possesses diminished modeling prowess, translating to a performance dip relative to \textbf{\text{mono-base}}. Yet, even with this inherent constraint, it remains competitive with the original HuBERT — a noteworthy feat considering the model operates with 20\% fewer parameters and realizes a 21\% enhancement in inference speed.

\subsection{Target Units for Prediction}
\label{appendix: units}

\begin{table}[t]
\centering
\small
\caption{Ablation study on different target units within the \textit{base} context. The experiments are conducted on ASR fine-tuning experiments over LibriSpeech subsets. \texttt{HuBERT-base-40} represents a model trained on 40ms resolution, whereas \texttt{HuBERT-base}\textsuperscript{0} denotes the model's first iteration trained with MFCC clusters. KM symbolizes the $K$-means algorithm with $K = 1000$, and Encodec units are denoted as Encodec-\{Frequency\}-\{No. Stream\}.}
\begin{tabular}{lcccc}
\toprule
Model & High-resolution & Low-resolution & dev-clean & test-clean \\
\midrule 
\texttt{HuBERT-base}  & KM(\texttt{HuBERT-base}\textsuperscript{0}) & - &  9.62 & 9.71   \\
\texttt{HuBERT-base}\textsuperscript{+}  & KM(\texttt{HuBERT-base}) & -  & 9.51 & 9.72  \\
\midrule
\textbf{\texttt{mono-base}}  & KM(\texttt{HuBERT-base}) & $\mathrm{Skip}(\text{KM(\texttt{HuBERT-base})})$ & \textbf{8.51}  & \textbf{8.46}  \\
\midrule
\texttt{(B.7)-a} & KM(\texttt{HuBERT-base}) & KM(\texttt{HuBERT-base-40}) &  9.20 & 9.36 \\
\texttt{(B.7)-b} & Encodec-50-1 & $\mathrm{Skip}\text{(Encodec-50-1)}$ & 26.98 & 27.34 \\
\texttt{(B.7)-c} & Encodec-50-1 & Encodec-25-1 & 18.74 & 19.15 \\
\texttt{(B.7)-d} & Encodec-50-2 & $\mathrm{Skip}\text{(Encodec-50-1)}$ & 27.56 & 28.19 \\
\bottomrule
\end{tabular}
\label{tab: appendix-target-units}
\end{table}

As delineated in Section~\ref{ssec: pretraining}, our approach favored skip-downsampling the designated high-resolution units to obtain target low-resolution units for the intermediate masked prediction supervision. This strategy emerged as the most efficacious in training MR-HuBERT effectively. Nevertheless, we ventured into exploratory ablations using alternative units. Given that direct skip-downsampling isn't inherently data-driven, we experimented with units extracted from the pre-trained 40ms-resolution HuBERT model, \texttt{HuBERT-base-40}, in alignment with the model architecture introduced by \citet{shi23h_interspeech}. Additionally, we leveraged units from the increasingly prevalent Encodec approach as elucidated by \citep{defossez2022high}. It's worth noting that our preliminary observations revealed suboptimal performance for most models, leading us to restrict our analysis to just the 10-hour training scenarios. Nonetheless, we present these findings to offer a repository of insights for curious researchers.

Refer to Table~\ref{tab: appendix-target-units} for detailed results. Interestingly, harnessing units from \texttt{HuBERT-base-40} didn't elevate performance. This leads us to conjecture that MR-HuBERT may exhibit sensitivity to the homogeneity of prediction targets spanning diverse resolutions. In the case of Encodec, the outcomes were less than stellar, suggesting that a localized acoustic discrete representation might not be synergistic with the semantic learning intricacies inherent in masked unit prediction.

\subsection{Large Settings}
\label{appendix: large setting}
\begin{table}[t]
\centering
\caption{Ablation study configurations in \textit{large} settings. Frames/Step is shown in the format of {Maximum Number of Frames} * {Gradient Accumulation}. The Label column represents the model to extract hidden states for unit discovery. Audio Norm. is whether to conduct audio normalization to the raw audio.}
\resizebox {\linewidth} {!} {
\begin{tabular}{lccccccc}
\toprule
Model & Frames/Step & Label & Audio Norm. &  Layers & Note & Num. Param (M) & MACs (G) \\
\midrule 
\texttt{HuBERT-large} & 90k * 1 & \texttt{HuBERT-base} & True &  24 & - & 316 & 1116\\
\texttt{HuBERT-large}\textsuperscript{*} & 90k * 1 & \texttt{HuBERT-base} & True & 24 & - &317 & 1116  \\
\midrule
\textbf{\texttt{mono-large}} & 30k * 3 & \texttt{HuBERT-base} & True & (8, 8, 8)  & - &321 & 971  \\
\midrule
\texttt{(B.8)-a} & 60k * 1 & \texttt{HuBERT-base} & False & (8, 8, 8) & - &321 & 971 \\
\texttt{(B.8)-b} & 60k * 1 & \texttt{HuBERT-base} & True & (8, 8, 8) & - &321 & 971 \\
\texttt{(B.8)-c} & 60k * 1 & \texttt{HuBERT-large} & True & (8, 8, 8) & - &321 & 971 \\
\texttt{(B.8)-d} & 30k * 8 & \texttt{HuBERT-large} & True & (8, 8, 8) & - &321 & 971 \\
\texttt{(B.8)-e} & 90k * 1 & \texttt{HuBERT-base} & True & (8, 8, 8) & - &321 & 971 \\
\texttt{(B.8)-f} & 90k * 1 & \texttt{HuBERT-large} & True & (8, 8, 8) & - &321 & 971 \\
\texttt{(B.8)-g} & 90k * 1 & \texttt{HuBERT-base} & True & (10, 4, 10) & - &321 & 1049 \\
\texttt{(B.8)-h} & 90k * 1 & \texttt{HuBERT-large} & True & (10, 4, 10) & - & 321 & 1049 \\
\texttt{(B.8)-i} & 80k * 1 & \texttt{HuBERT-base} & True & (8, 8, 8) & Simple Sampling &319 & 965 \\
\texttt{(B.8)-j} & 80k * 1 & \texttt{HuBERT-large} & True & (8, 8, 8) & Simple Sampling &319 & 965 \\
\bottomrule
\end{tabular}
}
\label{tab: appendix-large-setting-config}
\end{table}

\begin{table}[t]
\centering
\small
\caption{Ablation study in \textit{large} settings. The experiments are conducted on ASR fine-tuning experiments over LibriSpeech subsets.}
\begin{tabular}{lcccc}
\toprule
Model  & dev-clean & dev-other & test-clean & test-other\\
\midrule \midrule
\multicolumn{5}{c}{\textit{\textbf{1-hour labeled}}} \\
\texttt{HuBERT-large}  & 14.42  & 18.80  & 14.40  & 19.29  \\
\texttt{HuBERT-large}\textsuperscript{*}  & 15.09  & 18.20   & 14.90  & 18.05  \\
\midrule
\textbf{\texttt{mono-large}} & 6.44  & 10.94  & 6.37 & 11.41  \\
\midrule
\texttt{(B.8)-a} & 20.62 & 23.43 & 20.66 & 23.45 \\
\texttt{(B.8)-b} & 7.31 & 12.58 & 7.32 & 13.39 \\
\texttt{(B.8)-c} & 7.15 & 12.30 & 7.37 & 12.89 \\
\texttt{(B.8)-d} & 6.53 & 11.79 & 6.64 & 12.14 \\
\texttt{(B.8)-e} & 6.40 & 10.89 & 6.25 & 11.03 \\
\texttt{(B.8)-f} & 6.83 & 12.26 & 6.97  & 12.77 \\
\texttt{(B.8)-g} & \textbf{6.21} & \textbf{10.21} & \textbf{6.11} & \textbf{10.63} \\
\texttt{(B.8)-h} & 6.83 & 12.52 & 6.81 & 12.63 \\
\texttt{(B.8)-i} & 6.42 & 11.29 & 6.50 & 11.91 \\
\texttt{(B.8)-j} & 6.78 & 12.06 & 6.92 & 12.53 \\
\midrule \midrule

\multicolumn{5}{c}{\textit{\textbf{10-hour labeled}}} \\
\texttt{HuBERT-large} & 5.68 & 8.67 & 5.75  & 8.96  \\
\texttt{HuBERT-large}\textsuperscript{*} & 5.61  & 8.68 & 5.57  & 9.02  \\
\midrule
\textbf{\texttt{mono-large}} &  5.58  & 8.57 & 5.52  & 8.74 \\
\midrule
\texttt{(B.8)-a} & 6.07 & 8.97 & 5.89 & 9.37 \\
\texttt{(B.8)-b} & 5.93 & 8.80 & 5.87 & 9.26 \\
\texttt{(B.8)-c} & 5.79 & 8.83 & 5.79 & 9.03 \\
\texttt{(B.8)-d} & \textbf{5.48} & 8.34 & 5.48 & 8.66 \\
\texttt{(B.8)-e} & 5.73 & 8.62 & 5.62 & 8.91 \\
\texttt{(B.8)-f} & 5.68 & 8.64 & 5.52 & 8.77 \\
\texttt{(B.8)-g} & 5.58 & \textbf{8.17} & \textbf{5.41} & 8.66 \\
\texttt{(B.8)-h} & 5.49 & 8.28 & 5.45 & \textbf{8.60} \\
\texttt{(B.8)-i} & 5.77 & 8.75 & 5.63 & 8.99\\
\texttt{(B.8)-j} & 5.66 & 8.59 & 5.64 & 9.14 \\
\midrule \midrule

\multicolumn{5}{c}{\textit{\textbf{100-hour labeled}}} \\
\texttt{HuBERT-large}  & 3.11  & 6.01 & 3.14  & 6.15  \\
\texttt{HuBERT-large}\textsuperscript{*}  & 3.03 & 6.30  & 3.12 & 6.14  \\
\midrule
\textbf{\texttt{mono-large}}  & 3.06  & 6.04 & 3.01  & 5.98 \\
\midrule
\texttt{(B.8)-a} & 3.18 & 6.31 & 3.17 & 6.30 \\
\texttt{(B.8)-b} & 3.09 & 6.01 & 3.13 & 6.13  \\
\texttt{(B.8)-c} & 3.13 & 6.11 & 3.18 & 6.17 \\
\texttt{(B.8)-d} & \textbf{2.83} & 5.86 & 2.98 & 5.91 \\
\texttt{(B.8)-e} & 3.05 & 6.27 & 3.15 & 6.02 \\
\texttt{(B.8)-f} & 2.90 & 5.90 & 3.01 & \textbf{5.74} \\
\texttt{(B.8)-g} & 2.90 & \textbf{5.64} & \textbf{2.93} & 5.88 \\
\texttt{(B.8)-h} & 2.89 & 5.71 & 3.01 & 5.69 \\
\texttt{(B.8)-i} & 3.09 & 6.22 & 3.16 & 6.13 \\
\texttt{(B.8)-j} & 2.98 & 5.94 & 3.09 & 6.02 \\

\bottomrule
\end{tabular}
\label{tab: appendix-large-setting}
\vspace{-15pt}
\end{table}

In the context of \textit{large} settings, MR-HuBERT continues to be examined. Table~\ref{tab: appendix-large-setting-config} delineates ten candidate configurations in the \textit{large} settings. Consistently, all models are trained for 400k steps, analogous to \texttt{\textbf{mono-base}} and \textbf{\texttt{mono-large}}. These configurations not only probe further into the ablation conditions established in the \textit{base} settings but also explore factors specifically impacting the performance of MR-HuBERT in the \textit{large} settings. These encompass audio normalization to the raw audio, variations in batch size, and the adoption of different target unit sequences either from \texttt{HuBERT-base} or \texttt{HuBERT-large}\footnote{Layer 9 and Layer 15 are respectively chosen for \texttt{HuBERT-base} and \texttt{HuBERT-large} for unit discovery. Post this, units are derived from the $K$-means method, with $K=1000$.}. Owing to memory constraints on V100-32GB, four models, specifically \texttt{(B.8)-e}-\texttt{(B.8)-h}, are trained on 128 A100-80GB GPUs.

The results for the ASR experiments in \textit{large} settings are encapsulated in Table~\ref{tab: appendix-large-setting}. A distilled account of key findings is as follows:
\begin{itemize}
\item \textbf{Best performing system}: A mix of results can be discerned across LibriSpeech's four evaluation sets. However, on average, the model \texttt{(B.8)-g} stands out, chiefly due to its layer distribution modification: transitioning from the default (8, 8, 8) to (10, 4, 10). This resonates with findings in Appendix~\ref{appendix: layer size}, suggesting that depth isn't imperative for low-resolution modeling. Nonetheless, curtailing low-resolution layers inadvertently affects inference efficiency, as evidenced by the elevated MACs in Table~\ref{tab: appendix-large-setting-config}.
\item \textbf{Units from large models}: Predominantly, models trained on units from \texttt{HuBERT-large} outperform those reliant on \texttt{HuBERT-base} units. This aligns with the intuitive premise that \texttt{HuBERT-large} labels could potentially enrich the MR-HuBERT learning iteration.
\item \textbf{Batch size matters}: Corroborating the assertions of \citet{hsu2021hubert}, large batch sizes appear favorable for HuBERT training. A juxtaposition of \texttt{(B.8)-b} to \texttt{(B.8)-f} indicates that augmenting the batch size can potentially bolster MR-HUBERT's performance.
\item \textbf{Do use audio normalization}: Historically, audio normalization is typically applied in \textit{large} settings of speech self-supervised learning, while it's omitted in the \textit{base} settings. Our \texttt{(B.8)-a} model substantiates that audio normalization is quintessential for the successful training of \textit{large} setting models on vast unlabeled datasets.
\item \textbf{Simplified sampling is not recommended}: As elaborated in Appendix~\ref{appendix: simplified updownsampling}, models employing simplified sampling modules demonstrate performance metrics closely mirroring those integrating our flexible sampling modules. However, in \textit{large} settings, this parallelism breaks, revealing consistent enhancements when utilizing our tailored flexible sampling modules over the simplified versions.
\end{itemize}

\section{Inference Speed}

Although MACs offer a theoretical estimate of execution time, they are not always a reliable indicator of actual inference speed, particularly given the parallel processing capabilities of GPUs. To address this, we conduct empirical tests to compare theoretical predictions with real-world performance. We measure the inference speed in terms of 'tokens\_per\_second' using Fairseq on the Librispeech dev-clean set. This measurement is the average of ten times to account for variability in real-time execution.

Our findings, detailed in Table~\ref{tab: real-time}, reveal that MR-HuBERT models demonstrate a significant and consistent increase in speed compared to HuBERT models in both \textit{base} and \textit{large} settings. Notably, the model \texttt{(B.2)-c}, equipped with three resolutions, emerges as the fastest in terms of inference speed. This empirical evidence suggests a strong alignment between the MACs calculations presented earlier and the actual performance observed in real-world scenarios.

\begin{table}[t]
\centering
\caption{Real-time measurements on Librispeech dev-clean set.}
\begin{tabular}{lcc}
\toprule
Model  & MACs ($\downarrow$) & token\_per\_second ($\uparrow$) \\
\midrule
\texttt{HuBERT-base} & 431 & 5833 \\
\texttt{HuBERT-large} & 1116 & 2220 \\
\midrule
\textbf{\texttt{mono-base}} & 394 & 6310 \\
\textbf{\texttt{mono-large}} & 971 & 2505 \\
\midrule
\texttt{(B.1)-a} & 394 & 6293 \\
\texttt{(B.1)-b} & 416 & 5911 \\
\texttt{(B.1)-c} & 394 & 6299 \\
\texttt{(B.2)-a} & 353 & 6925 \\
\texttt{(B.2)-b} & 331 & 7332 \\
\texttt{(B.2)-c} & \textbf{316} & \textbf{7580} \\
\texttt{(B.3)-a} & 390 & 6435 \\
\texttt{(B.4)-a} & 394 & 6322 \\
\texttt{(B.4)-b} & 390 & 6450 \\
\texttt{(B.5)-a} & 439 & 5229 \\
\texttt{(B.6)-a} & 339 & 7096 \\
\texttt{(B.6)-b} & 373 & 6670 \\
\bottomrule
\end{tabular}
\label{tab: real-time}
\end{table}

\section{More in SUPERB Benchmark}
\label{appendix: superb details}

\subsection{SUPERB Score in SUPERB Benchmark}
\label{appendix: superb score}

The SUPERB score (i.e., $\mathrm{SUPERB}_s$ is a sophisticated metric designed to provide a standardized assessment across various tasks, each potentially with its own scoring system \citep{feng2023superb}. By employing linear interpolation between Mel filter banks feature ($\mathrm{FBank}$) scores and state-of-the-art (SOTA) representation scores, it normalizes scores across different scales. If a single task has multiple metrics, an intra-task average is computed, ensuring that tasks with a myriad of metrics don't dominate the overall score. Subsequently, an inter-task average is derived, guaranteeing each task's equal contribution to the final score. A scaling factor of 1000 amplifies readability. 
 For consistency, the score in this paper benchmarks against a static snapshot of the SUPERB leaderboard from August 15, 2023, as detailed in Table~\ref{tab: appendix-superb-calculation}. Thoughtfully, SUPERB score's design considers task difficulty, granting more weight to tasks where even small advancements signify significant progress. This approach ensures a balanced evaluation across varying tasks, highlighting the metric's comprehensive and fair nature.

\begin{table}[t]
\centering
\caption{Information to calculate SUPERB score in Section~\ref{ssec: superb evaluation}. All the results are from the SUPERB leaderboard on August 15, 2023.}
\resizebox {\linewidth} {!} {
\begin{tabular}{l|ccccccc|ccc}
\toprule
\multirow{2}{*}{Model} & \multicolumn{7}{c}{Understanding} &  \multicolumn{3}{|c}{Enhancement} \\
 & PR($\downarrow$) & ASR($\downarrow$) & IC($\uparrow$)  & KS($\uparrow$) & SF-F1($\uparrow$) & SF-CER($\downarrow$) & ST($\uparrow$)  & SE-STOI($\uparrow$) & SE-PESQ($\uparrow$) & SS($\uparrow$) \\
\midrule 
$\mathrm{FBank}$ & 82.00 & 23.18 & 10.44  & 8.63 & 69.64 & 52.92 & 2.32  & 0.94 & 2.55 & 9.23 \\
SOTA & 3.09 & 3.36 & 99.34  & 97.89 & 92.25 & 17.61 & 25.52  & 0.95 & 3.06 & 11.19 \\
\bottomrule
\end{tabular}
}
\label{tab: appendix-superb-calculation}
\end{table}

Let $\psi_{\tau, i}$ be the $i$th metrics for task $\tau$, $\psi_{\tau, i}(f)$ be the corresponding score of upstream model $f$, $\mathcal{T}$ be the set of tasks, and $I_\tau$ be the set of metrics for task $\tau$. Then, the detailed formulation is as:
\begin{equation}
    \mathrm{SUPERB}_s(f) = \frac{1000}{|\mathcal{T}|}\Sigma_\tau^\mathcal{T} \frac{1}{|I_\tau|}\Sigma_i^{I_\tau} \frac{\psi_{\tau,i}(f) -\psi_{\tau,i}(\mathrm{FBank})}{\psi_{\tau,i}(\mathrm{SOTA}) - \psi_{\tau,i}(\mathrm{FBank})}.
\end{equation}

\subsection{Voice Conversion in SUPERB Benchmark}

In voice conversion, self-supervised learning representations have become increasingly popular as intermediate features for speech generation, as demonstrated by notable works such as \citep{wang2022drvc, s3prl-vc-journal, huang2021s3prl, huang2021any, wu2022multilingual, choi2021neural, huang2023singing}. Drawing inspiration from \citet{tsai-etal-2022-superb}, we also extended our research to voice conversion tasks to examine the efficacy of our approach.

To achieve this, we largely followed the blueprint provided by the S3PRL recipe on the Voice Conversion Challenge 2020 (VCC2020) as detailed by \citep{yi20_vccbc}. In particular, our experiments employed the \texttt{Taco2-AR} model as the primary downstream mechanism, a model introduced by \citep{liu2020non}. The final waveform synthesis was facilitated by a pre-trained parallel WaveGAN-based vocoder, a method pioneered by \citep{yamamoto2020parallel}.

For our evaluation metrics, we leaned on Mean Cepstrum Distortion (MCD), WER for ASR, and ACC for SV, utilizing pre-trained models available within the S3PRL toolkit. Echoing the methodology behind the SUPERB score articulated in Appendix~\ref{appendix: superb score}, we derived a comprehensive score by averaging across all evaluation metrics.

The outcomes of these experiments are presented in Table~\ref{tab: appendix-superb-vc}. As an important side note, rather than directly referencing numbers from \citet{tsai-etal-2022-superb}, we opted to rerun the experiments for \texttt{HuBERT-base} and \texttt{HuBERT-large}. This decision stemmed from challenges faced in replicating the original outcomes, potentially due to variations in ASR checkpoints or tweaks in hyperparameter settings. According to the results, we observe marginal improvements in the \textit{base} setting, but worse performance in the \textit{large} setting. Our hypothesis is that the data might suffer from overfitting issues with the enhanced modeling power of the large model. We plan to delve deeper into this in subsequent research, with the aim to better harness the capabilities of MR-HuBERT for voice conversion.

\begin{table}[t]
\centering
\small
\caption{Voice conversion evaluation for the proposed method.}
\begin{tabular}{lcccc}
\toprule
Model  & MCD($\downarrow$) & ASR-WER($\downarrow$) & SV-ACC($\uparrow$) & $\mathrm{SUPERB}_{\text{vc}}$\\
\midrule
$\mathrm{FBank}$ & 8.47 & 38.30 & 77.25 & 0.0 \\
SOTA & 7.08 & 8.00 & 100.00 & 1000.0 \\
\midrule \midrule
\texttt{HuBERT-base} &  7.47 & 10.93 & 97.50 & 854.6 \\
\texttt{HuBERT-base}\textsuperscript{+} & 7.32 & \textbf{10.60} & 99.00 & 903.4 \\
\texttt{HuBERT-large} & 7.23 & 10.98 & \textbf{99.25} & 915.7 \\
\texttt{HuBERT-large}\textsuperscript{*} & 7.24 & 11.53 & \textbf{99.25} & \textbf{934.6} \\
\midrule
\textbf{\texttt{mono-base}} & \textbf{7.18} & 11.15 & \textbf{99.25} & 921.3 \\
\textbf{\texttt{mono-large}} & 7.56 & 11.93 & 98.50 & 851.3 \\
\bottomrule
\end{tabular}
\label{tab: appendix-superb-vc}
\end{table}

\subsection{Ablation Models in SUPERB Benchmark}
\label{appendix: ablation-models-superb}

In our aforementioned ablation studies, the evaluation was limited to the ASR performance of each model. This scope might not offer a comprehensive assessment, especially when considering the diverse objectives of different tasks. Hence, we extended our evaluation to encompass most models in the SUPERB benchmark, as detailed in Appendix~\ref{appendix: ablation}. The exhaustive results are cataloged in Table~\ref{tab: appendix-whole-superb}. Below, we provide concise discussions for each task:

\begin{itemize}
\item \textbf{PR, KS, SF, ST, and SS}: Across these five tasks, which target understanding and enhancement, respectively, MR-HuBERT consistently outshines HuBERT. There's a noticeable performance uplift across both \textit{base} and \textit{large} settings, corroborated by nearly all configurations in Appendix~\ref{appendix: ablation}.

\item \textbf{ASR}: In \textit{base} settings, models tend to surpass the baselines for ASR. However, the performance landscape shifts in the \textit{large} settings, often not in favor. Multiple factors could be responsible — perhaps the challenges of applying CTC to low-resolution, repeated features, or constraints from frozen representations. Given these observations as well as the exploration in Appendix~\ref{appendix: ablation}, a more sophisticated fusion strategy might be beneficial when leveraging MR-HuBERT as an upstream, or fine-tuning could be explored for speech recognition tasks.

\item \textbf{IC}: The \textit{base} models benefit from low-resolution data, yielding better intent classification accuracy. In contrast, despite one \textit{large} model setting a benchmark for accuracy, many configurations don't yield improvements. A plausible cause, discerned from training curves, could be overfitting on a limited dataset. A comprehensive study on larger intent classification datasets, such as SLURP \citep{bastianelli-etal-2020-slurp}, might offer clearer insights.


\item \textbf{SE}: In \textit{base} settings, MR-HuBERT consistently registers worse PESQ for SE, while the trend inverts in \textit{large} settings. We theorize that MR-HuBERT initially emphasizes semantic information. But as model size increases, its augmented high-resolution encoders facilitate finer local information processing. When these high-resolution encoders robustly learn local patterns, the model's generalization capabilities arguably supersede single-resolution counterparts, like the baseline HuBERT. This conjecture is supported by the SS task, where the \textit{large} MR-HuBERT demonstrates a significant edge over baselines, in contrast to the \textit{base} setting.



\end{itemize}

\begin{table}[t]
\centering
\caption{Ablation study on SUPERB Benchmark}
\resizebox {\linewidth} {!} {
\begin{tabular}{l|ccccccc|ccc}
\toprule
\multirow{2}{*}{Model} & \multicolumn{7}{c}{Understanding}  &  \multicolumn{3}{|c}{Enhancement} \\
 & PR($\downarrow$) & ASR($\downarrow$) & IC($\uparrow$) & KS($\uparrow$) & SF-F1($\uparrow$) & SF-CER($\downarrow$) & ST($\uparrow$) & SE-STOI($\uparrow$) & SE-PESQ($\uparrow$) & SS($\uparrow$) \\
\midrule  \midrule
\multicolumn{11}{c}{\textit{\textbf{Baseline}} (Section~\ref{ssec: pretraining})} \\
\texttt{HuBERT-base} & 5.40 & 6.42 & 98.34  & 96.30 & 88.53 & 25.20 & 15.53  & \textbf{0.94} & 2.58 & 9.36 \\
\texttt{HuBERT-base}\textsuperscript{+} & 4.56 & 6.34 & 98.39  & 96.46 & 89.12 & 23.10 & 16.33 & 0.93 & 2.55 & 9.72 \\
\texttt{HuBERT-large} & 3.54 & 3.62 & 98.76  & 95.29 & 89.81 & 21.76 & 20.01  & \textbf{0.94} & 2.64 & 10.45 \\
\texttt{HuBERT-large}\textsuperscript{*} & 3.59 & \textbf{3.53} & 98.73  & 97.70 & 89.88 & 22.51 & 20.02 & \textbf{0.94} & 2.65 & 10.61 \\
\midrule
\midrule
\multicolumn{11}{c}{\textit{\textbf{Proposed Method}} (Section~\ref{ssec: pretraining})} \\
\textbf{\texttt{mono-base}} & 4.16 & 5.76 & 98.68  & 96.49 & 88.96 & 23.59 & 16.94 & \textbf{0.94} & 2.55 & 9.92 \\
\textbf{\texttt{mono-large}} & \textbf{3.15} & 3.78 & 98.76  & \textbf{97.76} & 90.57 & \textbf{20.60} & 21.05  & \textbf{0.94} & \textbf{2.67} & 10.97 \\ 
\midrule
\midrule
\multicolumn{11}{c}{\textit{\textbf{Layer Size}} (Appendix~\ref{appendix: layer size})} \\
\texttt{(B.1)-a} & 4.20 & 5.87 & 98.71  & 96.59 & 89.42 & 22.95 & 16.96  & \textbf{0.94} & 2.51 & 9.64\\
\texttt{(B.1)-b} & 4.05 & 5.67 & 98.76  & 96.20 & 89.29 & 22.52 & 16.85  & \textbf{0.94} & 2.54 & 9.63 \\
\texttt{(B.1)-c} & 4.36 & 5.98 & 98.89  & 96.66 & 89.24 & 24.15 & 16.65  & \textbf{0.94} & 2.52 & 9.78 \\
\midrule
\midrule
\multicolumn{11}{c}{\textit{\textbf{Multi-Resolution Analysis}} (Appendix~\ref{appendix: three-resolution})} \\
\texttt{(B.2)-a} & 4.36 & 6.02 & 98.84  & 96.11 & 88.63 & 24.09 & 17.02  & \textbf{0.94} & 2.52 & 9.65 \\
\texttt{(B.2)-b} & 4.30 & 6.28 & 98.20  & 96.17 & 89.06 & 23.99 & 16.42  & \textbf{0.94} & 2.53 & 9.73 \\
\texttt{(B.2)-c} & 4.60 & 6.87 & 98.81 & 95.85 & 87.15 & 27.61 & 16.35 & \textbf{0.94} & 2.55 & 9.78 \\
\midrule
\midrule
\multicolumn{11}{c}{\textit{\textbf{Simpler Upsampling \& Downsampling Modules}}(Appendix~\ref{appendix: simplified updownsampling})} \\
\texttt{(B.3)-a} & 4.20 & 5.68 & 98.68  & 96.07 & 88.87 & 24.18 & 16.93 & \textbf{0.94} & 2.54 & 9.77 \\
\midrule 
\midrule
\multicolumn{11}{c}{\textit{\textbf{Single Prediction Target}} (Appendix~\ref{appendix: single prediction target})} \\
\texttt{(B.4)-a} & 4.40 & 6.31 & 98.18  & 96.62 & 89.48 & 23.74 & 16.45 & \textbf{0.94} & 2.56 & 9.94 \\
\texttt{(B.4)-b} & 4.50 & 6.30 & 98.84  & 96.59 & 89.08 & 23.78 & 16.53 & \textbf{0.94} & 2.55 & 9.78 \\
\midrule 
\midrule
\multicolumn{11}{c}{\textit{\textbf{Single Resolution}} (Appendix~\ref{appendix: single resolution})} \\
\texttt{(B.5)-a} & 4.20 & 5.78 & 98.34  & 96.36 & 88.23 & 24.53 & 16.57 & \textbf{0.94} & 2.54 & 9.76 \\
\midrule \midrule
\multicolumn{11}{c}{\textit{\textbf{Compact Model}} (Appendix~\ref{appendix: small setting})} \\
\texttt{(B.6)-a} & 4.39 & 6.25 & 98.50  & 96.17 & 88.10 & 25.41 & 15.65 & \textbf{0.94} & 2.50 & 9.80 \\
\texttt{(B.6)-b} & 5.09 & 6.06 & 97.94  & 95.46 & 88.67 & 24.48 & 15.51 & \textbf{0.94} & 2.53 & 9.80 \\
\midrule \midrule
\multicolumn{11}{c}{\textit{\textbf{Large Settings}} (Appendix~\ref{appendix: large setting})} \\
\texttt{(B.8)-a} & 3.36 & 3.96 & 98.81 & 96.98 & 90.36 & 21.60 & 19.86 & \textbf{0.94} & 2.66 & 10.45 \\
\texttt{(B.8)-b} & 3.29 & 4.05 & 97.73 & 97.44 & 90.27 & 21.80 & 20.17 & \textbf{0.94} & 2.65 & 10.53\\
\texttt{(B.8)-c} & 3.37 & 4.01 & 98.68 & 97.60 & 89.95 & 21.74 & 19.60 & \textbf{0.94} & 2.66 & 10.80 \\
\texttt{(B.8)-d} & 3.21 & 3.68 & 98.76 & 97.60 & 90.36 & 21.25 & \textbf{21.52} & \textbf{0.94} & \textbf{2.67} & \textbf{11.25} \\
\texttt{(B.8)-e} & 3.46 & 4.06 & 98.39  & 97.34 & 90.57 & 21.26 & 20.23  & \textbf{0.94} & 2.65 & 10.91\\
\texttt{(B.8)-f} & 3.40 & 4.02 & \textbf{99.05}  & 97.50 & 89.89 & 21.54 & 20.56  & \textbf{0.94} & 2.66 & 10.93 \\
\texttt{(B.8)-g} & 3.21 & 3.81 & 98.42  & 97.31 & 90.22 & 21.36 & 20.50 & \textbf{0.94} & \textbf{2.67} & 10.83\\
\texttt{(B.8)-h} & 3.18 & 3.92 & 98.39  & 97.14 & \textbf{90.64} & 20.65 & 20.43  & \textbf{0.94} & 2.66 & 10.72 \\
\texttt{(B.8)-i} & 3.41 & 4.15 & 98.39  & 97.14 & 89.95 & 22.28 & 19.89  & \textbf{0.94} & 2.65 & 10.97 \\
\texttt{(B.8)-j} & 4.93 & 3.99 & 98.34  & 97.27 & 89.64 & 22.45 & 20.17  & \textbf{0.94} & 2.66 & 10.91 \\
\bottomrule
\end{tabular}
}
\label{tab: appendix-whole-superb}
\end{table}

\begin{table}[t]
\centering
\small
\caption{Ablation study in categorical SUPERB score on SUPERB Benchmark}
\begin{tabular}{l|cc|c}
\toprule
Model & Understanding  &  Enhancement & General \\
\midrule  \midrule
\multicolumn{4}{c}{\textit{\textbf{Baseline}} (Section~\ref{ssec: pretraining})} \\
\texttt{HuBERT-base} & 861.2 & 98.20 & 670.4 \\
\texttt{HuBERT-base}\textsuperscript{+} & 876.9  & 150.2 & 695.2 \\
\texttt{HuBERT-large} & 932.6  & 456.0 & 813.4 \\
\texttt{HuBERT-large}\textsuperscript{*} & 936.2  & 501.5 & 827.5 \\
\midrule
\midrule
\multicolumn{4}{c}{\textit{\textbf{Proposed Method}} (Section~\ref{ssec: pretraining})} \\
\textbf{\texttt{mono-base}} & 885.8  & 195.0 & 708.7 \\
\textbf{\texttt{mono-large}} & 949.7  & 609.5 & 864.6 \\ 
\midrule
\midrule
\multicolumn{4}{c}{\textit{\textbf{Layer Size}} (Appendix~\ref{appendix: layer size})} \\
\texttt{(B.1)-a} & 888.4 & 80.4 & 686.4 \\
\texttt{(B.1)-b} & 889.5 &  127.7 & 699.1 \\
\texttt{(B.1)-c} & 881.9 & 134.8 & 695.1 \\
\midrule
\midrule
\multicolumn{4}{c}{\textit{\textbf{Multi-Resolution Analysis}} (Appendix~\ref{appendix: three-resolution})} \\
\texttt{(B.2)-a} & 881.0 & 101.7 & 686.2\\
\texttt{(B.2)-b} & 875.4 & 140.6 & 691.7 \\
\texttt{(B.2)-c} & 854.2 & 145.4 & 677.0 \\
\midrule
\midrule
\multicolumn{4}{c}{\textit{\textbf{Simpler Upsampling \& Downsampling Modules}}(Appendix~\ref{appendix: simplified updownsampling})} \\
\texttt{(B.3)-a} & 883.8 & 147.0 & 699.6 \\
\midrule 
\midrule
\multicolumn{4}{c}{\textit{\textbf{Single Prediction Target}} (Appendix~\ref{appendix: single prediction target})} \\
\texttt{(B.4)-a} & 878.0 & 219.5 & 713.4 \\
\texttt{(B.4)-b} & 878.1 & 184.1 & 704.6 \\
\midrule 
\midrule
\multicolumn{4}{c}{\textit{\textbf{Single Resolution}} (Appendix~\ref{appendix: single resolution})} \\
\texttt{(B.5)-a} & 877.1 & 163.3 & 698.7  \\
\midrule \midrule
\multicolumn{4}{c}{\textit{\textbf{Compact Model}} (Appendix~\ref{appendix: small setting})} \\
\texttt{(B.6)-a} & 863.6 & 133.0 &  680.9 \\
\texttt{(B.6)-b} & 864.6 & 156.8 & 687.7 \\
\midrule \midrule
\multicolumn{4}{c}{\textit{\textbf{Large Settings}} (Appendix~\ref{appendix: large setting})} \\
\texttt{(B.8)-a} & 934.6 & 487.4 & 822.8 \\
\texttt{(B.8)-b} & 934.3 & 484.3 & 821.8 \\
\texttt{(B.8)-c} & 931.4 & 565.2 & 839.9 \\
\texttt{(B.8)-d} & \textbf{951.1} & \textbf{686.4} & \textbf{885.0} \\
\texttt{(B.8)-e} & 937.7 & 596.8 & 852.5 \\
\texttt{(B.8)-f} & 938.9 & 584.3 & 850.3 \\
\texttt{(B.8)-g} & 940.8 & 576.9 & 849.8 \\
\texttt{(B.8)-h} & 942.2 & 543.8 & 842.6 \\
\texttt{(B.8)-i} & 929.6 & 598.5 & 846.8 \\
\texttt{(B.8)-j} & 928.3 & 598.4 & 845.8 \\
\bottomrule
\end{tabular}
\label{tab: appendix-whole-superb-cate}
\end{table}

While the preceding discussion predominantly centers on individual tasks, we consolidate categorical SUPERB scores in Table~\ref{tab: appendix-whole-superb-cate}. In aggregate terms, the apex model—contrary to the ASR fine-tuning experiments delineated in Appendix~\ref{appendix: ablation}—is \texttt{(B.8)-d}, which leverages labels from \texttt{HuBERT-large} and employs the maximum batch size of (30k * 8 * 128) frames (amounting to approximately 1920 seconds or 0.53 hours) per step.





\begin{figure}[t]
    \centering
    \tiny
    \begin{subfigure}[h]{\linewidth}
        \centering
        \begin{tikzpicture}[x=15mm,y=3mm]

\def\mymatrix{
{0.013009216,0.011555474,0.026162451,0.03970657,0.061050285,0.06932682,0.33035374,0.42755324},
{0.006602347,0.007570311,0.025713148,0.03533721,0.032334857,0.04237374,0.21160312,0.18932247},
{0.009563642,0.005457424,0.025988705,0.03946214,0.017451951,0.024606153,0.12694466,0.117053986},
{0.025129337,0.004143363,0.024581967,0.03805113,0.010592697,0.021918418,0.02702072,0.097877614},
{0.041423164,0.003652869,0.024507938,0.04237036,0.008837508,0.025775637,0.011079414,0.09619936},
{0.00012519,0.003822461,0.027796531,0.057739906,0.004972473,0.004673101,0.03382047,0.002569463},
{0.001942232,0.017863607,0.04149045,0.089573026,0.01585074,0.004806199,0.024041243,0.00517333},
{0.002429567,0.2188888,0.076045625,0.092312895,0.21646203,0.007196756,0.019587852,0.004293244},
{0.05924191,0.20291834,0.0927987,0.07627058,0.30453146,0.03654611,0.021016182,0.001539485},
{0.18058069,0.005206108,0.061283723,0.043985806,0.006953783,0.15270035,0.012344852,0.03157247},
{0.003966404,0.006563371,0.079589255,0.086671054,0.006147042,0.002703915,0.043480158,0.000141508},
{0.048962366,0.08699738,0.10388507,0.09006493,0.048875313,0.036424734,0.037210494,0.006303677},
{0.11240917,0.35514104,0.10852428,0.08497241,0.2417758,0.2636586,0.02958155,0.004041701},
{0.41734982,0.06608127,0.12279536,0.09106555,0.019080842,0.2903806,0.033522192,0.00170063},
{0.0772649,0.004138225,0.15883681,0.092416376,0.005083182,0.016908886,0.03839337,0.014657789},
}

\foreach \a [count=\i] in {PR,ASR,IC,KS,SF,ST,SE,SS} {
    \node[minimum width=15mm, minimum height=3mm] at (\i, 0) {\a};
}

\foreach \row [count=\y] in \mymatrix {
    \ifnum \y<16 
        \node[minimum width=15mm, minimum height=3mm] at (0,\y) {\y}; 
    \fi
    \foreach \value [count=\x] in \row {
        \pgfmathsetmacro\colorvalue{\value*300}
        \node[fill=yellow!\colorvalue!purple, minimum width=15mm, minimum height=3mm, text=white] at (\x,\y) {\pgfmathprintnumber[precision=4,fixed,zerofill]{\value}};
    }
}

\end{tikzpicture}
        \caption{\texttt{\textbf{mono-base}} (\textit{base} MR-HuBERT)}
        \label{fig: mono-base-layer-weight}
        \vspace{+10pt}
    \end{subfigure}
    \hfill
    \begin{subfigure}[h]{\linewidth}
        \centering
        \begin{tikzpicture}[x=15mm,y=3mm]

\def\mymatrix{
{0.009590062,0.014060478,0.01425023,0.04813291,0.046406474,0.07501776,0.23218067,0.36241823},
{0.007382365,0.008651595,0.014048547,0.02534612,0.03254369,0.03493266,0.13805646,0.15051846},
{0.000757392,0.004378268,0.017994527,0.03808522,0.014286022,0.01830709,0.13259235,0.10034375},
{4.90E-05,0.00251464,0.01390283,0.027608877,0.005171637,0.014634023,0.13020782,0.13084239},
{2.41E-05,0.001099406,0.010716648,0.03356243,0.002068263,0.012085595,0.07510678,0.07760321},
{1.80E-06,0.00114063,0.013191582,0.030878615,0.001556155,0.008885079,0.050495792,0.07278628},
{8.43E-05,0.004439817,0.014889402,0.039117333,0.005338541,0.007128295,0.0357657,0.037503876},
{0.010895385,0.06453234,0.06348733,0.10909898,0.07505468,0.009578668,0.032385264,0.018964482},
{0.06596085,0.20009229,0.17251521,0.11811831,0.19013652,0.046044134,0.027422009,0.006584437},
{0.13152687,0.29366347,0.27563792,0.15416194,0.38778943,0.1887135,0.028409408,0.005751078},
{0.34874502,0.38530856,0.17789465,0.13431904,0.23151031,0.42710444,0.03137273,0.009536441},
{0.42496514,0.0193854,0.11214129,0.105706714,0.006796226,0.15031682,0.033239763,0.027046558},
{1.78E-05,0.000733205,0.09932979,0.1358635,0.001342037,0.007251937,0.052765273,0.000100822},
}

\foreach \a [count=\i] in {PR,ASR,IC,KS,SF,ST,SE,SS} {
    \node[minimum width=15mm, minimum height=3mm] at (\i, 0) {\a};
}

\foreach \row [count=\y] in \mymatrix {
    \ifnum \y<14 
        \node[minimum width=15mm, minimum height=3mm] at (0,\y) {\y}; 
    \fi
    \foreach \value [count=\x] in \row {
        \pgfmathsetmacro\colorvalue{\value*300}
        \node[fill=yellow!\colorvalue!purple, minimum width=15mm, minimum height=3mm, text=white] at (\x,\y) {\pgfmathprintnumber[precision=4,fixed,zerofill]{\value}};
    }
}

\end{tikzpicture}
        \caption{\texttt{HuBERT-base}}
        \label{fig: hubert-base-layer-weight}
    \end{subfigure}
    \caption{Layer-weight analysis on SUPERB tasks over two \textit{base} models. The weights are the layer-wise weights after the Softmax function, which are trained together with downstream models as detailed in Section~\ref{ssec: superb evaluation}.}
    \label{fig : base-layer-weight-compare}
\end{figure}
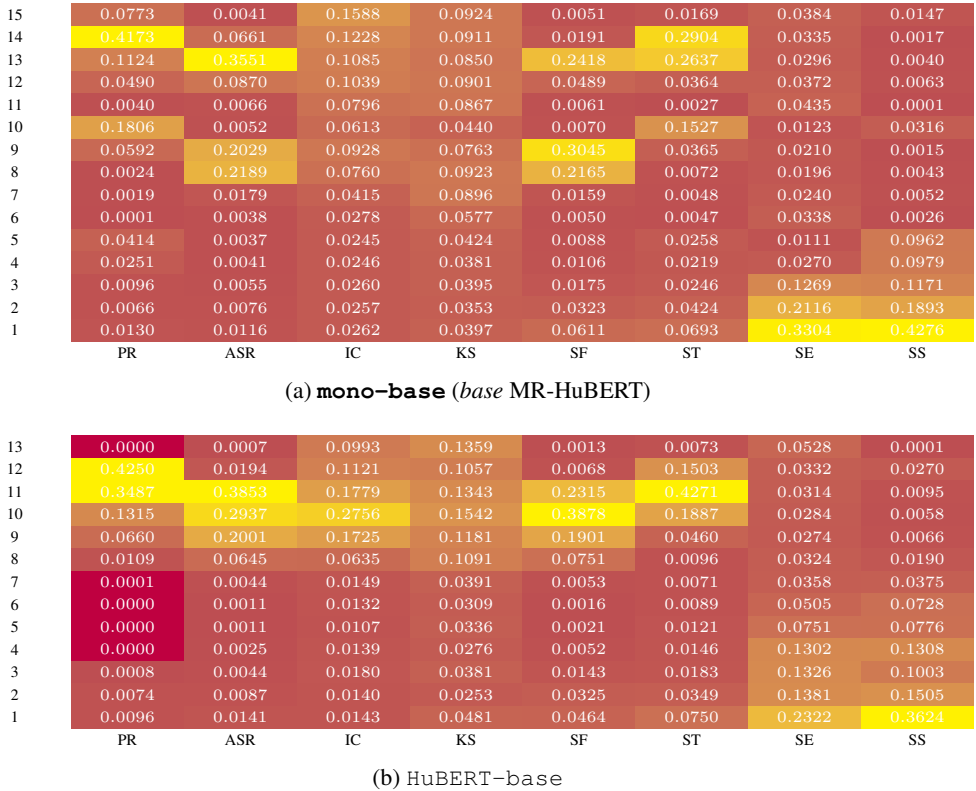

\begin{figure}[t]
    \centering
    \tiny
    \begin{subfigure}[h]{\linewidth}
        \centering
        \begin{tikzpicture}[x=15mm,y=3mm]

\def\mymatrix{
{0.000596848,0.009380179,0.025249183,0.03156229,0.010321653,0.003274405,0.07355955,0.028085582},
{2.36E-05,0.006799264,0.01852732,0.031970218,0.006357076,0.001505521,0.076034345,0.051538724},
{1.77E-07,0.003529122,0.016073558,0.033248823,0.002550928,0.000897049,0.094771005,0.09229099},
{1.78E-08,0.002225264,0.016906964,0.033714272,0.001426533,0.000720065,0.06729387,0.016144456},
{6.76E-09,0.001216153,0.012506147,0.03379509,0.000574427,0.000518184,0.09017882,0.03995649},
{3.98E-09,0.001084583,0.012207524,0.034999527,0.000529652,0.000464365,0.059495714,0.011710747},
{2.14E-09,0.001559101,0.014666071,0.039070103,0.000804393,0.000361177,0.061276916,3.74E-08},
{1.39E-09,0.001499034,0.016633675,0.04381287,0.000735526,0.000316936,0.062758215,6.05E-09},
{0.35562265,0.09115018,0.002430844,0.030682268,0.2845398,0.21942966,0.033467535,0.64351493},
{0.012828523,0.0573961,0.002447427,0.030360928,0.096173614,0.17659537,0.025060333,0.07237613},
{6.91E-08,0.000721662,0.005513409,0.02943354,0.000337952,0.000673696,0.013120694,1.10E-07},
{2.90E-08,0.000287259,0.005297306,0.030473359,9.64E-05,0.000362024,0.018046994,6.59E-08},
{1.68E-08,0.00067553,0.011366825,0.034080014,0.000122887,0.000318487,0.02182781,4.49E-08},
{1.75E-08,0.08875706,0.031664662,0.039868884,0.05939356,0.000352822,0.018856967,3.02E-08},
{1.66E-08,0.108500384,0.08274454,0.042413622,0.11884745,0.000832603,0.022131415,2.06E-08},
{1.16E-08,0.03252102,0.073832676,0.044114973,0.0257869,0.010433286,0.020111123,1.36E-08},
{2.91E-08,0.018593566,0.108982675,0.047067158,0.005126104,0.003837699,0.019589433,8.04E-09},
{0.47736636,0.16518582,0.028559553,0.03478688,0.15411897,0.36362842,0.0492052,0.04054903},
{6.75E-05,0.017531637,0.009965862,0.032372884,0.00489469,0.003739763,0.014659497,5.21E-08},
{9.73E-05,0.04245517,0.014628162,0.032797657,0.023351826,0.001943197,0.016898,9.71E-08},
{4.64E-08,0.122080915,0.030748421,0.034459084,0.08427861,0.004710977,0.016401963,1.06E-07},
{1.11E-08,0.07229211,0.027383529,0.036019824,0.059669916,0.016669696,0.015144309,9.99E-08},
{1.00E-08,0.04471235,0.042311408,0.04062838,0.016456109,0.10735415,0.014831585,8.78E-08},
{0.055349648,0.052478235,0.08044362,0.044084594,0.009955999,0.048034415,0.018781278,7.41E-08},
{0.07055843,0.018432366,0.06888003,0.04627233,0.00339853,0.022354482,0.017831596,1.37E-07},
{0.013183656,0.020271026,0.091023706,0.047736317,0.010379617,0.000816388,0.018473895,6.91E-08},
{0.014304983,0.01866494,0.14900494,0.04017411,0.019770956,0.009855116,0.040191863,0.003831849},
}

\foreach \a [count=\i] in {PR,ASR,IC,KS,SF,ST,SE,SS} {
    \node[minimum width=15mm, minimum height=3mm] at (\i, 0) {\a};
}

\foreach \row [count=\y] in \mymatrix {
    \ifnum \y<28 
        \node[minimum width=15mm, minimum height=3mm] at (0,\y) {\y}; 
    \fi
    \foreach \value [count=\x] in \row {
        \pgfmathsetmacro\colorvalue{\value*300}
        \node[fill=yellow!\colorvalue!purple, minimum width=15mm, minimum height=3mm, text=white] at (\x,\y) {\pgfmathprintnumber[precision=4,fixed,zerofill]{\value}};
    }
}

\end{tikzpicture}
        \vspace{-5pt}
        \caption{\texttt{\textbf{mono-large}} (\textit{large} MR-HuBERT)}
        \vspace{+5pt}
        \label{fig: mono-large-layer-weight}
    \end{subfigure}
    \hfill
    \begin{subfigure}[h]{\linewidth}
        \centering
        \begin{tikzpicture}[x=15mm,y=3mm]

\def\mymatrix{
{0.000842894,0.006187008,0.031366978,0.04310961,0.014190417,0.003531616,0.11361877,0.03464036},
{0.00027982,0.00353359,0.027512304,0.03211602,0.006232067,0.001511696,0.057538413,0.05235402},
{5.84E-05,0.001518401,0.027887218,0.03270651,0.003395474,0.000695098,0.07863104,0.059363984},
{1.53E-05,0.000829904,0.028694961,0.035797242,0.002364678,0.000551622,0.095830515,0.019618444},
{3.85E-06,0.000473178,0.027560527,0.031811256,0.001668883,0.000429796,0.07392272,0.031209957},
{6.69E-07,0.000244065,0.028705338,0.03334806,0.001084913,0.000352565,0.10487871,0.014097298},
{2.27E-07,0.000119398,0.02713994,0.029528106,0.000758753,0.000323197,0.082599804,0.024994826},
{5.20E-07,6.48E-05,0.028632274,0.028077284,0.000585833,0.000286968,0.041346896,0.009496566},
{1.75E-07,2.92E-05,0.029427474,0.028372029,0.00033753,0.00025928,0.0398077,0.008019053},
{1.08E-07,3.02E-05,0.02994621,0.034428205,0.000352178,0.000235432,0.03826197,0.004214089},
{2.86E-08,5.24E-05,0.027923632,0.03760926,0.000447991,0.000207439,0.031198487,0.000223411},
{2.54E-08,0.000361657,0.028287217,0.03877619,0.001246461,0.000189237,0.025956249,1.04E-05},
{2.72E-08,0.00294164,0.029055016,0.035685945,0.00444707,0.00018056,0.022639768,5.61E-07},
{4.37E-08,0.033908576,0.029647173,0.032419823,0.021638637,0.000171728,0.021440882,8.31E-08},
{5.67E-08,0.001579361,0.03166256,0.03271167,0.002821212,0.000163576,0.020001704,3.89E-08},
{2.73E-07,0.004005908,0.035464972,0.030311013,0.004648509,0.000168272,0.017545057,1.05E-08},
{1.55E-05,0.020619601,0.046832185,0.02809891,0.021380054,0.00019473,0.015602091,5.99E-09},
{1.61E-05,0.027540974,0.05488473,0.030495351,0.034722038,0.000295402,0.014982164,4.74E-09},
{6.50E-08,0.010131043,0.062390663,0.036065053,0.02109618,0.001057436,0.01535818,3.53E-09},
{7.27E-09,0.006985915,0.067182794,0.044467907,0.020051358,0.003494812,0.016904447,2.68E-09},
{1.52E-08,0.014376818,0.07254486,0.056075078,0.018015184,0.028327432,0.019466763,2.00E-09},
{0.005011491,0.019309187,0.0719004,0.06317745,0.010068512,0.027470412,0.014664339,1.46E-09},
{0.015993182,0.011040634,0.060269553,0.06165875,0.009980612,0.003004572,0.012885287,3.53E-09},
{2.34E-05,0.003533196,0.065418035,0.09659571,0.004935959,0.000330023,0.009687558,4.83E-10},
{0.9777378,0.8305835,0.029662963,0.046557605,0.79352945,0.92656714,0.015230508,0.7417569},
}

\foreach \a [count=\i] in {PR,ASR,IC,KS,SF,ST,SE,SS} {
    \node[minimum width=10mm, minimum height=3mm] at (\i, 0) {\a};
}

\foreach \row [count=\y] in \mymatrix {
    \ifnum \y<26
        \node[minimum width=15mm, minimum height=3mm] at (0,\y) {\y}; 
    \fi
    \foreach \value [count=\x] in \row {
        \pgfmathsetmacro\colorvalue{\value*300}
        \node[fill=yellow!\colorvalue!purple, minimum width=15mm, minimum height=3mm, text=white] at (\x,\y) {\pgfmathprintnumber[precision=4,fixed,zerofill]{\value}};
    }
}

\end{tikzpicture}
        \caption{\texttt{HuBERT-large}}
        \label{fig: hubert-large-layer-weight}
    \end{subfigure}
    \caption{Layer-weight analysis on SUPERB tasks over two \textit{large} models. The weights are the layer-wise weights after the Softmax function, which are trained together with downstream models as detailed in Section~\ref{ssec: superb evaluation}.}
    \label{fig : large-layer-weight-compare}
\end{figure}

\subsection{Layer Weights Analysis of SUPERB Benchmark}

As discussed in Appendix~\ref{appendix: ablation-models-superb}, we postulate that MR-HuBERT has implicitly prioritized different types of information across its resolutions. Intriguingly, the weighted summation approach in the SUPERB benchmark offers an insightful perspective into the layer-wise significance of the model for diverse downstream tasks. Prior works have employed these weights to ascertain the contribution of individual layers to specific downstream tasks \citep{chang2021exploration, chen2022unispeech, hung22_interspeech, chen22g_interspeech, mlsuperb, lin2023utility, shi23h_interspeech, otake2023parameter, chen2022wavlm}. Given that the weights of each layer participate in the backpropagation process, we surmise these weights can elucidate how each layer contributes to the final prediction in relation to the training objectives of each task.

Observing distinct behaviors between the models in both \textit{base} and \textit{large} configurations, we conduct separate comparisons for these two settings:

\noindent \textbf{\textit{Base} setting}: The juxtaposition of \textbf{\texttt{mono-base}} with \texttt{HuBERT-base} is illustrated in Figure~\ref{fig : base-layer-weight-compare}.\footnote{To clarify the distinction in layer numbers, MR-HuBERT encompasses not just the Transformer layers but also the outputs of the sampling module. Consequently, a two-resolution MR-HuBERT introduces two extra layers into the weighted summation computation during SUPERB downstream task training. Specifically, for \textit{\textbf{mono-base}}, the low-resolution layers span from layer 6 to layer 10.} In this comparison, both models manifest analogous behaviors. Broadly, echoing previous findings \citep{chen2022wavlm, chang2021exploration, chen2022unispeech}, layer weights are notably task-dependent: understanding tasks predominantly engage the later layers while enhancement tasks favor the initial layers. 

Yet, distinct layer weight distributions are palpable:
\begin{itemize}
    \item For the ASR task, while \texttt{HuBERT} predominantly targets its bottom layers (layers 9-11), \textbf{\texttt{mono-base}} allocates over 40\% of its attention to low-resolution layers 8 and 9. This inclination is explicable given the rich semantic content of low-resolution layers. This trait elucidates the pronounced contribution of layers 11-12 in MR-HuBERT for the PR task. Another distinction emerges in the SF task, where the semantic-centric slots in MR-HuBERT exhibit a propensity for low-resolution representations. 
    \item As tasks shift their focus away from native high-resolution representations, MR-HuBERT adeptly diminishes extraneous information, hinting at a potential for implicit speech disentanglement. For tasks like SE and SS, which emphasize local acoustics, the associated downstream models discern that the information from low-resolution layers (i.e., layers 5-9) isn't advantageous and pivot their attention to earlier layers—a contrast from the more varied layer focus observed in HuBERT.
\end{itemize}

\noindent \textbf{Comparison in the \textit{large} setting}: The behavior of models in the \textit{large} setting contrasts significantly with that in the \textit{base} setting. Figure~\ref{fig : large-layer-weight-compare} illustrates the comparison for \textit{large} models.\footnote{Recall from our discussion on the \textit{base} setting, MR-HuBERT incorporates two additional layers in the final prediction, positioning the low-resolution representations between layer 10 to layer 18.} We begin by evaluating each model individually before delving into a comparative analysis:

Three primary patterns emerge for \textbf{\texttt{mono-large}} when applied to SUPERB evaluation (See Figure~\ref{fig: mono-large-layer-weight}):
\begin{itemize}
\item \textbf{High-resolution encoder emphasis}: For tasks like SE and SS, which are associated closely with original audio signals, the first high-resolution encoder predominantly contributes.
\item \textbf{Low-resolution encoder emphasis}: Understanding tasks such as PR, ASR, SF, and ST predominantly lean on the second low-resolution encoder. Nonetheless, some information from the high-resolution encoders also plays a role, particularly when predictions align sequentially and emphasize semantic content.
\item \textbf{Equitable encoder distribution}: Tasks like IC and KS exhibit a balanced weight distribution across various encoders. Intriguingly, all these tasks revolve around speech classification.
\end{itemize}

For \texttt{HuBERT-large}, we discern three distinct trends (refer to Figure~\ref{fig: hubert-large-layer-weight}):
\begin{itemize}
\item \textbf{Top Layer Emphasis}: Tasks such as SE are heavily reliant on the top layers.
\item \textbf{Bottom Layer Emphasis}: Tasks including PR, ASR, SF, ST, and SS predominantly focus on the bottom layers.
\item \textbf{Diverse Layer Influence}: Tasks like IC and KS exhibit varied focus across different layers.
\end{itemize}

As for comparison, MR-HuBERT showcases a more nuanced understanding of speech signal intricacies, suggesting an implicit speech disentanglement. Conversely, HuBERT displays a rather arbitrary layer weight distribution. For instance, there's a pronounced emphasis on the final layer output for both understanding and speech separation tasks. The weight distribution patterns of MR-HuBERT hint at its potential to seamlessly transition into a more interpretable framework for speech representation studies.

\section{Delving Deeper into the ML-SUPERB Benchmark}
\label{appendix: ml-superb details}

\subsection{ML-SUPERB Score in ML-SUPERB Benchmark}

The ML-SUPERB score is derived as a linear-scaled average score of tasks spanning two specific leaderboards: the 10-minute and 1-hour leaderboards. Its computation is akin to that of the SUPERB score. Here, the scaling boundaries are defined by the $\mathrm{FBank}$ and the SOTA models. To ensure uniformity and to provide a holistic view of individual model performance, we reference the same leaderboard from the original ML-SUPERB paper when calculating the ML-SUPERB score \citep{mlsuperb}.

\begin{table}[t]
\centering
\small
\caption{Comparison of models within the ML-SUPERB evaluation.}

    \begin{tabular}{l|c|cc}
        \toprule
        \multirow{2}{*}{Model} & \multirow{2}{*}{Num. Params (M)} &  \multicolumn{2}{c}{Pre-Training}   \\
         &  &  Num. Hours &  Num. Languages \\
\midrule
\multicolumn{4}{c}{\textit{\textbf{wav2vec2-based}}} \\
\rowcolor{mono} \texttt{wav2vec2-base} \citep{baevski2020wav2vec}  & 95 & 1k & 1 \\
\rowcolor{mono} \texttt{wav2vec2-large} \citep{baevski2020wav2vec}  & 317 & 60k & 1 \\
\rowcolor{mono} \texttt{robust-wav2vec2-large} \citep{hsu21_interspeech}  & 317 & 65k & 1\\
\rowcolor{regional} \texttt{wav2vec2-base-23} \citep{wang2021voxpopuli}  & 95 & 100k & 23 \\
\rowcolor{regional} \texttt{wav2vec2-large-23} \citep{wang2021voxpopuli} & 317 & 100k & 23 \\
\rowcolor{multi} \texttt{XLSR-53} \citep{conneau2020unsupervised}  & 317 & 56k & 53 \\
\rowcolor{multi} \texttt{XLSR-128} \citep{babu2021xls}  & 317  & 400k & 128 \\
        \midrule
\multicolumn{4}{c}{\textit{\textbf{HuBERT-based}}} \\
\rowcolor{mono} \texttt{HuBERT-base} \citep{hsu2021hubert} & 95 & 1k & 1  \\
\rowcolor{mono} \texttt{HuBERT-base}\textsuperscript{+}  & 95 & 1k & 1  \\
\rowcolor{mono} \texttt{HuBERT-large} \citep{hsu2021hubert}  & 317 & 60k & 1 \\
\rowcolor{mono} \texttt{HuBERT-large}\textsuperscript{*} \citep{hsu2021hubert}  & 317 & 60k & 1 \\
\rowcolor{mono} \texttt{HuBERT-base-cmn}\footnote{Model accessed from Huggingface at \scriptsize{\url{https://github.com/TencentGameMate/chinese_speech_pretrain}}}  & 95 & 10k & 1\\
\rowcolor{mono} \texttt{HuBERT-large-cmn}  & 317 & 10k & 1 \\
\rowcolor{regional} \texttt{mHuBERT-base} \citep{lee2022textless}  & 95 & 14k & 3 \\
\rowcolor{regional} \texttt{mHuBERT-base}\textsuperscript{*}  & 95 & 100k & 23 \\
\midrule
\rowcolor{mono} \texttt{\textbf{mono-base}}  & 97 & 1k & 1 \\
\rowcolor{mono} \texttt{\textbf{mono-large}} & 97 & 60k & 1 \\
\rowcolor{regional} \texttt{\textbf{multi-base}} & 97 & 100k & 23 \\

        \bottomrule
    \end{tabular}

\label{tab: appendix-ml-superb-model}
\end{table}

The ML-SUPERB benchmark encompasses a diverse spectrum of models, each pre-trained with distinct configurations \citep{mlsuperb}. To render a comprehensive view of how our method stacks up against the competition, we amalgamated our data tables with the original ML-SUPERB leaderboard. The consolidated table, Table~\ref{tab: appendix-ml-superb-model}, offers insights into specific model configurations, highlighting their model parameters, pre-training data size, and linguistic diversity during pre-training. Previous studies on multilingual modeling underscore the advantage of a broader language spectrum \citep{hou20_interspeech, watanabe2017language, zhang22da_interspeech, chen2023improving, toshniwal2018multilingual, li2019bytes, gaur2021mixture, lugosch2022pseudo, mlsuperb}. Keeping this in mind, we've distinguished models based on their linguistic expanse: monolingual (blue), regional-multilingual (teal), and global-multilingual (yellow).

\begin{table}[t]
\centering
\caption{Complete ML-SUPERB Benchmark Results.}
\resizebox {\linewidth} {!} {
\begin{tabular}{l|c|cc|c|ccc|c}
\toprule
\multirow{3}{*}{SSL} & Monolingual ASR & \multicolumn{2}{c|}{Multilingual ASR} & \multicolumn{1}{c|}{LID} & \multicolumn{3}{c|}{Multilingual ASR + LID} & \multirow{3}{*}{SUPERB$_{s}$} \\
&          &             Normal & Few-shot & Normal & \multicolumn{2}{c}{Normal} & \multicolumn{1}{c|}{Few-shot} \\
& CER/PER & CER & CER & ACC & ACC & CER & \multicolumn{1}{c|}{CER} &  \\
\midrule \midrule
\multicolumn{8}{c}{\textit{\textbf{ML-SUPERB Benchmark}} \citep{mlsuperb}} \\
$\mathrm{FBank}$ & 72.1 / 63.7 & 62.4 / 59.3 & 58.3 / 57.4 & 11.1 / 9.3 & 35.9 / 43.5 & 62.0 / 58.6 & 58.9 / 58.1 & 0 / 0 \\
\rowcolor{mono} \texttt{wav2vec2-base} & 44.2 / 35.9 & 43.0 / 35.5 & 45.7 / 44.3 & 54.4 / 80.8 & 66.9 / 83.6 & 40.6 / 32.1 & 44.2 / 42.6 & 755.2 / 827.2 \\
\rowcolor{mono} \texttt{wav2vec2-large} & 42.0 / 35.4 & 42.6 / 35.7 & 45.8 / 43.9 &  30.9 / 8.0 & 54.6 / 78.2 & 45.5 / 34.7 & 50.3 / 42.2 & 598.3 / 586.9  \\
\rowcolor{mono} \texttt{robust-wav2vec2-large} & 44.4 / 35.7 & 40.1 / 31.1 & 45.4 / 42.2 & 50.8 / 72.1 & 33.1 / 62.9 & 38.6 / 33.7 & 44.9 / 46.0 & 680.3 / 768.6  \\
\rowcolor{regional} \texttt{wav2vec2-base-23} & 49.2 / 35.1  & 37.7 / 32.0 & 43.4 / 42.2 & 58.7 / 71.9 & 45.1 / 66.3 & 37.2 / 30.9 & 44.3 / 43.0 & 735.7 / 798.0 \\
\rowcolor{regional} \texttt{wav2vec2-large-23} & 42.0 / 34.2 & 42.1 / 35.3 & 44.3 / 42.4 & 1.1 / 64.2 & 21.8 / 49.7 & 43.4 / 35.2 & 46.1 / 43.1 & 433.8 / 724.9 \\
\rowcolor{multi} \texttt{XLSR-53} & 49.5 / 34.9 & 33.9 / 26.9 & 43.6 / 40.6 & 6.6 / 87.1 & 45.6 / 76.9 & 33.4 / 28.6 & 43.2 / 44.6 & 528.8 / 894.0 \\
\rowcolor{multi} \texttt{XLSR-128}  & 39.7 / \textbf{30.6} & \textbf{29.2} / \textbf{22.0} & 40.9 / 39.3 & \textbf{66.9} / \textbf{87.9} & 55.6 / 85.6 & \textbf{28.4} / \textbf{22.9} & 42.1 / 42.4 & 947.5 / \textbf{996.0} \\
\rowcolor{mono} \texttt{HuBERT-base}  & 42.8 / 35.3 & 39.8 / 31.4 & 44.5 / 42.7 & 61.2 / 86.1 & 71.5 / 86.0 &  39.2 / 30.9 & 43.8 / 41.8 & 831.9 / 884.9 \\
\rowcolor{mono} \texttt{HuBERT-large} & \textbf{38.2} / 32.2 & 44.4 / 37.7 & 48.2 / 43.5 &  46.5 / 64.1 & 55.4 / 77.7 & 45.6 / 35.1 & 49.3 / 42.2 & 678.7 / 783.6 \\
\rowcolor{mono} \texttt{HuBERT-base-cmn} & 43.1 / 35.3 & 40.8 / 31.4 & 45.4 / 42.7 & 49.3 / 86.1 & \textbf{75.1} / 86.1 & 37.7 / 30.9 & 43.5 / 41.8 & 779.0 / 810.2 \\
\rowcolor{mono} \texttt{HuBERT-large-cmn}  & 39.4 / 32.2 & 42.6 / 37.7 & 45.8 / 43.5 & 39.5 / 64.1 & 66.4 / 77.7 & 41.9 / 35.1 & 45.2 / 42.2 & 715.4 / 783.6 \\
\rowcolor{regional} \texttt{mHuBERT-base}  & 41.0 / 33.0 & 40.5 / 33.4 & 45.6 / 43.6 & 52.4 / 72.5  & 46.6 / 70.9 & 36.8 / 29.7 & 44.2 / 43.1 & 746.2 / 812.7 \\
\midrule \midrule
\multicolumn{8}{c}{\textit{\textbf{Additional Baseline}} (Section~\ref{ssec: pretraining})} \\
\rowcolor{mono} \texttt{HuBERT-base}\textsuperscript{+} & 42.9 / 35.3 & 41.5 / 31.2 & 45.8 / 42.8 & 63.8 / 81.9 & 70.1 / 85.8 & 39.6 / 31.3 & 44.6 / 40.7 & 819.1 / 875.8 \\
\rowcolor{mono} \texttt{HuBERT-large}\textsuperscript{*} & 41.2 / 32.6 & 42.8 / 32.8 & 45.6 / 42.5 & 42.3 / 58.9 & 59.2 / 84.7 & 42.3 / 29.8 & 44.1 / 41.4 & 704.5 / 817.6 \\
\rowcolor{regional} \texttt{mHuBERT-base}\textsuperscript{*} & 40.1 / 32.3 & 36.3 / 27.3 & \textbf{38.6} / 39.0 & 64.0 / 82.0 & 70.4 / 84.6 & 35.4 / 27.1 & \textbf{39.0} / 37.0 & 950.8 / 964.5 \\
\midrule \midrule
\multicolumn{8}{c}{\textit{\textbf{Proposed Method}} (Section~\ref{ssec: pretraining})} \\
\rowcolor{mono} \textbf{\texttt{mono-base}} & 42.8 / 34.6 & 40.2 / 30.6 & 45.0 / 42.2 & 67.2 / 86.3 & 68.7 / \textbf{86.9} & 40.3 / 30.6 & 44.1 / 41.6 & 843.5 / 899.9 \\
\rowcolor{mono} \textbf{\texttt{mono-large}} & 40.5 / 32.0  & 38.9 / 29.4 & 42.7 / 40.5 & 45.1 / 75.4 & 67.6 / 85.9 & 39.0 / 29.7 & 43.8 / 40.8 & 785.2 / 905.4  \\
\rowcolor{regional} \textbf{\texttt{multi-base}} & 38.3 / \textbf{30.6}  & 34.1 / 27.5  & 39.6 / \textbf{38.9} & 64.0 / 85.1 & 69.9 / 84.4 & 34.4 / 28.0 & 40.9 / \textbf{36.6} & \textbf{957.2} / 986.8  \\
\bottomrule
\end{tabular}
}

\label{tab: appendix-ml-superb-full}
\end{table}

Table~\ref{tab: appendix-ml-superb-full} provides an in-depth overview of performance metrics across benchmark tasks. To sum it up, our MR-HuBERT makes a commendable mark amidst the broader ML-SUPERB landscape. Within the monolingual category, our model conspicuously outpaces competitors—be it wav2vec2-based or HuBERT-aligned. Intriguingly, it even surpasses several multilingual counterparts, including the likes of \texttt{wav2vec2-base-23}, \texttt{wav2vec2-large-23}, and \texttt{XLSR-53}. This is particularly noteworthy given that these models benefit from vast datasets and broader linguistic diversity.

Navigating to the multilingual segment, our MR-HuBERT \textbf{\texttt{multi-base}} showcases a performance nearly on par with the frontrunner, \texttt{XLSR-128}, excelling in the 10-minute benchmark while slightly trailing in the 1-hour category. These outcomes are indeed remarkable, especially when accounting for our model's leaner parameters, compact pre-training data size, and reduced linguistic breadth. We anticipate MR-HuBERT to be instrumental in sculpting the future of multilingual modeling.

\section{Limitations}
\label{appendix: limitation}

\noindent \textbf{Dependency on Prior Models}: Instead of training from scratch, MR-HuBERT is predominantly trained using additional iterations from HuBERT discrete units. The potential of training MR-HuBERT from scratch, without leveraging previously trained models, remains unexplored.

\noindent \textbf{Performance Gaps in Specific Tasks}: While MR-HuBERT exhibits superior results compared to HuBERT, it lags behind WavLM, especially in enhancement tasks within the SUPERB framework \citep{chen2022wavlm}. The disparity might stem from differences in the training data and conditions. Notably, WavLM benefits from training on augmented unlabeled datasets that incorporate noise and other speech augmentations. Merging the MR-HuBERT framework with WavLM's training approach is a promising direction that warrants further investigation.

\noindent \textbf{Applicability to Non-Speech Audio Tasks}: Since MR-HuBERT's training centers around speech data, its efficacy diminishes for non-speech audio tasks, such as music or generic audio processing \citep{li2023mert, turian2022hear, liu2022audio, yuan2023marble, ma2023effectiveness}. This limitation surfaces when trying to deploy MR-HuBERT in contexts divergent from speech. Delving into a more holistic representation is crucial to achieve peak performance in a broad spectrum of audio tasks.

\section{Potential Extensions}
\label{appendix: future works}

In our examination, MR-HuBERT emerges as a promising alternative to existing speech pre-trained models. The outcomes highlight not only its immediate relevance but also hint at a host of future research directions:

\begin{itemize}
\item \textbf{Integration with Other Frameworks:} While MR-HuBERT primarily hinges on the HuBERT-style training, its multi-resolution architecture can potentially be fused with a variety of self-supervised frameworks, such as wav2vec2, WavLM, w2v-bert, w2v-bert2, and data2vec \citep{baevski2020wav2vec, chen2022wavlm, chung2021w2v, barrault2023seamlessm4t, baevski2022data2vec}.

\item \textbf{Diverse Resolutions:} Our experimental paradigm predominantly hinged on two-resolution MR-HuBERT, albeit with a cursory glance at a three-resolution approach. Delving deeper into varying resolution combinations might unearth optimal configurations tailored to specific use cases, such as higher resolutions for detailed acoustic analysis or lower resolutions for environmental information.

\item \textbf{Richer Representation:} HuBERT is renowned for its wide usage for extracting discrete semantic representations, facilitating tasks like resynthesis, voice conversion, and speech-to-speech translation \citep{li2023textless, huang2021any, sicherman2023analysing, polyak21_interspeech, lee2022direct, lakhotia2021generative, shi2021discretization, lin22c_interspeech, lian2022utts, nguyen2023expresso, shi2023enhancing, choi2023intelligible, inaguma-etal-2023-unity, barrault2023seamlessm4t, yan-etal-2023-espnet, huang2023singing}. As MR-HuBERT melds low-resolution layers for enriched semantics with high-resolution layers for nuanced acoustics, it can offer a more holistic representation. This multi-faceted view could be pivotal in enhancing speech quality in generative tasks.

\item \textbf{Speech Disentanglement:} Our insights, as dissected in Appendix~\ref{appendix: superb details}, highlight an implicit speech disentanglement capability in the \textit{large} MR-HuBERT model. Scaling up the model could amplify this characteristic. Furthermore, incorporating adversarial elements can engender explicit disentanglement, proving invaluable for tasks that necessitate isolating semantic or acoustic information from speech signals. We believe the architecture would be even better integrated with existing disentanglement approaches for self-supervised learning \citep{qian2022contentvec, chang23_interspeech}.

\end{itemize}

\end{document}